\RequirePackage[l2tabu, orthodox]{nag}

\documentclass[10pt,journal,compsoc]{IEEEtran}
\ifCLASSOPTIONcompsoc
  \usepackage[nocompress]{cite}
\else
  \usepackage{cite}
\fi
\usepackage[T1]{fontenc}
\usepackage[utf8]{inputenc}

\usepackage{amsmath,amssymb,amsfonts}
\usepackage{graphicx}
\usepackage[all]{nowidow}
\usepackage{colortbl}
\usepackage{xspace}
\usepackage[all]{nowidow}
\usepackage[inline]{enumitem}
\usepackage[scaled]{beramono}
\usepackage{booktabs} % For formal tables
\usepackage[colorinlistoftodos]{todonotes}
\usepackage[tight,footnotesize]{subfigure}
\usepackage{multirow}
\usepackage{balance}
\usepackage{listings}
\usepackage{microtype}
\usepackage{csquotes}

\usepackage{algorithm}% http://ctan.org/pkg/algorithms
\usepackage{algpseudocode}% http://ctan.org/pkg/algorithmicx

\usepackage{multirow}
\usepackage[many]{tcolorbox}
\usepackage{collectbox}
\usepackage{color,hyperref}
\usepackage{cleveref}
\usepackage{tikz}
\usepackage{collcell}
\usepackage[utf8]{inputenc}
\usepackage{xstring}
\usetikzlibrary{shadows}

\definecolor{darkblue}{rgb}{0.1,0.6,1.0}
\definecolor{red}{rgb}{1.0, 0.01, 0.24}
\definecolor{gray}{rgb}{0.8, 0.8, 0.8}
\hypersetup{colorlinks,breaklinks,
            linkcolor=darkblue,urlcolor=darkblue,
            anchorcolor=darkblue,citecolor=darkblue}

\ifCLASSINFOpdf
\else
\fi
\lstdefinelanguage{m3}{
	basicstyle=\ttfamily\scriptsize,
	keywordstyle=\bfseries,
	keywords={m3,declarations,methodInvocation},
	literate={<-}{$\leftarrow$}{1},
	tabsize=2,
	alsoletter={-}
}



\newcommand{\code}[1]{{\small \texttt{#1}}}

\newcommand*\circled[1]{\tikz[baseline=(char.base)]{\color{black} 
		\node[shape=circle,draw=cyan,fill=black!10!white,inner sep=.3pt] (char) {{{\texttt\textbf #1}}};}}

\usepackage{xspace}
\newcommand*{\ie}{i.e.,\@\xspace}
\newcommand*{\eg}{e.g.,\@\xspace}
\newcommand*{\cf}{cf.\@\xspace}
\newcommand*{\FC}{FOCUS\@\xspace}
\newcommand*{\UM}{UP-Miner\@\xspace}

\newcommand{\GH}{\textsc{GitHub}\xspace}

\makeatletter
\newcommand*{\etc}{%
	\@ifnextchar{.}%
	{etc}%
	{etc.\@\xspace}%
}
\makeatother
\newcommand*{\etal}{\emph{et~al.}\@\xspace}

\newcommand*{\numParticipants}{16\@\xspace}

\newcommand{\rqfirst}{\textbf{RQ$_1$}: \emph{How does \FC compare with \UM and PAM?}~}
\newcommand{\rqsecond}{\textbf{RQ$_2$}: \emph{How successful is \FC at providing recommendations at different stages of a development process?}~} %What are the completeness and accuracy of \FC recommendations?

\newcommand{\rqthird}{\textbf{RQ$_3$}: \emph{Is there a significant correlation 
between the cardinality of a category and accuracy?}~}%How significant is the %correlation between similarity and accuracy
\newcommand{\rqfourth}{\textbf{RQ$_4$}: \emph{Can \FC recommend relevant code snippets?}~}%Are the recommended code snippets useful for developers? Are the recommended API code snippets relevant?
\newcommand{\rqfifth}{\textbf{RQ$_5$}: \emph{How are \FC recommendations perceived by software engineers during a development task?}~} %Is \FC positively received by developers? %How is \FC received by developers?
\definecolor{verylightgray}{gray}{0.98}

\newtcolorbox{shadedbox}{
	drop shadow southeast,
	breakable,
	enhanced jigsaw,
	colback=white,
	boxrule=0.80pt,
	left=0.3em,
	right=0.3em,
	top=0.1em,
	bottom=0.05em
}

\definecolor{mygreen}{rgb}{0,0.6,0}
\definecolor{mygray}{rgb}{0.95,0.95,0.95}
\definecolor{myred}{rgb}{0.5,0,0}

\lstdefinestyle{JavaStyle} {
	backgroundcolor=\color{verylightgray},   
	commentstyle=\color{mygreen}, 
	breakatwhitespace=false,
	keywordstyle=\color{violet},
	language=Java,
	stringstyle=\color{blue},
	basicstyle=\tiny\ttfamily,
	showstringspaces=false}

\newcommand*{\etclst}{\colorbox{blue!10}{\textbf{\vphantom{a}\dots}}}

\newcommand*{\MinNumber}{0.00}%
\newcommand*{\MidNumber}{50.00} %
\newcommand*{\MaxNumber}{100.00}%
\newcommand*\mynum{}% used inside \ApplyGradient; just checking if it exists already
\definecolor{ao(english)}{rgb}{0.0, 0.5, 0.0}
\definecolor{cadmiumgreen}{rgb}{0.0, 0.42, 0.24}
\definecolor{emerald}{rgb}{0.31, 0.78, 0.47}

\newcommand{\ApplyGradient}[1]{%
	\IfDecimal{#1}{%
		\edef\mynum{#1}%
		\ifdim #1 pt > \MaxNumber pt\relax
		\edef\mynum{\MaxNumber}%
		\else
		\ifdim #1 pt < \MinNumber pt\relax
		\edef\mynum{\MinNumber}%
		\fi
		\fi
		\ifdim \mynum pt > \MidNumber pt
		\pgfmathsetmacro{\PercentColor}{max(min(100.0*(\mynum - \MidNumber)/(\MaxNumber-\MidNumber),100.0),0.00)}%
		\xdef\PercentColorr{\PercentColor}% must be global here if used in \cellcolor
		\cellcolor{white!\PercentColorr!emerald}#1%<- to have sth. of a `gradient'
		\else
		\pgfmathsetmacro{\PercentColor}{max(min(100.0*(\MidNumber - \mynum)/(\MidNumber-\MinNumber),100.0),0.00)}%
		\xdef\PercentColorr{\PercentColor}% must be global here if used in \cellcolor
		\cellcolor{white!\PercentColorr!emerald}#1%
		\fi	% \MinNumber < #1 <= \MaxNumber
	}{\textbf{#1}}% else it's not a decimal
}

\newcolumntype{R}{>{\collectcell\ApplyGradient}c<{\endcollectcell}}

\begin{document}

%\title{Supporting software development by recommending API function calls and usage patterns}

\title{Recommending API Function Calls and Code Snippets to Support Software Development}


\author{Phuong~T.~Nguyen,~%~\IEEEmembership{Member,~IEEE,}
		Juri~Di~Rocco,~
		Claudio Di Sipio,~
        Davide~Di~Ruscio,~%\IEEEmembership{Member,~IEEE,}
%		Lina~Ochoa,~%\IEEEmembership{Member,~IEEE,}
%		Thomas~Degueule,~%\IEEEmembership{Member,~IEEE,}
        and Massimiliano~Di~Penta~%\IEEEmembership{Life~Fellow,~IEEE}% <-this % stops a space
%		and~Tamás Gergely~
%the Department of Information Engineering, Computer Science and Mathematics, 
\IEEEcompsocitemizethanks{\IEEEcompsocthanksitem P.T. Nguyen, J. Di Rocco, C. Di Sipio, D. Di Ruscio are with Universit\`a degli Studi dell'Aquila, 67100-L'Aquila, Italy.\protect\\
E-mail: \{phuong.nguyen,juri.dirocco,claudio.disipio,davide.diruscio\}@univaq.it%see http://www.michaelshell.org/contact.html
%\IEEEcompsocthanksitem L. Ochoa and T. Degueule are with Centrum Wiskunde \& Informatica, Amsterdam, Netherlands. Email: \{firstname.lastname\}@cwi.nl
\IEEEcompsocthanksitem M. Di Penta is with Universit\`a degli Studi del Sannio, Benevento, Italy. Email: dipenta@unisannio.it}
%\IEEEcompsocthanksitem Tamás Gergely is with FrontEndART, Hungary. Email: tamas.gergely@frontendart.com}
}% <-this % stops an unwanted space
%\thanks{Manuscript received Month day, 2019; revised Month day, 2019.}

% <tamas.gergely@frontendart.com>

%Universit\`a degli Studi del Sannio} \\
%					Benevento, Italy \\
%					dipenta@unisannio.it


\markboth{ }
{Shell \MakeLowercase{\textit{et al.}}: Bare Demo of IEEEtran.cls for Computer 
Society Journals}



\IEEEtitleabstractindextext{%
\begin{abstract}
		Software development activity has reached a high degree of complexity, 
		guided by the heterogeneity of the components, data sources, and tasks. 
		The proliferation of open-source software (OSS) repositories has 
		stressed the need to reuse available software artifacts efficiently. To 
		this aim, it is necessary to explore approaches to mine data 
		from software repositories and leverage it to produce helpful 
		recommendations. We designed and implemented \FC as a novel approach to 
		provide developers with API calls and source code while they are 
		programming. %\FC uses a context-aware collaborative filtering 
		%technique to recommend relevant API function calls and real code 
		%snippets related to the tasks being developed. 
		The system works on the basis of a context-aware collaborative filtering technique to extract API usages from OSS projects. %API function calls and real code snippets related to the tasks being developed.
			In this work, we show the suitability of \FC for Android programming by evaluating it on a dataset of 2,600 mobile apps. The empirical evaluation results show that our approach outperforms two state-of-the-art API recommenders, \UM and PAM, in terms of prediction accuracy. We also point out that there is no significant relationship between the categories for apps defined in Google Play and their API usages. %success rate, accuracy, and execution time. %\MAX{todo update with results of other research questions}. 
		Finally, we show that participants of a  user study positively perceive 
		the API and source code recommended by \FC as relevant to the current 
		development context.
\end{abstract}

}

\maketitle

\IEEEdisplaynontitleabstractindextext

% \fi
\IEEEpeerreviewmaketitle

\IEEEraisesectionheading{\section{Introduction}\label{sec:introduction}}

% Context
% Learning APIs
%Developers therefore often face the need to learn new APIs.
%Leveraging the time-honored principles of modularity and reuse, modern software systems development typically entails the use of external libraries. 
% to integrate into their projects, libraries
%and employ recurrent sequence of API calls, known as API usage patterns developers look for, and try
% APIs, mining, existing tools, etc.
% We attempt to answer two 

When dealing with certain programming tasks, rather than implementing new 
systems from scratch, developers often make use of third-party libraries that 
provide the desired functionalities. Such libraries expose their functionality 
through Application Programming Interfaces (APIs) which govern the interaction 
between a client project and its incorporated libraries. To use a library in a 
proper way, it is necessary to use the correct sequence of API calls, known as 
API usage patterns. The knowledge needed to manipulate an API can be 
extracted from various sources: the API source code itself, the official 
website and documentation, Q\&A websites such as StackOverflow, forums and 
mailing lists, bug trackers, other projects using the same API, \etc. However, 
official documentation often merely reports the API description without 
providing non-trivial example usages. Besides, querying informal sources such 
as StackOverflow might become time-consuming and 
error-prone~\cite{robillard2009makes}. Also, API documentation may be 
ambiguous, incomplete, or erroneous~\cite{uddin2015api}, while API examples 
found on Q\&A websites may be of poor quality~\cite{nasehi2012makes}. In this 
respect, we come across the following motivating question: %. Open-source 
%software (OSS) 

%\vspace{.2cm}
%\framebox{
%	\parbox[t][1.1cm]{7.50cm}{
%%		\textbf{MQ$_1$}: 
%		\emph{``Which API methods should this piece of client code invoke, considering that it has already invoked these other API methods?''}
%	} 
%}
%\vspace{.2cm}
\begin{displayquote}	
\textquotedblleft \emph{Which API calls should this piece of code invoke, given that it has already invoked these API calls?}\textquotedblright
\end{displayquote}
%Previous research has shown how recommending API calls and code patterns facilitates the software development process. %Over the past decade, the problem of API learning has garnered considerable interest from the research community. 

The problem of recommending API function calls and usage patterns has 
garnered considerable efforts and attention of the research community in recent 
years \cite{Zhong2009MAPO,Moreno:2015:IUT:2818754.2818860}. Several 
techniques have been developed to automate the extraction of API \emph{usage 
patterns}~\cite{Robillard:2013:AAP:2498733.2498776} for reducing developers' 
burden when manually searching these sources, and for providing them with 
high-quality code examples. However, these techniques, based on 
clustering~\cite{Wang2013Mining, Zhong2009MAPO} or predictive 
modeling~\cite{Fowkes:2016:PPA:2950290.2950319}, still suffer from high 
redundancy and poor run-time performance. Moreover, most of the existing 
approaches are based on clustering on data from code snippets to recommend API 
usage, which sustains redundancy.

In an attempt to transcend the limitations, we proposed 
\FC~\cite{Nguyen:2019:FRS:3339505.3339636} as a novel approach to mining 
open-source software repositories to provide developers with API \emph{FunctiOn 
Calls and USage patterns}. We aim to suggest to developers highly relevant API 
usages that ease the development process. Our tool distinguishes itself from 
other tools that recommend API usages as it can provide both function calls and 
real code snippets that match well with the developer's context. \FC has been 
built based on a collaborative-filtering recommender 
system~\cite{Chen:2005:CCF:2154509.2154540}, whose fundamental principle is to recommend to users the items that have been bought by similar users in similar 
contexts. By considering API methods as products and client code as customers, 
we reformulate the problem of usage pattern recommendation in terms of a 
collaborative-filtering recommender system. We modeled the mutual relationships 
among projects using a tensor and mined API usage from the most similar 
projects.

%\revised{We propose \FC, a context-aware collaborative-filtering system that exploits cross relationships among OSS projects to suggest the inclusion of additional API invocations and concrete API usage patterns. Our tool distinguishes itself from other tools as it can recommend both function calls and real code snippets that match well with developer's context. Our evaluation on two datasets curated from the Maven repository shows that \FC achieves a good performance with respect to some quality indicators. The deployment of a context-aware recommender system to provide APIs recommendation is meaningful and promising.}

%Informally, the question the proposed system can answer is:
%\textbf{MQ1} Add a frame here to describe the motivation.
%
%\begin{quote}
%	\textit{``Which API methods should this piece of client code invoke, considering 
%	that it has already invoked these other API methods?"}
%\end{quote}

% Our specificities
Implementing a collaborative-filtering recommender system requires to assess the similarity of two customers, \ie~two projects.
Existing approaches assume that any two projects using an API of interest are equally valuable sources of knowledge.
On the contrary, we postulate that not all projects are equal when it comes to recommending usage patterns:~a project that is highly similar to the project currently being developed should provide higher quality patterns than a highly dissimilar one does.
Our recommender system attempts to narrow down the search scope by considering only the projects that are the most similar to the active project.
Thus, methods that are typically used conjointly by similar projects in similar contexts tend to be recommended first.

% \FC: context-aware collaborative-filtering 
%we presented the first version of \FC. 
%on different datasets comprising $610$ Java projects from GitHub and $3,600$ JAR archives from the Maven Central Repository. 

The first prototype of \FC~\cite{Nguyen:2019:FRS:3339505.3339636} has been successfully realized and 
integrated into the Eclipse IDE, and it is available for 
download.\footnote{\url{https://mdegroup.github.io/FOCUS-Appendix/}} An empirical 
evaluation has been conducted on a large number of Java projects extracted from 
\GH and the Maven Central repository to study \FC's performance, and to compare 
it with a state-of-the-art tool for API usage patterns mining, \ie 
PAM~\cite{Fowkes:2016:PPA:2950290.2950319}. We simulated different stages of a 
development process, by removing portions of client code and assessing how \FC 
can recommend snippets with API invocations to complete the code being 
developed. The experiments showed that \FC outperforms PAM, with regards to 
success rate, accuracy, and execution time.

% Evaluation / Results

%\vspace{.2cm}

%\mybox{
%	\textit{``How is the system beneficial?"}
%}

%\framebox{
%\parbox[t][2.0cm]{5.50cm}{
%%\addvspace{0.2cm} %\centering 
%How is the system beneficial?
%%$ p \;=\; -\, \dfrac{17}{23}\;;\; \quad q \;=\; \dfrac{10}{23} $ 
%} 
%}
%\revised{}

In this paper, we further extend the evaluation to study if \FC can 
assist mobile developers in finding relevant API function calls as well as real 
code snippets by means of an Android dataset. In this sense, the main 
contributions of our work are summarized as follows: \emph{(i)} We propose \FC 
as a practical solution to API recommendations employing a context-aware 
collaborative filtering recommender system; \emph{(ii)} Through a comprehensive 
evaluation on a dataset collected by mining \GH and Google Play, we show that \FC is applicable to Android 
programming, outperforming two state-of-the-art API recommenders; \emph{(iii)} We investigate how the calibration of the various \FC 
parameters can influence its performance; \emph{(iv)} We show that the system is 
capable of achieving good performance regardless of the availability of an extensive training set within the same app category; \emph{(v)}  
 By means of a clone detection evaluation and a user study, we show that \FC 
can be used to recommend real code snippets relevant for the program artifact being developed;
	%\MAX{add something about RQ4}
	%\emph{(v)} we perform a small-scale user study to validate the usability. 
	%\MAX{we need to drop the latter, right?}
	%To allow for future replication of our work, w
\emph{(vi)} Finally, we make available both the \FC tool and the datasets related 
to our paper to allow for future replication~\cite{focus-zenodo}. %More 
%importantly, we confirm that computing software similarity is important: ...} 
%%% 
%%of our work

The paper is organized as follows. Section~\ref{sec:Background} introduces a motivating example and background notions. Section~\ref{sec:ProposedApproach} brings in \FC, our proposed solution to API recommendation. The materials and methods used to evaluate the approach are presented in Section~\ref{sec:Evaluation}, while Section~\ref{sec:Results} analyzes the key results. The lessons learned and the threats to validity are discussed in Section~\ref{sec:Discussions}. We present related work and conclude the paper in Section~\ref{sec:RelatedWork} and Section~\ref{sec:Conclusions}, respectively.

%experiment settings as well as evaluation metrics
%\revised{Supports for mobile developers are highly desirable and stringent.}

%Furthermore, we concentrate on evaluating the usability of \FC by performing a large-scale user-study.

%\revised{The main contributions of the paper are summarized as follows:}
%recommender system for API mining, \FC,
% Outline

%	\setcounter{tocdepth}{2}
%	\tableofcontents

	\section{Background}
	\label{sec:Background}
	We describe a motivating example to introduce the problem addressed by \FC  in Section~\ref{sec:MotivatingExample}. Section~\ref{sec:API} gives an overview of the main components of the proposed solution; Afterwards, we introduce %in Section~\ref{sec:CACF} 
	the main notions underpinning our approach, by reviewing the previous work by \emph{Schafer \etal}~\cite{Schafer2007Collaborative} and \emph{Chen}~\cite{Chen:2005:CCF:2154509.2154540}.
%\MAX{Consider splitting section 2 into 2 separate sections, the first one titled ``motivating example'' and containing section 2.1, and then ``Definitions and background'' with 2.2. and 2.3}
%\begin{figure}[t!]

%	\centering
%	\includegraphics[width=\columnwidth]{figs/codeExample-original.pdf} \\
%	\scriptsize{(a) Initial version} \\ \vspace{.3cm}
%	\includegraphics[width=\columnwidth]{figs/codeExample-rec.pdf} \\
%	\scriptsize{(b) Final version}\\
%	\caption{Motivating example}
%	\vspace{-.2cm}
%	\label{fig:codeExample}
%\end{figure}

%\begin{lstlisting}[caption={Motivating example.}, label=lst:AugmentedCode, 
%style=JavaStyle,captionpos=b,xleftmargin=2.0em,frame=single,framexleftmargin=1.2em]
%public List<Boekrekening> findBoekrekeningen() {
%  CriteriaBuilder cb = entityManager.getCriteriaBuilder();
%  CriteriaQuery<Boekrekening> criteriaQueryBoekrekening = 
%  							cb.createQuery(Boekrekening.class);
%  Root<BoekrekeningPO> boekrekeningFrom = 
%  							criteriaQueryBoekrekening.from(BoekrekeningPO.class);
%  criteriaQueryBoekrekening.select(boekrekeningFrom);
%  criteriaQueryBoekrekening.orderBy(cb.asc(boekrekeningFrom.get(BoekrekeningPO_.rekeningnr)));
%
%  return entityManager.createQuery(criteriaQueryBoekrekening).getResultList();
%}	
%\end{lstlisting}

%. Fig.~\ref{fig:originalCode} depicts the situation

\subsection{Motivating Example} \label{sec:MotivatingExample}
\begin{figure*}[h!]
	\vspace{-.2cm}
	\centering    
	\begin{tabular}{c c}		
		\subfigure[Initial version]{\label{fig:originalCode}\includegraphics[width=88mm]{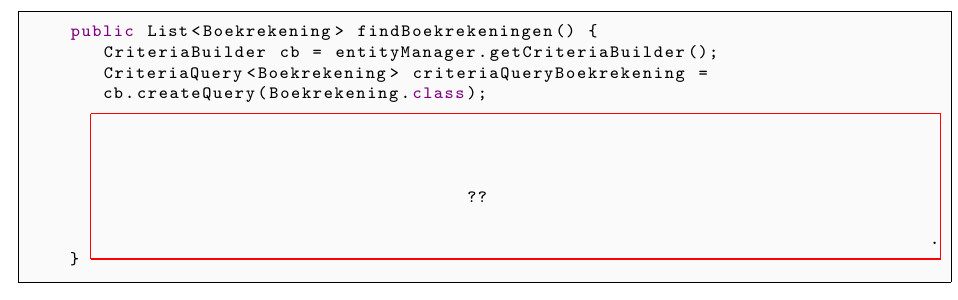}}  &
		\subfigure[Final version]{\label{fig:recCode}\includegraphics[width=88mm]{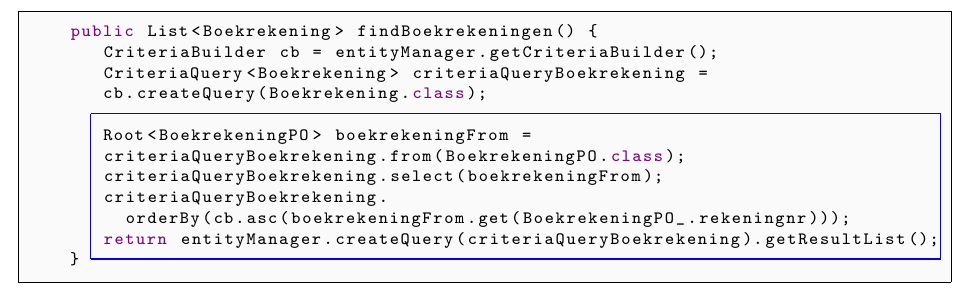}} 
	\end{tabular}
	\vspace{-.3cm}
	\caption{Motivating example.}
	\vspace{-.2cm}
\end{figure*}

The typical setting considered in the paper is shown in 
Fig.~\ref{fig:originalCode}: a programmer is implementing some methods to 
satisfy the requirements of the system being developed. The development is at 
an early stage, and the developer already used some methods of the chosen API 
to realize the required functionality. However, she is not sure how to proceed 
from this point. Under such circumstances, the programmer may browse different 
sources of information, including Stack Overflow, video tutorials, official API 
documentation, \etc.

%The snippet in 

Figure~\ref{fig:recCode} depicts the final version of the snippet in 
Fig.~\ref{fig:originalCode}. In the framed code, the 
\textit{findBoekrekeningen} method queries the available entities and retrieves 
those of type \textit{Boekrekening}. To this end, the \textit{Criteria API} 
library\footnote{\url{https://docs.oracle.com/javaee/6/tutorial/doc/gjivm.html}}
 is used as it provides useful interfaces for querying system entities 
according to the defined criteria. \FC has been conceptualized to do the 
exactly same thing: it is able to suggest to developers recommendations 
consisting of a list of API method calls that should be used next. Furthermore, 
it also recommends real code snippets that can be used as a reference to 
support developers in finalizing the method definition under development. %, 
%\eg code snippets that could support developers in finalizing the method 
%definition.
%\vspace{-.1cm}

% for completing the development of the method
%suggest. In this work, we propose an approach aiming to provide
%\subsection{API Function Calls and Usage Patterns} \label{sec:API}

\subsection{Definitions} \label{sec:API} 

``\textit{Collaborative Filtering} (CF) is the process of filtering or 
evaluating items through the opinions of other 
people''~\cite{Schafer2007Collaborative}.
In a CF system, a \textit{user} who buys or uses an \textit{item} attributes a 
rating to it based on her experience and perceived value.
Therefore, a \textit{rating} is the association of a user and an item through a 
value in a given unit (usually in scalar, binary, or unary form).
The set of all ratings of a given user is also known as a \textit{user 
profile}~\cite{Chen:2005:CCF:2154509.2154540}.
Moreover, the set of all ratings given in a system by existing users can be 
represented in a so-called \textit{rating matrix}, where a row represents a 
user, and a column represents an item. 

The expected outcome of a CF system is a set of predicted ratings (aka. \textit{recommendations}) for a specific user and a subset of items~\cite{Schafer2007Collaborative}.
The recommender system considers the most similar users (aka. \textit{neighbors}) to the \textit{active} user to suggest new ratings.
A similarity function $sim_{usr}(u_a, u_j)$ computes the \textit{weight} of the active user profile $u_a$ against each of the user profiles $u_j$ in the system. 
Finally, to suggest a recommendation for an item $i$ based on this subset of similar profiles, the CF system computes a weighted average $r(u_a, i)$ of the existing ratings, where $r(u_a, i)$ varies with the value of $sim_{usr}(u_a, u_j)$ obtained for all neighbors~\cite{Chen:2005:CCF:2154509.2154540, Schafer2007Collaborative}.

\textit{Context-aware CF} systems compute recommendations based not only on neighbors' profiles but also on the \textit{context} where the recommendation is demanded.
Each rating is associated with a context~\cite{Chen:2005:CCF:2154509.2154540}.
Therefore, for a tuple $C$ modeling different contexts, a \textit{context similarity} metric $sim_{ctx}(c_a, c_i)$, for $c_a, c_i \in C$ is computed to identify relevant ratings according to a given context.
Then, the weighted average is reformulated as $r(u_a, i, c_a)$~\cite{Chen:2005:CCF:2154509.2154540}.

\vspace{.2cm}
	
%	\section{\revised{\FC: Recommending Relevant API Calls and Source Code}}%
	\section{Proposed Approach}
	\label{sec:ProposedApproach}
	To tackle the problem of recommending API function calls and usage patterns, we 
leverage the wisdom of the crowd and existing recommender system techniques.
In particular, we hypothesize that API calls and usages can be mined from 
existing codebases, prioritizing the projects that are similar to the one from 
where the recommendation is demanded. We start with a definition of the main components of our approach in Section~\ref{sec:InputData}. Afterwards, Section~\ref{sec:Architecture} presents in detail the conceived architecture.

\subsection{Input data} \label{sec:InputData}

A \textit{software project} is a standalone source code unit that performs a set of tasks. Furthermore, an \textit{API} is like a black-box, \ie an interface that abstracts the piece of functionality offered by a project by hiding its implementation details. This interface is meant to support reuse and modularity~\cite{Parnas1971Information,robillard2009makes}. An API $X$ built in an object-oriented programming language, \eg~the \textit{Criteria API} in Fig.~\ref{fig:originalCode}, consists of a set $T_X$ 
of public types, \eg~\textit{CriteriaBuilder} and \textit{CriteriaQuery}.
Each type in $T_{X}$ consists of a set of public methods and fields that are available to 
client projects, \eg the method \textit{createQuery} of the type 
\textit{CriteriaQuery}.

A \textit{method declaration} consists of a name, a (possibly empty) list of types of parameters, a return type, and a (possibly empty) body, for example the 
\textit{findBoekrekeningen} method in Fig.~\ref{fig:recCode}. Given a set of declarations $D$ in a project $P$, an API \textit{method invocation} $i$ is 
a call made from a declaration $d \in D$ to another declaration $m$. Similarly, an API \textit{field access} is an access to a field $f \in F$ from a declaration $d$ in $P$.
API method invocations $MI$ and field accesses $FA$ in $P$ form the set of \text{API usages} $U = MI \cup FA$. Finally, an \textit{API usage pattern} (or code snippet) is a sequence $( u_1, 
u_2, ..., u_n )$, $\forall u_k \in U$. For the sake of presentation, in the scope of this paper the following terms are used interchangeably: \emph{method declaration} vs. \emph{declaration} and \emph{API} vs. \emph{invocation}. For each declaration, we extract its method name, a list of types of the parameters, and a list of API function calls. In this way, a project is represented as a set of declarations from its constituent classes.

%~\cite{Nguyen2014Statistical}.

%More specifically, 

Our tool makes use of a context-aware collaborative-filtering technique to search for invocations from highly relevant projects. This allows us to consider both project and declaration similarities to recommend APIs and code snippets. Following the terminology of recommender systems~\cite{Chen:2005:CCF:2154509.2154540}, we treat \textit{projects} as  the enclosing \textit{contexts}, \textit{method declarations} as \textit{users}, and \textit{method invocations} as \textit{items}. Intuitively, we recommend a method invocation for a declaration in a given  project, which is analogous to recommending an item to a customer in a specific context. For instance, the set of method invocations and the usage pattern (\cf framed  code in Fig.~\ref{fig:recCode}) recommended for the declaration \code{findBoekrekeningen} can be obtained from a set of similar projects and declarations in a codebase. The \textit{collaborative} aspect of the approach enables to extract recommendations from the most similar projects, while the \textit{context-awareness} aspect enables to narrow down the search space further to similar declarations.

\begin{figure*}%[b!]
	\centering
	\includegraphics[width=0.680\textwidth]{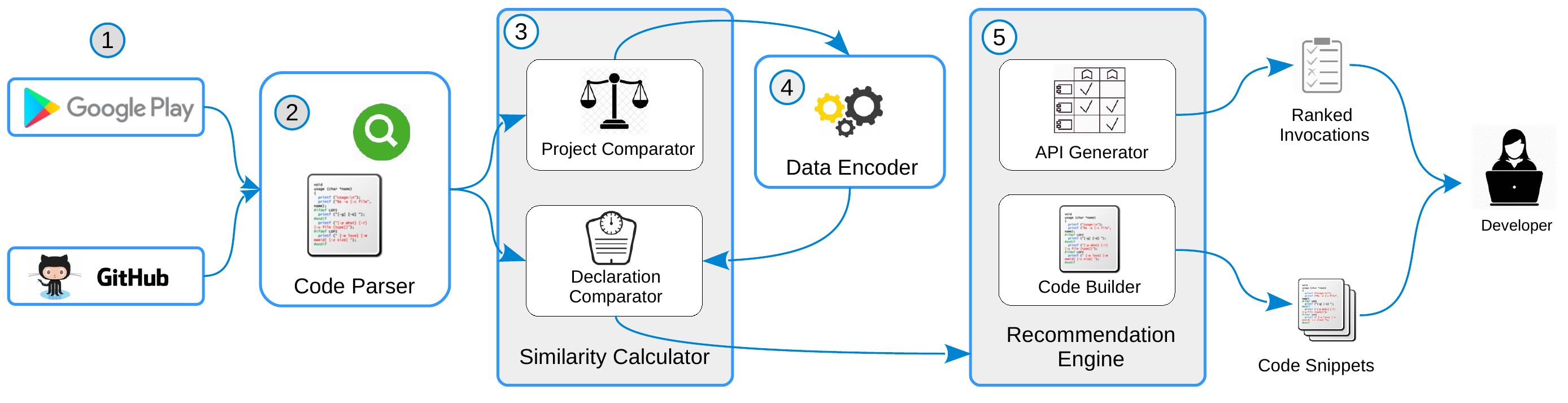} %FocusArchV2.pdf
	\vspace{-.2cm}
	\caption{Overview of the \FC architecture.} %apiAdvisor
	\vspace{-.4cm}
	\label{fig:Architecture}
\end{figure*}

\vspace{.2cm}
\subsection{Architecture} \label{sec:Architecture}
The architecture of \FC is depicted in Fig.~\ref{fig:Architecture}. 
To provide its recommendations, \FC considers a set of \textit{OSS Repositories}~\circled{1}. The \textit{Code Parser} \circled{2} component extracts method declarations and invocations from the source code or bytecode of these projects. 
\textit{Project Comparator}, a subcomponent of \textit{Similarity Calculator} \circled{3}, measures the similarity between projects in the repositories and the project under development. 
Using the set of projects and the information extracted by \emph{Code Parser}, the \textit{Data Encoder} \circled{4} component computes rating matrices which are introduced later in this section.
Afterwards, \textit{Declaration Comparator} computes the similarities between declarations.
From the similarity scores, \textit{Recommendation Engine} \circled{5} generates recommendations, either as a ranked list of API function calls using \textit{API Generator}, or as usage patterns using \textit{Code Builder}, which are presented to the developer.
In the remainder of this section, we present in greater details each of these components.

\subsubsection{Code Parser}
\FC is dependent on Rascal M$^3$~\cite{Basten2015M3} to function. Rascal M$^3$ is an intermediate model that performs static analysis on source code to extract method declarations and invocations from a set of projects. This model is an extensible and composable algebraic data type that captures both language-agnostic and Java-specific facts in immutable binary relations. These relations represent program information such as existing \textit{declarations}, \textit{method invocations}, \textit{field accesses}, \textit{interface implementations}, \textit{class extensions}, among others~\cite{Basten2015M3}. To gather relevant data, Rascal M$^3$ leverages the Eclipse JDT Core Component\footnote{\url{https://www.eclipse.org/jdt/core/}} to build and traverse the abstract syntax trees of the target Java projects.

%In the context of \FC, 
We consider the data provided by the \textit{declarations} and \textit{methodInvocation} relations of the M$^3$ model~\cite{Basten2015M3}.
Both of them contain a set of pairs $\langle v_1, v_2\rangle$, where $v_1$ and $v_2$ are values representing \textit{locations}. % in the Rascal environment.
These locations are uniform resource identifiers that represent artifact identities (aka. logical locations) or physical pointers on the file system to the corresponding artifacts (aka. physical locations).
The \textit{declarations} relation maps the logical location of an artifact (\eg~a method) to its physical location.
The \textit{methodInvocation} relation maps the logical location of a \emph{caller} to that of a \emph{callee}. %\revised{Given a declaration, we extract the following components: the method name, a list of types of the parameters, and a list of API function calls.} 
%Interested readers are referred to a dedicated paper for the technical details of the inference of Java M$^3$ models~\cite{Basten2015M3}.

%\vspace{-.2cm}

%Example

Listing~\ref{lst:m3-excerpt} depicts an excerpt of the M$^3$ model extracted from the code presented in Fig.~\ref{fig:originalCode}. 
The \textit{declarations} relation links the logical location of 
the method \code{findBoekrekeningen}, to its corresponding physical location in 
the file system.
The \textit{methodInvocation} relation states that the 
\code{getCriteriaBuilder} method of the \code{EntityManager} type is 
invoked by the \code{findBoekrekeningen} method in the current project.

\begin{lstlisting}[caption={Excerpt of the M$^3$ model extracted from Fig.~\ref{fig:originalCode}.}, label=lst:m3-excerpt, language=m3, captionpos=t, 
escapeinside={\%*}{*)}, float=h,frame=single,captionpos=b]
m3.declarations = {
<|java+method://StandaardBoekrekeningService/findBoekrekeningen|,
|file://%*\etclst*)/StandaardBoekrekeningService.java(501,531,<17,4>,<33,5>)|>, 
%[...]}
m3.methodInvocation = {
<|java+method://StandaardBoekrekeningService/findBoekrekeningen|,
|java+method://EntityManager/getCriteriaBuilder|>, [...]}
\end{lstlisting}
\vspace{-.2cm}

%Rascal M$^3$ follows a conservative approach when building the \textit{methodInvocation} relation:~method invocations are only included when the callee's owner type can be resolved.
%To mitigate this issue, the classpath of each project must be reconstructed carefully.
%Thus, given that a plethora of Java projects rely on dependency managers and modular systems such as Apache Maven~\cite{maven} and OSGi~\cite{osgi}, we use both Maven and Tycho~\cite{tycho} to build Maven and OSGi-based projects and extract their classpaths. 
%Plain libraries included as JAR archives in the project are also considered.
%In the end, we discard method invocations that cannot be properly resolved.
%
%When facing conditional or loop statements, Rascal M$^3$ adds to the \textit{methodInvocation} relation method calls present in all branches of the control flow.
%In the presence of inheritance, Rascal M$^3$ first considers the methods of type $T$ related to the invocation. 
%If the corresponding method is only declared in $T$ or if it overrides a superclass, the fully qualified name points to the method in $T$.
%Otherwise, if the method is inherited from a superclass $T'$ and it is not overridden in $T$, then the relation is associated to $T'$.
%Besides, method invocations of a super class are also considered.
%When considering interfaces, if there is an object $o$ of type $T$ and $T$ implements interface $I$, the method invocation through $o$ points to $I$ or $T$ according to the static type of $o$.
%Finally, we do not consider method invocations through dynamic mechanisms such as Java reflection.

\subsubsection{Data Encoder}
\begin{figure}[h!]
	\centering
	\vspace{-.4cm}
	\includegraphics[width=0.21\textwidth]{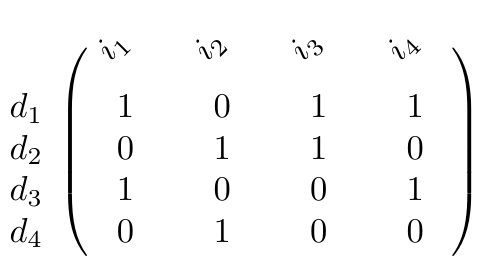}
	\vspace{-.2cm}
	\caption{Rating matrix for a project with 4 declarations and 4 invocations.}
	\vspace{-.2cm}
	\label{fig:MD}
\end{figure}

Once all the method declarations and invocations have been parsed 
with 
Rascal, \FC represents the relationships among them using a \emph{rating 
matrix}. Given a project, each row in the matrix corresponds to a 
declaration, 
and each column corresponds to an API call. A cell is set to $1$ if the 
declaration in the corresponding row contains the invocation in the 
column, 
otherwise it is set to $0$. In Fig.~\ref{fig:MD}, we show an example of the rating matrix for an explanatory project $p_{1}$ with four declarations 
$p_{1} 
\ni (d_{1}, d_{2}, d_{3}, d_{4})$ and four invocations $(i_{1}, i_{2}, 
i_{3}, 
i_{4})$. In practice, a matrix is generally big to contain a large number 
of 
methods and invocations.

\begin{figure*}[h!]
	\vspace{-.2cm}
	\centering    
	\begin{tabular}{c c}		
		\subfigure[3D context-based rating matrix]{\label{fig:3DRepresentation}\includegraphics[width=75mm]{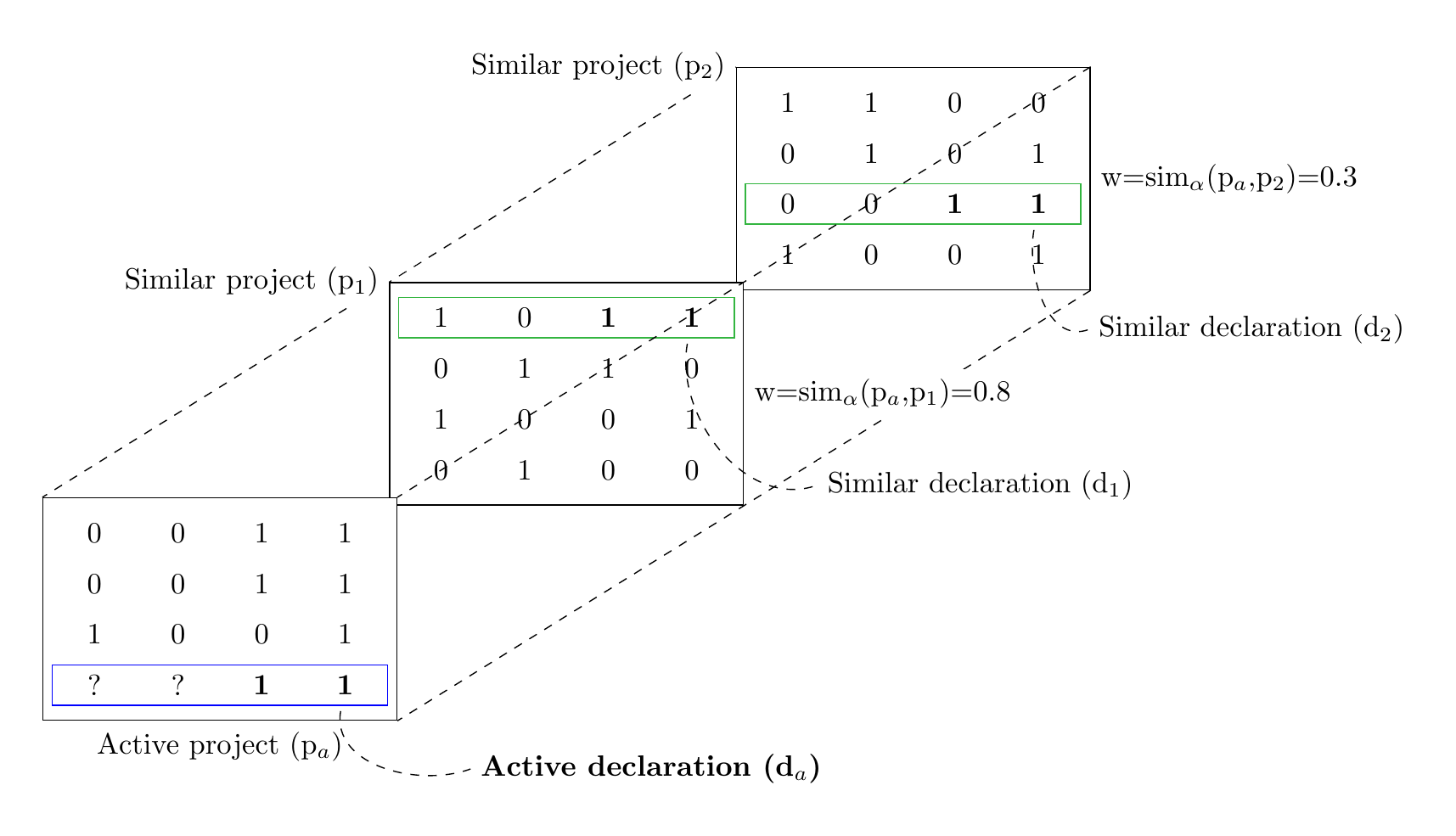}}  &
		\subfigure[Graph representation of projects and invocations]{\label{fig:Similarity}\includegraphics[width=70mm]{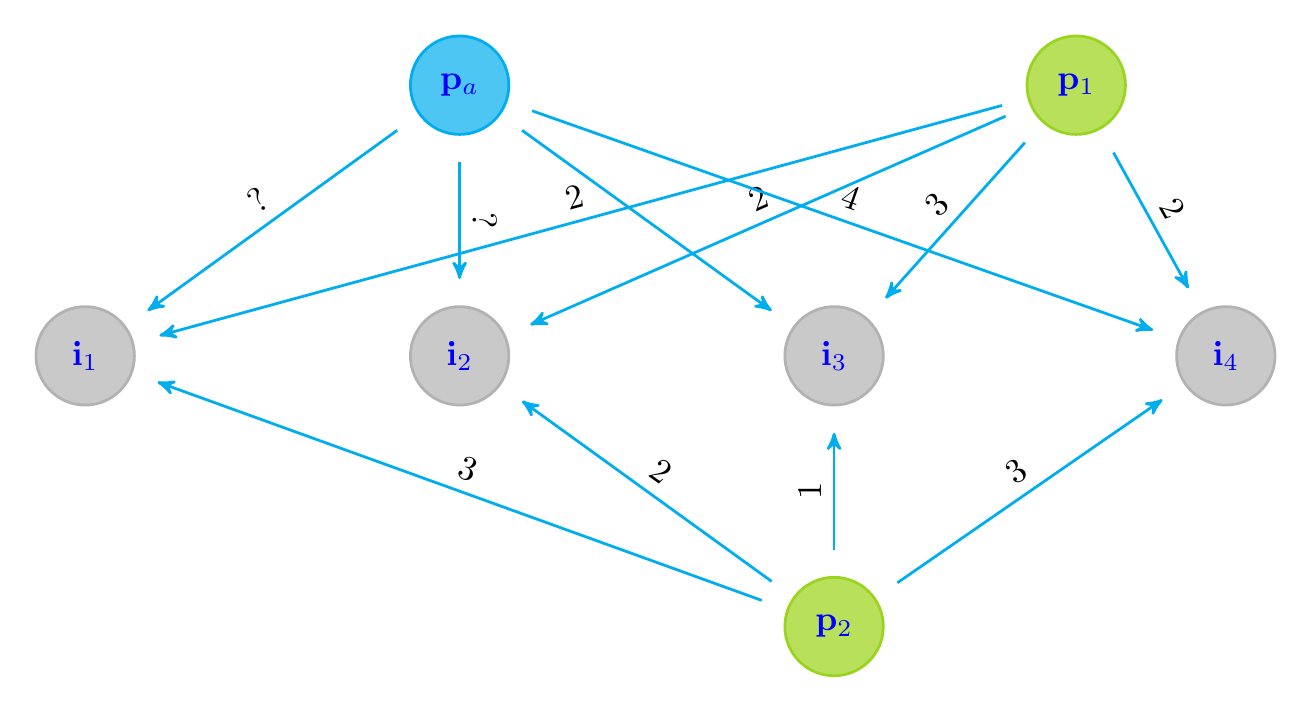}} 
	\end{tabular}
	\vspace{-.2cm}
	\caption{Matrix and graph representation for a set of three OSS projects.}
	\vspace{-.2cm}
\end{figure*}

We conceptualized a 3D context-based ratings matrix to model the intrinsic relationships among various projects, declarations, and invocations. The third dimension of this matrix represents a project, which is analogous to the so-called context in context-aware CF systems.
For example, Fig.~\ref{fig:3DRepresentation} depicts three projects $P = (p_{a}, p_{1}, p_{2})$ represented by three slices with four method declarations and four method invocations. 
Project $p_1$ has already been introduced in Fig.~\ref{fig:MD} and, for the sake of readability, the column and row labels are omitted from all the slices in Fig.~\ref{fig:3DRepresentation}. 
There, $p_{a}$ is the \emph{active project} and it has an \emph{active declaration} $d_{a}$.
\emph{Active} here means the artifact (project or declaration), being considered or developed. 
Both $p_{1}$ and $p_{2}$ are complete projects similar to the active project $p_a$.
The former projects, \ie $p_{1}$ and $p_{2}$ are also called \emph{background data} since they are already available and serve as a base for the recommendation process. In practice, the more background projects we have, the better is the chance that we recommend relevant API invocations.

% higher the number of complete projects considered as background data, the higher the probability to recommend relevant invocations.

%\begin{figure}[tb]
%	\centering
%	\vspace{-.2cm}
%	\includegraphics[width=0.50\textwidth]{figs/3DMatrix3.pdf}
%	\vspace{-.4cm}
%	\caption{3D context-based rating matrix.}
%	\vspace{-.3cm}
%	\label{fig:3DRepresentation}
%	\vspace{-.2cm}
%\end{figure}

\subsubsection{Similarity Calculator} \label{sec:SimilarityCalculator}
By exploiting the context-aware CF technique, the presence of additional invocations is deduced from similar declarations and projects. 
Given an active declaration in an active project, it is essential to find the subset of the most similar projects, and then the most similar declarations in that set of projects. 
To compute similarities, we devised %from \cite{8498236} 
a weighted directed graph that models the relationships among projects and invocations. 
Each node in the graph represents either a project or an invocation. 
If project $p$ contains invocation $i$, then there is a directed edge from $p$ to $i$.
The weight of an edge $p \rightarrow i$ represents the number of times a project $p$ performs the invocation $i$. 
Fig.~\ref{fig:Similarity} depicts the graph for the set of projects in Fig.~\ref{fig:3DRepresentation}. 
For instance, $p_a$ has four declarations and all of them invoke $i_4$.
As a result, the edge $p_a \rightarrow i_4$ has a weight of $4$. 
In the graph, a question mark represents missing information.
For the active declaration in $p_a$, it is not known yet whether invocations $i_1$ and $i_2$ should be included.

%\begin{figure}[h!]
%	\centering
%	\vspace{-.2cm}
%	\includegraphics[width=0.40\textwidth]{figs/Similarity.pdf}
%	\vspace{-.3cm}
%	\caption{Graph representation of projects and invocations.}
%	\vspace{-.3cm}
%	\label{fig:Similarity}
%\end{figure}

%The similarity between two project nodes $p$ and $q$ is computed by considering their feature sets. %\cite{DiNoia:2012:LOD:2362499.2362501}. 
%Given that $p$ has a set of neighbor nodes 
Considering $(i_{1},i_{2},..,i_{l})$ as a set of neighbor nodes of $p$, the feature set of $p$ is the vector $\overrightarrow{\phi}=(\phi_{1},\phi_{2},..,\phi_{l})$, with $\phi_{k}$ being the weight of node $i_{k}$. Each constituent weight is computed as the \emph{term-frequency inverse document frequency} value, \ie $\phi_{k} = f_{i_{k}}*log(\frac{ \left | P \right |}{a_{i_{k}}})$, where $f_{i_{k}}$ is the weight of the edge $p \rightarrow i_k$; $\left | P \right |$ is the number of all considered projects; and $a_{i_{k}}$ is the number of projects connected to $i_{k}$. Eventually, the similarity between $p$ and $q$ is computed as the cosine between their corresponding feature vectors $\overrightarrow{\phi}=\{\phi_{k}\}_{k=1,..,l}$ and $\overrightarrow{\omega}=\{\omega_{j}\}_{j=1,..,m}$, given below:
\vspace{-.1cm}

\begin{equation} \label{eqn:VsmSim}
sim_\alpha(p,q)=\frac{\sum_{t=1}^{n}\phi_{t}\times \omega_{t}}{\sqrt{\sum_{t=1}^{n}(\phi_{t})^{2} }\times \sqrt{\sum_{t=1}^{n}(\omega_{t})^{2}}} 
\end{equation}

Given that $\mathbb{F}(d)$ and $\mathbb{F}(e)$ are the sets of invocations for declarations $d$ and $e$, respectively, then the similarities between $d$ and $e$ are calculated using the Jaccard similarity index as follows: %~\cite{jaccard}
\vspace{-.1cm}	
\begin{equation} \label{eqn:Jaccard}
sim_\beta(d,e)=\frac{|\mathbb{F}(d)\bigcap \mathbb{F}(e)|}{|\mathbb{F}(d)\bigcup \mathbb{F}(e)|} 
\end{equation}
\subsubsection{API Generator} \label{sec:APIGenerator}
This component is a part of \textit{Recommendation Engine}, and it is used to generate a ranked list of API function calls.	
As shown in Fig.~\ref{fig:3DRepresentation}, the active project $p_{a}$ already includes three declarations, and at the time of consideration, the developer is working on the fourth declaration, corresponding to the last row of the matrix. 
$p_{a}$ has only two invocations, represented in the last two columns of the matrix, \ie cells marked with $1$. 
The first two cells are filled with a question mark ($?$), implying that it is not clear if these two invocations should also be integrated into $p_{a}$. 
\textit{API Generator} predicts additional invocations for the active declaration by computing the missing ratings exploiting the following collaborative-filtering formula \cite{Chen:2005:CCF:2154509.2154540}: % using the following formula
%\vspace{-.1cm}
\begin{equation} \label{eqn:missingRating}
r_{d,i,p} = \overline{r}_{d} + \frac{\sum_{e \in topsim(d)}(R_{e,i,p}-\overline{r}_{e}) \cdot sim_{\beta}(d,e)}{\sum_{e \in topsim(d)}sim_{\beta}(d,e)}
\end{equation}
Equation~\ref{eqn:missingRating} is used to compute a score for the cell representing method invocation $i$, declaration $d$ of project $p$, where $topsim(d)$ is the set of top similar declarations of $d$;
$sim_{\beta}(d,e)$ is the similarity between $d$ and a declaration $e$, computed using Eq.~\eqref{eqn:Jaccard}; 
$\overline{r}_{d}$ and $\overline{r}_{e}$ are the mean ratings of $d$ and $e$, respectively;
and $R_{e,i,p}$ is the combined rating of $d$ for $i$ in all the similar projects, computed as follows~\cite{Chen:2005:CCF:2154509.2154540}:
\vspace{-.1cm}
\begin{equation} \label{eqn:combinedRating}
R_{e,i,p}=\frac{\sum_{q \in topsim(p)}r_{e,i,q} \cdot sim_{\alpha}(p,q)}{\sum_{q \in topsim(p)}sim_{\alpha}(p,q)} %\sum_{x \in C}
\end{equation}

\noindent where $topsim(p)$ is the set of top similar projects of $p$, $k$=$\left |  topsim(p) \right |$ is the number of neighbor projects;
and $sim_{\alpha}(p,q)$ is the similarity between $p$ and a project $q$, computed using Eq.~\ref{eqn:VsmSim}. Equation~\ref{eqn:combinedRating} implies that a higher weight is given to projects with higher similarity. In practice, it is reasonable since, given a project, its similar projects contain more relevant API calls than less similar projects. Using Eq.~\ref{eqn:missingRating} we compute all the missing ratings in the active declaration and get a ranked list of invocations with scores in descending order, which is then suggested to the developer. % as recommendations.
In Eq.~\ref{eqn:combinedRating}, a set of $k$ projects is used to compute the ranking, and no matter how large $k$ is, eventually we obtain %a ranking as 
a real score for each API. Therefore, the final list always contains N items, regardless of $k$.

In our implementation, we employed a sparse matrix to store the 3D tensor. This allows us to optimize both the storage and computation, and thus increasing the number of neighbor projects for the recommendation. By the current version, \FC is able to efficiently compute the recommendations, and maintain a trade-off between computational complexity and effectiveness.
%In this way, we can optimize the storage and thus the calculation. 

\subsubsection{Code Builder} \label{sec:CodeBuilder}
This sub-component %is a part of \textit{Recommendation Engine}, and it 
is responsible for recommending real code snippets to developers. From the ranked list, \emph{top-N} invocations are selected as query to search the corpus for relevant declarations.
%\JDR{In particular, the query consist of \emph{top-N} recommended API function calls on top of the ones extracted from  the input method.} 
To limit the search scope, we consider only the most similar projects.
% by means of the \textit{Similarity Calculator}.
%The Jaccard index is used to compute similarities between the query invocations and a given declaration.
%\JDR{We compute the Jaccard index between the resulting query and each method of $topsim(p)$ similar projects to the one that the developer is implementing.} 
Using the Jaccard index as the similarity metric, for each query, we search for declarations that contain as many invocations of the query as possible. Once the corresponding declarations are identified, their source code is retrieved using the \emph{declarations} relation of the Rascal M$^3$ model.
Thanks to its modularity, Rascal is able to decompile and analyze projects written in different programming languages~\cite{BASTEN20157}, \eg Java~\cite{Basten2015M3}, C/C++~\cite{rodin_aarssen_2017_891122}, PHP~\cite{hills2014php}.
%In the same way, 
Rascal also allows us to compute M$^3$ model from both source code folders and binaries, \eg JAR files independently. Thus we implemented a dedicated function %(\ie ~\code{getCodeFromM3}) 
that extracts the real source code of a method declaration by means of the computed M$^3$ model and the project location. Finally, the resulting code snippet is suggested to the developer.
%, and the relation to a method declaration.%(Listing~\ref{lst:m3-excerpt} shows an instance of  \code{m3.declaration} relation).
%}%To this aim, we rely on Rascal to retrieve the code snippet of the declaration from the specified location, which is then recommended to the developer.

\subsection{\FC in Action} This section describes two use cases that illustrate how \FC works in practice. Section~\ref{sec:CodeRecommendation} presents the final result produced by \FC for the motivating example in Section~\ref{sec:MotivatingExample}, while Section~\ref{sec:IDE} describes the \FC IDE through a real development scenario, where we recommend both a list of API function calls and real source code.

\vspace{-.2cm}
\subsubsection{Code Recommendation} \label{sec:CodeRecommendation}
%To illustrate how \FC recommends real code snippets, 
In Fig.~\ref{fig:originalCode}, given that \textit{findBoekrekeningen} is the 
active declaration, the invocations it contains are used together with the 
other declarations in the current project as the query to feed the 
recommendation engine. The produced outcome is a ranked list of real code 
snippets, and we show the top one, named \textit{findByIdentifier}, in 
Listing~\ref{lst:RecommendedSnippet}.

%it.To this end, 
%In this sense, \FC.
%\begin{figure}[tb]
%	\centering
%	\includegraphics[width=\columnwidth]{figs/codeExample-recommendation.png}
%	\caption{Real source code recommended by \FC}
%	\vspace{-.3cm}
%	\label{fig:RecommendedSnippet}
%	\vspace{-.3cm}
%\end{figure}

\begin{lstlisting}[caption={Recommended source code for the snippet in Fig.~\ref{fig:originalCode}.}, label=lst:RecommendedSnippet, 
style=JavaStyle,captionpos=b,xleftmargin=1.8em,frame=single,framexleftmargin=1.2em]
public List<QuestionsStaged> findByIdentifier(String identifier) {
   log.fine("getting Session instance by identifier: " + identifier);
   try {
       CriteriaBuilder cb = entityManager.getCriteriaBuilder();
       CriteriaQuery<QuestionsStaged> criteria = cb.createQuery(QuestionsStaged.class);
       Root<QuestionsStaged> qs = criteria.from(QuestionsStaged.class);
       criteria.select(qs).where(cb.equal(qs.get("identifier"), identifier));
       log.fine("get identifier successful");
       return entityManager.createQuery(criteria).getResultList();

   } catch (RuntimeException re) {
       log.severe("get identifier failed" + re);
       throw re;
   }
}	
\end{lstlisting}

\begin{figure*}[t]
	\centering
	\includegraphics[width=\linewidth]{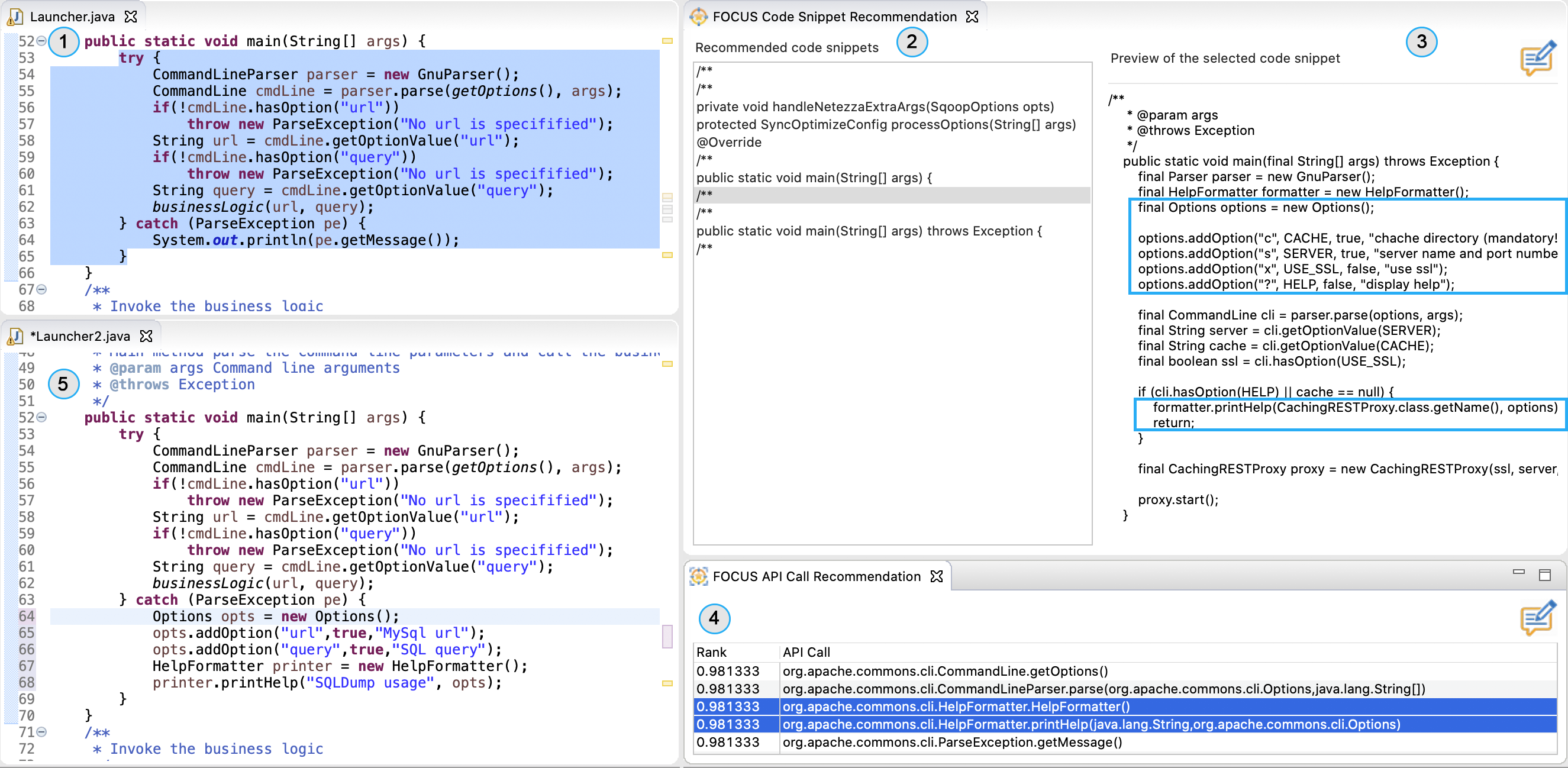}
	\vspace{-.3cm}
	\caption{\FC IDE.}
	\label{fig:focus-in-action}
	\vspace{-.3cm}
\end{figure*}

By comparing the recommended code and the 
original one in Fig.~\ref{fig:recCode}, we realize that though they are not the 
same, they indeed share several method calls and a shared intent:~both snippets 
exploit a \textit{CriteriaBuilder} object to build, perform a query, and 
eventually retrieve some results. Furthermore, the outcome of both declarations 
is of the \textit{List} type. More importantly, compared to the original code 
in Fig.~\ref{fig:recCode}, the recommended snippet appears to be of a higher 
quality and robustness. %since it contains a \textit{try}/\textit{catch} 
%construct to handle possible exceptions, and this is useful for the whole 
%function.
%\MAX{I fear this could be confusing, like \FC explicitly searches for better quality code with exception handling. Maybe we can explicitly say that, since in this case the developer has found better quality (more robust) code, she opts for it.}
We conclude that for the motivating example, \FC is helpful since the recommended code together with the corresponding list of function calls, \ie~\textit{get}, \textit{equal}, \textit{where}, \textit{select}, \etc, provides the developer with practical instructions on how to use the API at hand to implement the desired functionality.

\vspace{-.2cm}
\subsubsection{Using \FC in the Eclipse IDE} \label{sec:IDE}

As shown in Fig.~\ref{fig:focus-in-action}, \FC has been integrated into the Eclipse IDE.\footnote{Instruction to install the IDE: \url{https://bit.ly/3joJpnT}} The figure depicts a real development scenario where a developer is implementing the \textit{SQLDump} project\footnote{\url{https://github.com/aparsons/SQLDump}} by improving the existing code with recommendations provided by \FC. 
\textit{SQLDump} is a simple command-line utility that exploits the \textit{apache-cli} library\footnote{\url{http://commons.apache.org/proper/commons-cli/}} to execute an SQL query and export results as a CSV file. 
%	is used to extract parameters from the command line.
	%We illustrate how \FC helps the developer evolve the \code{main} method of the class \code{Launcher} as follows. 
%	The method takes all needed parameters from the console and calls the businesses logic to execute SQL query and export it as a \textit{csv} file. 
The first implementation of the \textit{main} method prints parameter errors to the console by using Java I/O facilities, \ie \textit{System.out.println} \circled{1}. \FC suggests to the developer both code snippets \circled{2} and \circled{3}, and a ranked list of predicted APIs \circled{4} that are relevant to the code being developed.
Furthermore, it recommends a possible improvement that includes the usage of the \textit{HelperFormatter} class \circled{5}: %Compared to the original code, in this snippet, 
the \textit{catch} statement block is completely defined and %more mature: 
the \textit{System.out.println} invocation is replaced by \textit{HelperFormatter} provided by \textit{apache-cli}. Meanwhile \textit{printHelp} is a method of \textit{HelperFormatter} that prints both possible parameter errors as well as an introduction on how to run \textit{SQLDump} from command line. As a result, with the help of \FC, the developer can learn how to use the method both from the code snippets \circled{3} and the list of API calls \circled{4}.% (see the highlighted lines of code)

%discovers the \code{printHelp} method from API function calls recommendations \circled{4} and she 

	%	\section{Method}
	%	\label{sec:Method}
	%	\input{src/Method}
	
	\section{Evaluation}
	\label{sec:Evaluation}
	The \emph{goal} of this study is to evaluate \FC and compare it with two state-of-the-art tools, \ie \UM~\cite{Wang2013Mining} and PAM~\cite{Fowkes:2016:PPA:2950290.2950319}, with the \emph{purpose} of determining the extent to which it can provide a developer with accurate and useful recommendations, featuring code snippets containing API usage patterns relevant for the developers' context. The {\em quality focus}  relates to the API recommendation accuracy and completeness,  the time required to provide a recommendation, and the extent to which developers perceive the recommendation useful.

PAM has been chosen as baseline for comparison, since it is among the state-of-the-art tools in API recommendation: it has been shown~\cite{Fowkes:2016:PPA:2950290.2950319} to outperform other similar tools such as MAPO~\cite{Zhong2009MAPO} and UP-Miner~\cite{Wang2013Mining}. To conduct the comparison with PAM, we exploited its original source code which has been made available online by its authors.\footnote{\url{https://github.com/mast-group/api-mining}} Furthermore, to facilitate future replications, we published all the artifacts together with the tools used in our evaluation in \GH~\cite{focus-zenodo}.
% \cite{PAM}

%Though this category of coding shares many commonalities with traditional programming, it has its own characteristics, which essentially pose various challenges to developers~\cite{joorabchi2013real}. In this context, supports for Android programmers with respect to different tasks are highly desirable. In fact, recommending API calls and usage patterns is deemed to be important in Android programming~\cite{10.1145/3293882.3330571}.}

%\revised{In recent years, the proliferation of portable devices, especially smartphones, has led to the necessity of mobile programming, and among others, programming on the Android operating system \cite{10.1145/1882362.1882443}. 
	
 %Thus, we came across the following motivating question.}% as follows
%\vspace{.2cm}
%\framebox{
%	\parbox[t][0.80cm]{7.50cm}{
%		%\revised{\textbf{MQ$_2$}: \emph{``Can \FC assist mobile apps developers?''}} %Can we automatize the classification process
%		\revised{\textbf{MQ$_2$}: \emph{``Can \FC provide Android developers with relevant API calls?''}}
%	} 
%}
%\begin{displayquote}
%\revised{\textquotedblleft \emph{Can \FC provide Android developers with relevant API calls?}\textquotedblright}
%\end{displayquote}
%\MAX{I would just say that Android is context, and explain why we choose it, without asking the question}

%~\cite{Fowkes:2016:PPA:2950290.2950319}
% are available online
%For the sake of reproducibility and ease of reference, 

After formulating the research questions in Section \ref{sec:rqs}, the following subsections describe datasets, analysis methodology, and the evaluation metrics used to evaluate \FC. % are explained in detail in

\subsection{Research Questions}
\label{sec:rqs}
Our study aims to address the following research questions:

%Our research questions are as follows:

%\newcommand{\rqone}{To what extent is \FC able to provide accurate and complete recommendations?}
%\newcommand{\rqtwo}{What are the timing performances of \FC in building its models and in providing recommendations?}
%\newcommand{\rqthree}{How does \FC perform compared with PAM?}

%\smallskip\noindent
%\textbf{RQ$_1$} {\em \rqone} 
\vspace{.1cm}
\noindent
$\rhd$~\rqfirst Both \UM~\cite{Wang2013Mining} and PAM~\cite{Fowkes:2016:PPA:2950290.2950319} are well-founded API recommendation tools. \UM has been shown to outperform MAPO~\cite{Zhong2009MAPO}, while PAM gains a superior performance compared to both \UM and MAPO. In our previous work~\cite{Nguyen:2019:FRS:3339505.3339636}, we showed that \FC outperforms PAM on different datasets collected from \GH and MVN. In this work, we compare \FC with \UM and PAM on an Android dataset to further study their performance on a new application domain.

%This research question directly compares \FC with PAM in terms of recommendation accuracy.
% Moreover, we also compare \FC with another baseline, namely \UM~\cite{Wang2013Mining}

%\smallskip\noindent
%\textbf{RQ$_2$} {\em \rqtwo } 
\vspace{.1cm}
\noindent
$\rhd$~\rqsecond
For a recommender system, it is essential to be able to return relevant recommendations, indicating by a high number of true positives as well as a low number of both false positives and false negatives. This research question evaluates to which extent our tool can provide accurate and complete results. 

%the capability of FOCUS
%Having too many false positives would end up being counterproductive, whereas having too many false negatives would mean that the tool is not able to provide recommendations in many cases where this is needed. 
%This research question aims at assessing whether, from a timing point of view, FOCUS---compared to PAM---could be used in practice. We evaluate the time required by both tools to provide a recommendation. We mainly focus on the recommendation time because,  while it is acceptable that the model training phase is relatively slow (\ie the model could be built offline), the recommendation time has to be fast enough to make the tool applicable in practice. 

%\smallskip\noindent
%\textbf{RQ$_3$} {\em \rqthree } 
\vspace{.1cm}
\noindent
$\rhd$~\rqthird
We examine whether given a testing app, having more apps of the same category is beneficial to the recommendation outcome.
% getting highly similar projects is beneficial to finding relevant API function calls. Computing similarity is important
 
\vspace{.1cm}
\noindent
$\rhd$~\rqfourth
We study if the recommended code snippets provided by \FC are relevant to support developers in fulfilling their tasks.

\vspace{.1cm}
\noindent
$\rhd$~\rqfifth Finally, we are interested in investigating whether \FC is useful from a developer point of view. To this end, we conducted a user study to evaluate the relevance of API calls and code snippets provided by \FC to support %to understand the usefulness of the provided recommendations to a developer working on a 
a particular development context. A group of \numParticipants Master's students in Computer Engineering has been involved to assess two real-world development scenarios.

%The goal of this study is to evaluate, from a developer’s perspec- tive, the relevance of the Stack Overflow discussions identified by Prompter, to understand to what extent the retrieved discussions provide useful information to a developer working on a particular code snippet.

%The context of the study consists of participants, i.e., various kinds of developers, among professionals and students, and objects, i.e., source code snippets and its related Stack Overflow discus- sion as identified by Prompter. The study addresses the following research question (RQ1): To what extent are the SO discussions identified by Prompter relevant for developers?
%We aim at investigating to what extent the Stack Overflow dis- cussions identified by Prompter contain information perceived as relevant by developers for a specific programming task. We asked 55 people (industrial developers, academics, and students) to complete a questionnaire aimed at evaluating the relevance of the Stack Over- flow discussions identified by Prompter, by analyzing a specific code snippet.

%for developers 

\subsection{First Evaluation: Simulating Developers' Behavior}

In the following, we describe the dataset used to address 
RQ$_1$-RQ$_4$, as well as the data extraction method. As it is explained in 
Section \ref{sec:userstudy}, for RQ$_5$ we rely on different datasets, 
because the aim is to let developers leverage \FC recommendations, and tasks 
should be simple enough for an experimental setting.

\vspace{-.3cm}
\subsubsection{Evaluation Dataset and Data Extraction}
\label{sec:Dataset}

While \FC is able to work with different data sources as well as programs written in various languages, the evaluation \emph{context} for this paper focuses on the applicability to a specific domain, \ie Android programming. Although Android development is per se not very different from the development of other kinds of applications, after the evaluation reported in our previous paper featuring heterogeneous Java programs \cite{Nguyen:2019:FRS:3339505.3339636}, the aim of this evaluation is to show how, by learning from a training set belonging to applications from the same ecosystem, \FC is capable of providing accurate recommendations.
We have chosen Android not only because of the large availability of data needed to perform an empirical evaluation, but also because recommending API calls and usage patterns is deemed to be important in Android programming~\cite{10.1145/3293882.3330571}.

\begin{figure}[t!]
	\centering
	\vspace{-.1cm}
	\includegraphics[width=0.88\columnwidth]{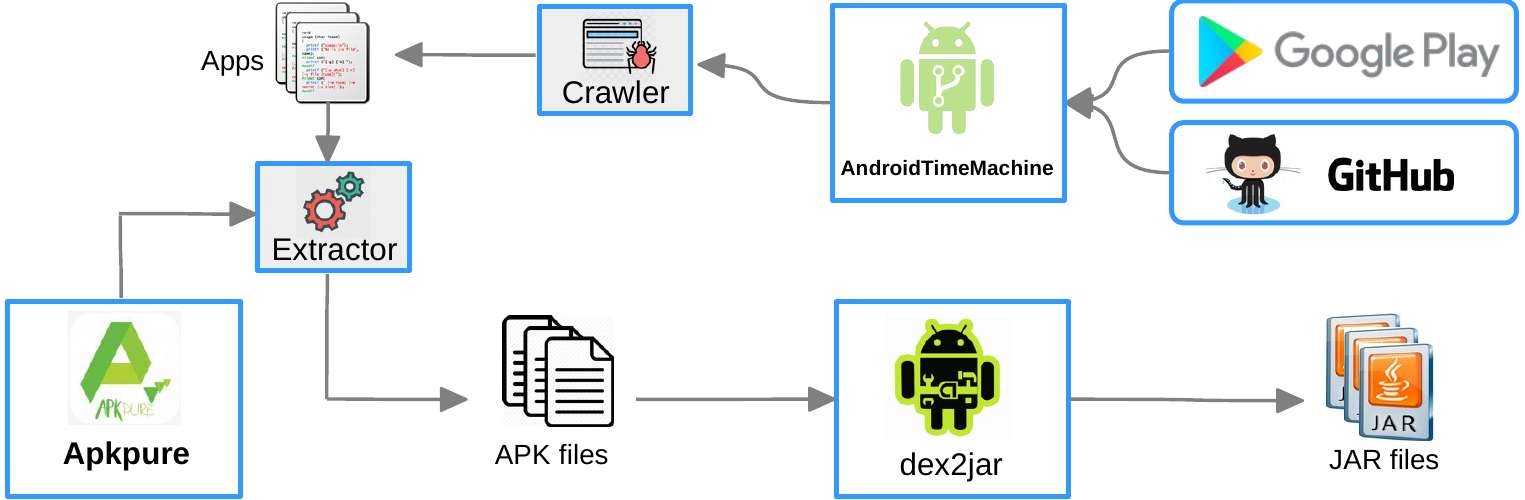}
	\caption{The data extraction process.}
	\vspace{-.3cm}
	\label{fig:DataExtraction}
\end{figure}

Since \FC accepts as input data extracted by Rascal, which in turn requires a specific format, %is able to parse Java and Kotlin source code, 
we devised our own method to acquire an Android dataset eligible for the evaluation. The extraction process needs to comply with some certain requirements, and it is illustrated in Fig.~\ref{fig:DataExtraction}. First, we exploited the \emph{AndroidTimeMachine} platform~\cite{8595172} to crawl open source projects. The platform fetches apps from the Google Play store\footnote{\url{https://play.google.com/}} and associates them with the open source counterparts hosted in GitHub. The crawling process resulted in a set of 7,968 open source Android apps. 
Most of the apps (82\%) in the dataset are written in Java; 4\% in Kotlin; 4\% in JavaScript, 2\% in C++, and 1\% in C\#. The remaining 7\% belong to other languages.

%Figure~\ref{fig:PieChart} summarizes the number of apps in percentage and the corresponding programming languages. The dominance of Java is seen from the figure: 82\% of the apps are written in Java, and the rest are written in various languages, \eg C++, C\#, or JavaScript.}

 % and the corresponding percentage of apps

%\begin{figure}[h!]
%	\centering
%	\vspace{-.2cm}
%	\includegraphics[width=0.68\linewidth]{figs/PieChart2.png}
%	\vspace{-.2cm}
%	\caption{\revised{A summary of the apps and programming languages.}}
%	\vspace{-.2cm}
%	\label{fig:PieChart}
%\end{figure}

As Rascal can parse certain programming languages, from the initial dataset we filtered out irrelevant projects to select only the Java and Kotlin ones, which account for the majority of the apps. % \MAX{does this depend on the analysis capability? clarify.
Afterwards, we retrieved the corresponding compiled APK files by querying the Apkpure platform\footnote{\url{https://apkpure.com/}} using some tailored Python scripts~\cite{8543433}. The process culminated in the final corpus consisting of 2,600 APK binary files (mined from Apkpure) together with additional metadata (mined from Google Play), including authors, categories, star rating, price, and the number of downloads. %Figure~\ref{fig:StarsAndDownloads} depicts a map of the apps with respect to the number of stars and downloads. As can be seen, the apps are scattered across the map: there are apps that locate on the lower part of the figure, indicating a low rating as well as a low number of downloads. 
By carefully inspecting the data, we realized that most of the apps are highly rated and they have a high number of downloads.

%locate on the upper part of the figure

We decompiled the APKs into the JAR format by means of the \textbf{dex2jar} tool \cite{dex2jar}. The JAR files were then fed as input for Rascal to convert them into the  M$^3$ format, which can eventually be consumed by \FC.

In total, there are 26,854 API functions in the whole dataset, and 
most of them 
are invoked by a small number of declarations (and thus projects\footnote{For 
the sake of presentation, from now on the two terms ``app'' and ``project'' are 
used interchangeably.}): 15,731 APIs are called in only one project. Only a 
tiny fraction of the APIs is extremely popular by being included in a large 
number of projects: ten APIs are called in more than 1,900 projects and 15,000 
declarations. The most popular API call is 
\emph{java/lang/StringBuilder/append(java.lang.String)} and it appears in 2,512 
projects and 54,828 declarations. 

Altogether, this reflects \emph{the long tail effect} which has 
already been encountered by third-party libraries 
recommendation~\cite{NGUYEN2019110460}. Such an effect can be expressed as 
follows: For many outcomes, about 80\% of consequences originate from 20\% of 
the causes~\cite{koch199980}. When we apply this to API recommendation, it is 
interpreted as: ``\emph{About 80\% of the APIs come from 20\% of the apps.}'' 
As it has been shown in various 
studies~\cite{NGUYEN2019110460,10.1145/2645710.2645744}, providing products in 
the long tail is beneficial to the final recommendations. In a similar fashion, 
we suppose that the ability to suggest APIs rarely included by apps, is of 
particular importance, as this may help discover useful APIs that have been 
normally obscured from search engines.

A summary of the categories and their corresponding number of items in 
the considered dataset is also provided. Due to the space limit, we cannot 
show and discuss all the figures here. Please refer to the online appendix for more details.\footnote{\url{https://mdegroup.github.io/FOCUS-Appendix/}}

With this dataset, we aim at evaluating if the proposed approach is able to support mobile developers in diverse application domains as well as with various levels of apps' maturity, thereby attempting to resemble real-world development scenarios. We use the collected dataset in RQ$_1$, RQ$_2$, RQ$_3$, and RQ$_4$ to evaluate \FC as well as to compare it with the two baselines.

Finally, the following main steps are conducted to create the required metadata, which can then be used to feed \FC.

\begin{itemize}[noitemsep]
	\item the corresponding Rascal M$^3$ model is generated for every project in the dataset;
	\item the corresponding ARFF representation\footnote{\url{https://www.cs.waikato.ac.nz/ml/weka/arff.html}} for each M$^3$ model is generated in order to be used as input for applying 
	\FC and PAM during the actual evaluation steps discussed in the next sections.
\end{itemize}

\vspace{-.25cm}
\subsubsection{How we simulate developers' behavior} 
\label{sec:methodology}

\begin{figure}[t!]
	\centering
	%\vspace{-.2cm}
	\includegraphics[width=0.680\linewidth]{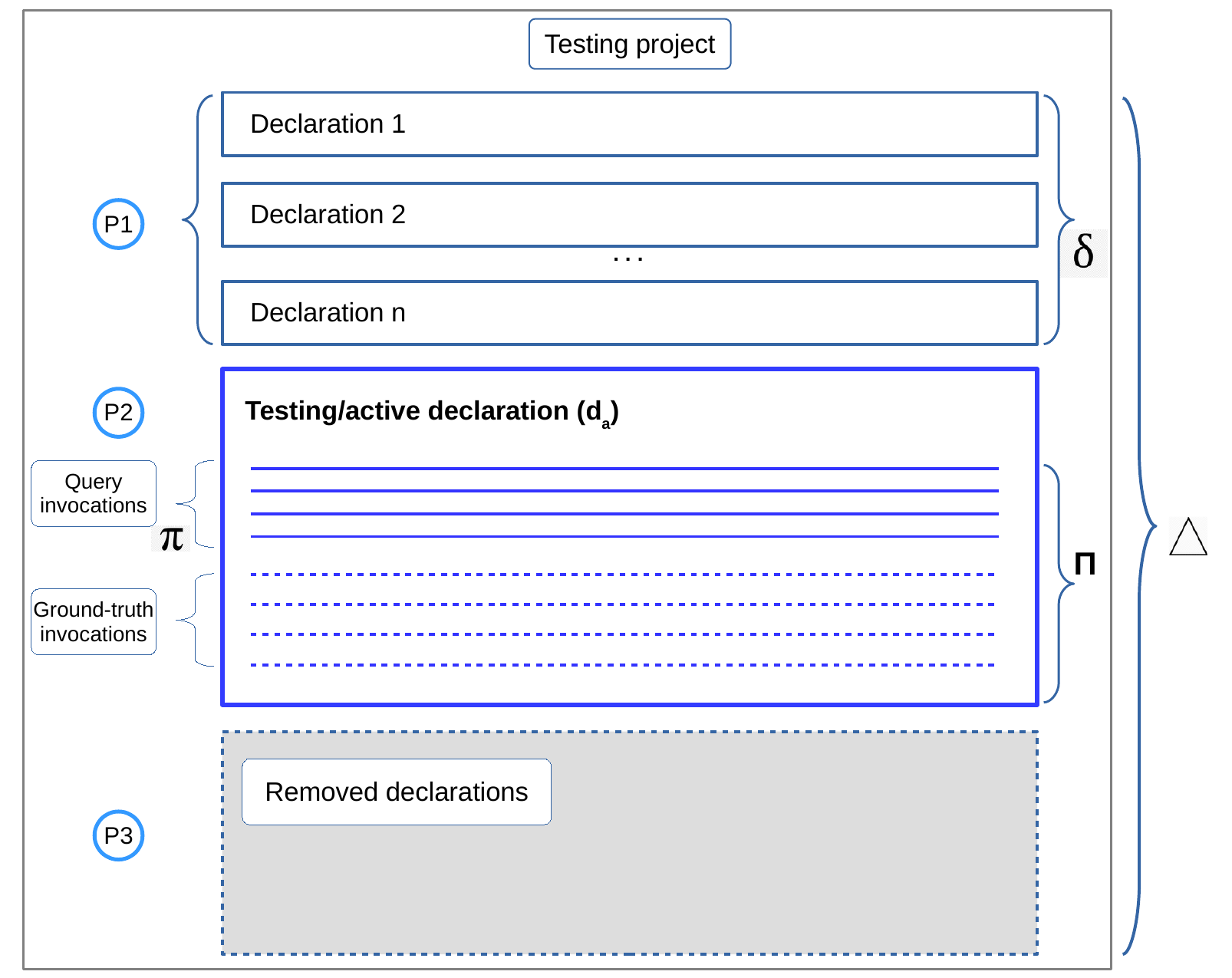}
	\vspace{-.2cm}
	\caption{The extraction of data for a testing project.}  %\PN{The figure needs to be improved.}
	\label{fig:CodeRemoval}
	\vspace{-.2cm}
\end{figure}

%In this paper, we study if \FC is applicable in real-world settings by being able to provide instant recommendations. 

To evaluate \FC in RQ$_1$-RQ$_4$,	
we simulate the behavior of a developer who is programming a project and needs 
practical recommendations to complete it. Figure~\ref{fig:CodeRemoval} provides 
an intuition on how the extraction of an  active/testing project $p_{a}$ is 
done. The project consists of a set of declarations and they are divided into 
three parts, namely \textbf{P1}, \textbf{P2}, and \textbf{P3}, which are 
explained as follows.
%\color{blue}
\begin{itemize}[noitemsep]
	\item \textbf{P1}: A set of complete declarations, \eg Declaration 1, Declaration 2, \etc.
	\item \textbf{P2}: A testing declaration, for this declaration, only a portion of code is available to feed the recommendation engine, while the rest is removed and saved as ground-truth data. This corresponds to the scenario in Fig.~\ref{fig:originalCode}, where the developer is implementing the \emph{active declaration} $d_a$, and she needs recommendations on the next APIs to be added; %This aims to model the scenario presented in Fig.~\ref{fig:originalCode}: the developer is programming $d_a$ and she needs sensible suggestions about which API calls can be added;
	\item \textbf{P3}: Removed declarations: A certain part consisting of some declarations is removed. This aims to simulate the scenario when the developer is only at an early stage of the project.
\end{itemize}	
%parameters:%We consider the following 
Correspondingly, there are the following parameters:
\begin{itemize}[noitemsep]
	\item $\Delta$ is the number declarations in $p_{a}$ ($\Delta > 0$);	
	\item %To mimic an authentic scenario, o
	Only $\delta$ declarations ($\delta < \Delta$) are used as input for recommendation and the rest is discarded; %This corresponds to the scenario when the developer finished $\delta$ declarations, and she is working on the \emph{active declaration} $d_a$;  	
	\item In total, $d_a$ has $\Pi$ invocations, however only the first $\pi$ invocations ($\pi < \Pi$) are selected as testing, and the rest is ground-truth data; %and the rest is removed and saved as ground-truth data; %This aims to model the scenario presented in Fig.~\ref{fig:originalCode}: the developer is programming $d_a$ and she needs sensible suggestions about which API calls can be added;
	\item $k$ is the number of neighbour projects (cf. Section~\ref{sec:APIGenerator});
	\item Given a ranked list of APIs, the developer typically pays attention to the \emph{top-N} items only, \ie $N$ is the \emph{cut-off value} for the list.
\end{itemize}
%\color{black}
%In the evaluation, the parameters $\delta$ and $\pi$ were varied to simulate different levels of the project's maturity as depicted in 
For $d_a$, only a half of the code lines of the method's body is selected to feed the recommendation engine. In fact, Rascal can parse only compilable code, thus there might be some compilation errors at some points, where the code is incomplete. As a result, in practice, we suppose that \FC can provide recommendations only when the developer temporarily stops at a certain point where the whole declaration becomes compilable. Thus, to increase the applicability of \FC, as a developer one should try to make the code compilable as soon as they can by closing open loops, try/catch blocks, return statements, etc. This is supported pretty well by IDEs such as Eclipse which automatically recommend and insert closed loops and try/catch blocks. In this respect, we suppose that in most cases, code is executable, though it is yet complete.

Table~\ref{tab:Configurations} shows four configurations, \ie \textbf{C1.1}, \textbf{C1.2}, \textbf{C2.1}, and \textbf{C2.2}, corresponding to different combinations of $\delta$ and $\pi$. Furthermore, \textbf{C1.1} and \textbf{C1.2} as well as \textbf{C2.1} and \textbf{C2.2} are pairwise relevant. For example, both \textbf{C1.1} and \textbf{C1.2} have the same number of method declarations ($\delta$), they differ in the number of invocations in the testing declaration ($\pi$).

%to compute recommendations, 
%Referring to Table~\ref{tab:Configurations}, we see that 
% for future comparison
%, to finish the declaration
%(\code{Split declaration})
%This is the case when developer has just written $\pi$ invocations in $d_a$.

\begin{table*}[t!]
	\centering	
	\vspace{-.2cm}	
	\caption{Experimental configurations.}
	\vspace{-.2cm}
	\scriptsize	
	\begin{tabular}{|p{0.6cm}|p{1.0cm}|p{0.6cm}|p{14.2cm}|}	\hline
		\rowcolor{verylightgray}
		\textbf{Conf.} & \textbf{$\delta$} & \textbf{$\pi$} & \textbf{Description}   \\ \hline			
		\textbf{C1.1} & $\Delta/2-1$ & $1$ & Nearly the first half of the 
		declarations is used and the second half is discarded. The last 
		declaration of the first half is selected as the active declaration 
		$d_a$. For $d_a$, only the \emph{first} invocation is provided as  
		query, and the rest is used as ground-truth data, \ie \texttt{GT(p)}. 
		This configuration represents an early stage of the development process 
		and, therefore, only limited context data is available to feed the 
		recommendation engine.  \\ \hline
		\textbf{C1.2} & $\Delta/2-1$ & $4$ & Similarly to \textbf{C1.1}, almost the first half of the declarations is retained and the second half is removed. $d_a$ is the last declaration of the first half declarations. For $d_a$, the first \emph{four} invocations are provided as query, and the rest is \texttt{GT(p)}.   \\ \hline
		\textbf{C2.1} & $\Delta-1$ & $1$ & The last method declaration is selected as testing, \ie $d_a$ and all the remaining declarations are used as training data. In $d_a$, the \emph{first} invocation is kept and all the others are taken out as ground-truth data \texttt{GT(p)}. This mimics a scenario where the developer is almost finished implementing $p$.  \\ \hline
		\textbf{C2.2} & $\Delta-1$ & $4$ & Similar to \textbf{C2.1}, $d_a$ is selected as the last method declaration, and all the remaining declarations are used as training data. The only difference with \textbf{C2.1} is that in $d_a$, the first \emph{four} invocations are used as query and all the remaining ones are used as \texttt{GT(p)}.  \\ \hline			
	\end{tabular}
	\label{tab:Configurations}	
%	\vspace{-.1cm}
\end{table*}

For the purpose of validation, the original dataset (cf. Section~\ref{sec:Dataset}) was split into two independent parts, namely a \emph{training set} and a \emph{testing set}. In practice, the training set represents the OSS projects that have been collected ex-ante, and they are available at the developer's disposal, ready to be exploited for any mining purposes. The testing set represents the project being developed, or \emph{the active project}. %In this way, our evaluation mimics a real development scheme:~\emph{the system should produce recommendations for the active project based on the data from a set of existing projects}. 
We opted for k-fold cross validation \cite{Kohavi:1995:SCB:1643031.1643047} as it has been widely chosen to study machine learning models. Depending on the availability of input data, the dataset with n elements is divided into $f$ equal parts, so-called \emph{folds}. For each validation round, one fold is used as testing data and the remaining $f$-1 folds are used as training data. In our evaluation, two values were selected, \ie~$f$=10 and $f$=n. The former corresponds to \emph{ten-fold cross validation} while the latter corresponds to \emph{leave-one-out cross validation} \cite{WONG20152839}, and they are exploited depending on the purpose as well as the availability of data. With ten-fold cross validation, we shuffle the list of the apps considered in the evaluation, and then randomly split them into ten equal parts. In the evaluation, we attempt to equally distribute the projects into the folds, so as to maintain a balance among the folds with respect to the projects' size. For every experiment, the execution is done ten times: each time one fold is used for testing, and the remaining nine folds are used as training data. Eventually, we averaged out the metrics obtained from the ten folds to get the final results.

\vspace{.1cm}

%This aims to equally distribute the projects into the folds. 

\subsubsection{Evaluation Metrics}
\label{sec:metrics}
For a testing project $p$, the outcome of a recommendation process is a ranked list of invocations, \ie~\code{REC(p)}. We believe that the ability to provide accurate invocations is important in the context of software development. Thus, we are interested in how well a system can recommend API invocations that eventually match with those stored in \code{GT(p)}. To measure the performance of \UM, PAM and \FC, we utilize \emph{Success rate}, \emph{Precision}, \emph{Recall}, \emph{Levenshtein distance}, and \emph{Time}. Given that $REC_{N}(p)$ is the set of \emph{top-N} items and $match_{N}(p)=  GT(p) \bigcap REC_{N}(p) $ is the set of items in the \emph{top-N} list that match with those in the ground-truth data, then the metrics are defined as follows.

%\cite{DiNoia:2012:LOD:2362499.2362501}. 
%Given a ranked list of recommendations, a developer typically pays attention to the \emph{top-N} items only. 
%\emph{Success rate} and \emph{accuracy} are computed by using $N$ as the \emph{cut-off value}. 

%\smallskip
%\noindent
%%$\blacksquare$~

\vspace{.1cm}
\noindent
$\rhd$~\textbf{Success rate.} %{\bf Accuracy:} 
For a set of \emph{P} testing projects, this metric measures the rate at which a recommender can return at least a match among the \emph{top-N} items for every project $p \in P$.
\vspace{-.2cm}
\begin{equation} \label{eqn:RecallRate}
%Success\ rate@N=\frac{ count_{p \in P}( \left |  match_{N}(p) \right | > 0 ) }{\left | P \right |} \times 100\%
success\ rate(p) =\frac{ \sum_{i}^{N} match_{i}(p) }{\sum_{i}^{N} \left| G_{i} \right|} \times 100\%
\end{equation}

%\noindent
%where \emph{count()} counts the number of times the boolean expression given as parameter evaluates to \emph{true}. 

\vspace{.1cm}
\noindent
$\rhd$~\textbf{Precision and recall.} %The metrics are employed to measure accuracy \cite{DiNoia:2012:LOD:2362499.2362501}. 
\emph{Precision} \emph{P@N} is the fraction of the \emph{top-N} recommended items to the total number of items in the ground-truth, while \emph{recall} \emph{R@N} is the ratio of the ground-truth items being found in the \emph{top-N} items: %,ISINKAYE2015261, %Davis:2006:RPR:1143844.1143874,
\vspace{-.1cm}

\begin{minipage}{0.47\linewidth}
	\begin{equation} \label{eqn:Precision}%\nonumber%
	P@N = \frac{ \left |  match_{N}(p) \right | }{N}
	\end{equation}
\end{minipage}% 
\begin{minipage}{0.47\linewidth}
	\begin{equation} \label{eqn:Recall}%\nonumber%
	R@N = \frac{ \left |  match_{N}(p) \right | }{\left | GT(p) \right |}
	\end{equation}
\end{minipage}%\par\vspace{\belowdisplayskip}

%\begin{equation} \label{eqn:Precision}
%P@N = \frac{ \left |  match_{N}(p) \right | }{N}
%\end{equation}
%
%and \emph{recall@N} is the ratio of the ground-truth items being found in the \emph{top-N} items: %,ISINKAYE2015261, 
%\vspace{-.1cm}
%\begin{equation} \label{eqn:Recall}
%R@N = \frac{ \left |  match_{N}(p) \right | }{\left | GT(p) \right |}
%\end{equation}

%\vspace{.1cm}
%\noindent
%$\rhd$~\textbf{F$_1$ score.} It is a harmonic average of precision and recall. %and computed as follows:%by means of the following formula:
%%\vspace{-.1cm}
%\begin{equation} \label{eqn:FMeasure}
%%F-Measure = \frac{ 2 \times precision \times recall }{precision + recall}
%F_{1}@N = \frac{2 \cdot P@N\cdot R@N}{P@N + R@N}
%\end{equation}

\vspace{.1cm}
\noindent
$\rhd$~\textbf{Levenshtein distance.}
%\MAX{to be completed based on what definition of Levensthein we actually use}
%==============================https://dzone.com/articles/the-levenshtein-algorithm-1==============================
Given two strings $s_1$ and $s_2$, the Levenshtein edit distance between them corresponds to the number of substitutions performed to transform $s_1$ to $s_2$. The metric is defined as follows.\footnote{\url{https://dzone.com/articles/the-levenshtein-algorithm-1}}
\begin{equation}\label{eqn:Levenshtein}
\resizebox{.9\linewidth}{!} {$
	L_{s{_1},s{_2}}(i,j) = 
	\begin{cases}
	max(i,j) & \text{if min(i,j)=0},\\
	%\text{0} &\quad\text{otherwise} \\
	min	
	\begin{cases}
	L_{s{_1},s{_2}} (i - 1,j) +1\\
	L_{s{_1},s{_2}} (i, j - 1) +1\\
	L_{s{_1},s{_2}} (i -1, j +1) +1\\
	\end{cases} & \text{otherwise.}
	\end{cases}
	$}
\end{equation}
where $i$ and $j$ are the terminal character position of strings $s_1$ and $s_2$, respectively.%; and $a_i$ and $b_j$ refer to the character in position $i$ of $s_1$ and the character in position $j$ of $s_2$, respectively

%Because the Eq.~\ref{eq:levenshtein} does not consider the length of the strings, we normalized the Levenshtein distance as following:
%\begin{equation}\label{eq:levenshtein_norm}
%lev^{norm}_{a,b} = \frac{lev_{a,b}}{max(|a|,|b|)}
%\end{equation}
%where $|a|$ and $|b|$ are the length of the strings $a$ and $b$ respectively, and $lev_{a,b}$ is calculated by Eq~\ref{eq:levenshtein} .

\vspace{.1cm}
\noindent
$\rhd$~\textbf{Recommendation time.} The time needed for the systems to generate predictions is measured using a laptop with Intel Core i5-7200U CPU @ 2.50GHz$\times$4, 8GB RAM, and Ubuntu 16.04. %has been exploited by both PAM and \FC to perform a prediction. % on a given infrastructure
\vspace{-.2cm}

%\begin{table}[t!]
%	\vspace{-.3cm}
%	\scriptsize
%	\caption{Categories and cardinality of the set of 500 apps.}
%	\centering
%	\vspace{-.2cm}
%	\begin{tabular}{|l|c|l|c|}\hline
%		\rowcolor{verylightgray}
%		%		& PAM & \multicolumn{4}{c|}{\FC} \\ \hline
%		\textbf{Category} & \textbf{Cardinality}  & \textbf{Category} & \textbf{Cardinality} \\ \hline
%		Tools & 151 & Board & 5 \\ \hline
%		Productivity & 44 & Health \& Fitness & 5 \\ \hline
%		Education & 28 & Video Players &  5 \\ \hline
%		Libraries \& Demo & 27 & Business & 4 \\ \hline
%		Entertainment & 23 & Medical &  4 \\ \hline 
%		Arcade & 23 & Role Playing &  4 \\ \hline
%		Music \& Audio & 20 & Adventure & 3 \\ \hline
%		Communication & 19 & Educational & 3 \\ \hline 
%		Personalization & 16 & Unknown & 3 \\ \hline
%		Casual & 13 & Action & 2 \\ \hline
%		Puzzle & 13 & Shopping & 2 \\ \hline
%		Maps \& Navigation & 12 & Racing & 2 \\ \hline
%		Books \& Reference & 10 & Strategy & 2 \\ \hline
%		Lifestyle & 9 & Events & 1 \\ \hline
%		Travel \& Local & 8 & News \& Magazines & 1 \\ \hline
%		Photography & 7 & Food \& Drink & 1 \\ \hline
%		Weather & 7 & Simulation &  1 \\ \hline
%		Sports & 6 & Music &  1 \\ \hline
%		Social & 6 & Comics & 1 \\ \hline
%		Card & 6 & --- & --- \\ \hline
%		%		\bottomrule
%	\end{tabular}
%	\label{tab:Dataset500}
%	\vspace{-.2cm}
%\end{table}

\subsubsection{How we address RQ$_1$-RQ$_4$}
\smallskip
\noindent
$\rhd$~\textbf{RQ$_1$.}
To address RQ$_1$, we compare the performance of \FC with that of \UM and PAM.
Our experience% with the previous work
~\cite{Nguyen:2019:FRS:3339505.3339636} reveals that PAM cannot scale well with large datasets, \ie it suffers from a high computational complexity. Meanwhile, \FC is more efficient as it is capable of incorporating a large number of background projects and swiftly producing recommendations. In particular, both systems were experimented on a mainstream laptop using a set of 549 training projects with 80MB in size to measure the execution time~\cite{Nguyen:2019:FRS:3339505.3339636}. On average, PAM requires 320 seconds to provide a recommendation, while \FC needs just 1.80 seconds. Through a careful observation on the Android dataset (cf. Section~\ref{sec:Dataset}), we realized that many of them are big in size, and a training set of 2,360 apps may add up to more than 2.0GB. This essentially means that it is infeasible to run PAM on the entire dataset, since the execution time may exponentially soar. Thus, for RQ$_1$ %to compare \FC with PAM, 
we can leverage only a portion of the original corpus. To be more precise, we selected 500 apps of average size. %whose categories are summarized in Table~\ref{tab:Dataset500}. 
There are 39 categories in total and most of them contain a small number of apps, while \emph{Tools} is still the biggest category with 151 apps, accounting for 30.20\% of the total amount. We opted for \emph{leave-one-out cross-validation} \cite{WONG20152839}, aiming to exhaustively exploit the background data.
We study the performance of \FC by considering all the four configurations listed in Table~\ref{tab:Configurations}, \ie \textbf{C1.1}, \textbf{C1.2}, \textbf{C2.1}, and \textbf{C2.2}. The cut-off value $N$ is used to investigate how accurately the system is able to provide recommendations with respect to different lengths of the ranked list. In RQ$_1$, we set $N$ to 30, attempting to study the three systems on a long list of recommendations.
%\MAX{can we justify this choice?}
We also consider, as can be seen in Eq.~\ref{eqn:combinedRating}, different values of the number of neighbor apps, \ie $k$=$\{1,2,3,4\}$. The evaluation was executed 500 times, by each validation, one app is used as testing and all the remaining 499 apps are used for training. To aim for a reliable comparison, we ran \UM and PAM using their original settings in our evaluation.

\smallskip
\noindent	
$\rhd$~\textbf{RQ$_2$.}	
For this research question, we made use of the whole corpus introduced in Section~\ref{sec:Dataset}, which contains all the 2,600 collected apps. Moreover, since we have a larger amount of data compared to RQ$_1$, we employ ten-fold cross-validation in this research question. We analyze the performance of \FC for combinations of: \emph{(i)} different configurations, \ie \textbf{C1.1}, \textbf{C1.2}, \textbf{C2.1}, and \textbf{C2.2}; \emph{(ii)} different values of $N$, \ie $N$=$\{1,5,10,15,20\}$; and \emph{(iii)} different values of $k$, \ie $k$=$\{1,2,3,4,6,10\}$. The rationale behind the selection of such specific values is as follows. We should incorporate a certain number of neighbor projects $k$ when computing recommendations, otherwise the matrix will become big (cf. Fig.~\ref{fig:3DRepresentation}), which possibly induces an expensive computational cost.
%At the same time, it is unnecessary to consider a large cut-off value \emph{N}, since in practice the developer may not skim through a long list of recommended API calls. 
While such a large number of $N$ seems to be unrealistic, in the scope 
of our evaluation, we have to consider it to ensure the generalizability of our 
final conclusions. In practice, a small enough number of $N$ items should be 
presented to the developers, so as to avoid overwhelming them.
We report, for different configurations and values of $N$ and $k$, the success rate, and performance gain. Also, we plot the precision/recall curves for different configurations and values of $k$.

%\revised{While such a large number of N seems to be unrealistic, in the scope of our evaluation, we have to consider it to ensure the generalizability of our final conclusions. In practice, a small enough number of N items should be presented to the developers, so as to avoid overwhelming them.} % making them confused

\smallskip
\noindent	
$\rhd$~\textbf{RQ$_3$.} To address RQ$_3$, we perform controlled experiments on the whole dataset described in Section~\ref{sec:Dataset}. Similar to RQ$_2$, we conducted the experiments following the ten-fold cross-validation methodology.
%\MAX{do we also use the whole dataset in RQ2? if so, we should just say in RQ1 (as I said above) that we use the smaller dataset}
% As seen in Fig.~\ref{fig:Statictics}, 
The apps collected in the corpus span over a total of 47 categories, such as \emph{Productivity}, \emph{Communication}, \emph{Music \& Audio}, or \emph{Business}. The cardinality (\ie the number of apps within a category) of the categories varies considerably: most of them contain a small number of apps, \ie ranging from 1 to 20 items for almost half of the topics. The biggest category with 659 apps is \emph{Tools}, while there are three categories with only two apps, \ie \emph{Trivia}, \emph{Music}, and \emph{Parenting}.

With this research question, we aim at examining if there is a strong positive correlation between two variables, \ie the cardinality of a category and the corresponding precision. %for that category.}
In other words, we hypothesize that apps belonging to populous categories might possibly get a better recommendation since they have more, presumably, \emph{relevant} background data, \ie projects coming from the same domains. This would have an impact in practice as follows: once the  developer specifies one or more domains for her app, we can search for recommendations just by looking for apps within the same domains, aiming to narrow down the search scope. This is useful since it contributes to a reduction in the overall execution time. However, this is a pure assumption, which needs to be carefully studied through concrete experiments.

For each category, we computed the precision for all of its constituent apps following Eq.~\ref{eqn:Precision}, and the precision of a category was averaged out over the apps. Eventually, the correlation between the cardinality and precision is computed using the Spearman's rank correlation coefficient and Kendall, \ie $\rho$ and $\tau$, respectively. The coefficients range from -1 (perfect negative correlation) to +1 (perfect positive correlation), while $\rho$=0 or $\tau$=0 implies that the variables are not correlated at all.
The reason why we compute both Spearman's and Kendall's correlation is because the number of categories is relatively small, and the Spearman's correlation may be more suitable in this case. We do not use the Pearson's correlation as we cannot assume the presence of a linear relationship between categories and precision.
	
\smallskip
\noindent	
$\rhd$~\textbf{RQ$_4$.}
%\revised{To address RQ$_4$ \MAX{to be completed}}
In this research question, we study if \FC is able to recommend %real \MAX{honestly I would drop the word ``real'' what is ``non-real'' source code?} 
source code relevant to the method declaration %\MAX{declaration?} 
under development, exploiting the ten-fold cross-validation technique. As an example, we assume that the developer is working on the incomplete code snippet depicted in Fig.~\ref{fig:originalCode}, and \FC is expected to suggest real code such as the one in Fig.~\ref{fig:recCode}, or the one in Listing~\ref{lst:RecommendedSnippet}.

To evaluate the similarity between two declarations, we compare their constituent APIs. This comparison is based on the observation coming from an existing work~\cite{McMillan:2012:DSS:2337223.2337267} that if projects or declarations share API calls implementing the same requirements, then they are considered to be more similar than those that do not have similar API usage. Following the same line of reasoning, we evaluate the similarity/relevance between two snippets by examining if they share common API function calls and have the same sequence of these calls.

%According to a previous work~\cite{McMillan:2012:DSS:2337223.2337267}, if projects share API calls implementing the same requirements, then the projects are considered to be more similar than those that do not have similar API usage. 
% By exploiting the curated corpus in Section~\ref{sec:Dataset}, we performed a series of experiments.
%, to see if it is still able to provide relevant code snippets

To address this research question, we leverage the dataset of 500 apps also used to address RQ$_1$.
%\MAX{we go back and forth between small and large datasets. Can we justify why?}	
We deliberately make use of such a small dataset due to the following reason: with this dataset, we analyze the ability of \FC to recommend relevant code snippets, given that there is a fairly small amount of training data. %As shown later in the paper, \FC can considerably improve its performance when more similar projects are available for training.
%\MAX{forward reference. Why we don't move this discussion in the results}
%The reason why we perform this analysis with a small dataset is because 
We conjecture that, as confirmed later in the paper, if \FC works effectively on a 
small dataset, it will perform well on 
bigger ones. To evaluate if a 
recommended snippet is relevant to the query, we measure the level of 
similarity between them using the Levenshtein edit 
distance~\cite{levenshtein1966bcc}, which has been used by prior work for 
similar purposes, \eg tracking source code 
clones~\cite{10.1007/s10664-009-9108-x}. Given the source code of a declaration 
d$_1$, we parse it using Rascal to get the API invocations. Afterwards, we 
encode each of the invocations using a unique character, resulting in a string 
s$_1$. Thus, the evaluation of the similarity between two declarations d$_1$ 
and d$_2$ boils down to comparing the corresponding strings s$_1$ and s$_2$, by 
counting the number of replacements needed to convert s$_1$ to s$_2$ using 
Eq.~\ref{eqn:Levenshtein}. Such a metric takes into account not only the common 
characters between s$_1$ and s$_2$, but also the order in which they appear. 
Correspondingly, this means that two code snippets are  
similar/relevant if they share common API function calls as well as have the same sequence of the calls. In this sense, the smaller the distance we get, the more similar the two snippets are, and vice versa.

To simplify the comparison performed in RQ$_4$, we only used Configuration \textbf{C1.2} (cf. Table~\ref{tab:Configurations}). The rationale behind the selection of the configuration is as follows: it represents a more authentic development scenario, corresponding to the situation where the developer already finishes a part of the declaration, and she expects to get recommendations. To be more concrete, given a testing project, we kept the first half of the declarations and removed the second half; the last declaration of the first half declarations is selected as the testing one $d_a$. For $d_a$, the first four invocations are provided as query, and the rest is \texttt{GT(p)}. Using the \textit{Code Builder} subcomponent (cf. Section~\ref{sec:CodeBuilder}), we extracted the real source code of a declaration by means of the computed M$^3$ model and the project location.
%\MAX{should we way why we have chosen C1.2 in particular?}

In fact, APK files do not contain source code, thus it is not possible to directly mine real code snippets from the apps. However, \FC allows us to extract the method canonical name of a recommended code snippet within the project scope. Moreover, since the dataset is extracted from \emph{AndroidTimeMachine}, there is a mapping between open-source Google store apps with their corresponding repositories. 
To locate the right pair of APK file and \GH repository, %we also attempt to get the right version of the code by 
we check the snapshot date when %AndroidTimeMachine created 
the mapping was created. In this way, we are able to trace back to the original source code for those apps that have a counterpart in \GH. Eventually, FOCUS is able to recommend source code, as long as the corresponding app is associated with a source project rooted in \GH.

%Finally, we want to see if \FC is useful from a developer point of view. To this end, we conducted a user study to study the relevance of API calls and code snippets provided by \FC to support a particular development context. A group of 20 master students in Computer Science has been involved to evaluate two real-world development scenarios.
%\revised{We perform a user study.}
%Before conducting the user study, 
%. we describe the user study conducted

\subsection{Second Evaluation: User Study}
\label{sec:userstudy}
In this section, we study \FC's usefulness of code and API recommendations by means of a task-based user study to address RQ$_5$. 

The \emph{goal} of this study is to evaluate \FC, with the purpose of understanding whether %the extent to which 
it could help developers with their implementation tasks. The \emph{quality 
target} of the study is the perceived usefulness that developers have of 
recommendations (code snippets and APIs) provided by \FC.
%We study \FC's usefulness of code and API recommendations by means of a task-based user study. 
%This aims to evaluate to what extent the use of our tool is helpful to programmers during a development task. 
The \emph{context} consists of participants, \ie \numParticipants Master's students in Computer Engineering, and objects, \ie programs involving command line argument parsing and HTML download/parsing.

%To collect information about the participants, we provided a questionnaire and asked the students to answer it. 
\begin{table}[t!]
%	\color{blue}
	\scriptsize
	\centering
	\caption{Task assignments to the evaluator groups.}
	\label{tab:evaluator_task_assignement}
	\begin{tabular}{|l|l|l|}
		\hline
		& \textbf{T1}          & \textbf{T2}          \\ \hline
		\textbf{Group I}   & Apache-cli, using \FC & gson                     \\ \hline
		\textbf{Group II}  & Apache-cli               & gson, using \FC       \\ \hline
		\textbf{Group III} & gson                     & Apache-cli, using \FC \\ \hline
		\textbf{Group IV}  & gson, using \FC       & Apache-cli               \\ \hline
	\end{tabular}
%	\color{black}
\end{table}

\subsubsection{Study Design and Tasks}
As shown in Table \ref{tab:evaluator_task_assignement}, the experimental design is a crossover design in which participants were split into four groups (each participant worked individually, but each group received the same treatment). Each participant had to carry out two implementation exercises, one using \FC recommendations and another without the availability of \FC. Different groups featured different ordering of the treatments, to mitigate any ordering/learning effect.

%\MAX{I suggest to put all the experimental material in a single place}
%We defined two tasks,\footnote{\url{https://docs.google.com/document/d/1iiU5tf9XZKaiFlp4FJt3YY3Sh9eyowdLp4ZDhlI1ILY/edit?usp=sharing}} and 
The two tasks focus on the usage of different libraries, 
\ie~\textit{commons-cli}\footnote{\url{https://commons.apache.org/proper/commons-cli/}}
 and \textit{jsoup},\footnote{\url{https://jsoup.org/}} and require the 
completion of three partially implemented methods.  \textit{commons-cli} 
provides APIs for parsing command-line options passed to programs, while 
\textit{jsoup} is a library for parsing and manipulating HTML pages using the 
best of DOM, CSS, and jquery-like methods.

% the method requirements at

\vspace{.1cm}
\begin{lstlisting}[caption={The commons-cli \textit{getOption()} partially implemented method.}, label=lst:getoption, 
	style=JavaStyle,captionpos=b,xleftmargin=1.8em,frame=single,framexleftmargin=1.2em]
/**
* Create apache-cli options for the following elements:
*  url (Mandatory),
*  username (Mandatory): -user <user>
*  password (Mandatory): -pass <password>
*  query (Mandatory): -sql <query>
*  CSV file path: -f -file <filepath>
*  includeHeaders: -headers
*  All the options contains and argument, with the exception of includeHeaders   
* @return Available command line options
*/
public static Options getOptions() {
	final Options options = new Options();	
	final Option urlOption = new Option("url", true, "database url <jdbc:subprotocol:subname>");	
	final OptionGroup urlGroup = new OptionGroup();
	urlGroup.setRequired(true);
	urlGroup.addOption(urlOption);
	options.addOptionGroup(urlGroup);
	
	//COMPLETE THE METHOD
	return options;
}
	
\end{lstlisting}

\begin{lstlisting}[caption={The unit tests for checking the correctness of the task.}, label=lst:unit_test, 
	style=JavaStyle,captionpos=b,xleftmargin=1.8em,frame=single,framexleftmargin=1.2em]
@Test
public void getOptionTest() throws IOException {
	Options options = Launcher.getOptions();
	assertEquals(6,options.getOptions().size());	
}

@Test
public void parseOKTest() throws Exception {
	String[] arguments = new String[]{"-url", "a", 
		"-pass", "pass",
		"-user", "user",
		"-sql", "sql"};
	assertEquals(4,Launcher.parse(arguments).size());
}

@Test
public void printUsageTest() throws IOException {
	assertNotEquals("", Launcher.printUsage());
}
\end{lstlisting}

For the tasks  with \textit{commons-cli}, the participants completed 
three methods by: \textit{(i)} implementing a method for specifying the 
command-line options (we provided the evaluator with the parameter list); 
\textit{(ii)} parsing the command line parameters and throwing an exception if 
the mandatory ones are missing; \textit{(iii)} handling parsing exception by 
printing possible options to the console. Listing~\ref{lst:getoption} shows an 
example of the partial implementation and the method requirements for 
specifying options. For a detailed description of the two performed tasks, due 
to space limit, interested readers are kindly referred to our online 
appendix.\footnote{\url{https://mdegroup.github.io/FOCUS-Appendix/tasks.html}}

For each method to be completed, we provided (for treatments having the availability of \FC) each evaluator with the \textit{top-5} snippets and \textit{top-20} method invocations recommended by \FC by giving the initial and partial method implementation as input.

%The participants were split into four groups as listed in Table~\ref{tab:evaluator_task_assignement}. Each group had to perform both tasks, one with and the other without \FC, however in reverse order. For instance, Group I and III performed exactly the same set of tasks, \ie Apache-cli using \FC (T1), and gson without \FC (T2), however Group I performed T1 before T2, while Group II conducted T1 after T2. This aims at eliminating any possible bias against a specific task caused by the familiarity with the tool.} 
%each participant worked with and without \FC in different orders, as well as on the two different tasks. 

\subsubsection{Study Operation}

Under the circumstance in which the experiment was conducted, it was neither possible to perform the experiment in a laboratory\footnote{Due to the COVID-19 emergency in 2020} nor to ask participants to return the results immediately. Instead, each participant could perform the tasks offline and return them to us. Before the study, we performed a laboratory introductory session in video conference, in which we introduced to participants the laboratory goals and tasks (without details about our research question, to avoid biasing them), and left them a detailed instruction documents.

%\revised{
During the tasks, participants could access any resource available on the Internet, besides \FC recommendations when available based on the study design.	
Once a participant finished the tasks, s/he had to complete a  questionnaire\footnote{\url{https://forms.gle/uoqSTaQ94PArdUST6}} consisting of the following questions: \textit{(i)} three general questions asking about their experience in programming and code search engine; \emph{(ii)} four questions, in a 5-level Likert scale \cite{Oppenheim:1992}, related to the understandability and complexity of the assigned tasks; and \textit{(iii)} four questions to evaluate the relevance and usefulness of the recommendations provided by \FC.%} %The last two groups of questions adopts a five-point Likert scale as the possible answer.} 

%\MAX{This paragraph should be moved later, maybe even in section 5, at the beginning of RQ5 results}
%\revised{Among them, 30\% and 50\% of the students have three years and more than four years of programming experience, respectively. Most of them use a code search engine in a daily basis. Concerning the tasks' comprehensiveness, 16 out of the students, corresponding to 80\%, agree that the tasks are clear and easy.}
 
%\revised{
Moreover, we asked the participants to submit their implementations. Such implementations have been used for understanding the correctness of the resulting code. For each method to be completed, we defined a specific JUnit\footnote{\url{http://junit.org/junit4}} unit test for checking their correctness. We did not provide the evaluator with the test methods to avoid bias towards the experiment. Listing~\ref{lst:unit_test} reports the simple testing methods used to check the correctness of the submitted task. Although the unit tests are rather simple, they have been able to effectively catch any possible implementation failures.%}

%\revised{
Then, we involved a senior developer experienced with Java 
programming, \emph{gsoup} and \emph{commons-CLI} libraries to further 
investigate the method implementations where the unit test fails. The senior 
developer checked the severity of the identified errors and discarded those 
that are not related to the usage of the involved library. For instance, some 
evaluators named the parameters differently, \eg they used \textit{password} 
instead of \textit{pass} or \textit{username} instead of \textit{user}. 
Consequently, the dedicated \textit{parseOKTest} test fails because of a wrong 
parameter naming. We marked this type of failure as a minor one, and we 
considered the implementation as correct for the evaluation scope.

\subsubsection{How we address RQ$_5$}
There are %we perform 
the following analyses to address \textbf{RQ$_5$}:
\begin{itemize}
\item We perform a Wilcoxon signed rank test \cite{wilcoxon} to determine whether there is any statistically significant difference between the number of passed tests for tasks implemented with and without \FC ($H_0$: \emph{there is no significant difference between the percentage of tests passed with and without the availability of \FC}). Also, we compute the Cliff's delta effect size \cite{Cliff:2005}.
\item As for the questionnaire results, we report them using diverging stacked bar charts and discuss them.
\end{itemize}
%\revised{Finally, we performed statistical test to measure the significance between the correctness of the task and the usage of \FC. The statistical study is reported in Section~\ref{sec:RQ4}.}

%provide developers with real 
%for a given declaration, 

%\smallskip
%\noindent	
%\MAX{this is not related to RQ5}
%$\rhd$~\revised{\textbf{Cross-validation.}
%It is worth mentioning that we use leave-one-out cross-validation only for RQ$_1$, thereby exhaustively exploiting the available data. For the remaining questions, \ie RQ$_2$, RQ$_3$, and RQ$_4$, we employ the ten-fold cross-validation method, \ie we shuffle the list of the apps considered in the evaluation, and then randomly split them into ten equal parts. This aims to equally distribute the projects into the folds. By a careful observation, we see that there is a balance among the folds with respect to the projects' size. For every experiment, the execution is done ten times: each time one fold is used for testing, and the remaining nine folds are used as training data. Eventually, we averaged out the metrics obtained from the ten folds to get the final results.}

\vspace{0.2cm}
	
	\section{Results}
	\label{sec:Results}

%it is impractical to introduce 
% analyzes
%\revised{This section analyzes the results we obtained from the experiments.} 
%\noindent{\bf RQ$_1$:} {\em \rqfirst}
% of 500 apps considered for this research question
% gives a summary of the categories and their cardinality for the subset

This section analyzes the experimental results obtained through the evaluation by referring to the four research questions mentioned in Section~\ref{sec:rqs}.

\subsection{\rqfirst}

\begin{table}[h!]
%	\footnotesize
	\scriptsize
	\vspace{-.3cm}
	\caption{Success rate of \UM, PAM and \FC.} %Comparison between
	\centering
	%	\begin{tabular}{|>{\bfseries} c | *{6}{R}|}		 \hline
%	\vspace{-.2cm}
	\begin{tabular}{|l|c|c|c|c|c|c|}\hline
		\rowcolor{verylightgray}
		& \textbf{\UM} & \textbf{PAM} & \multicolumn{4}{c|}{\textbf{\FC}} \\ \hline
%	    & \textbf{PAM~\cite{Fowkes:2016:PPA:2950290.2950319}} & \textbf{\UM~\cite{Wang2013Mining}} & \multicolumn{4}{c|}{\textbf{\FC}} \\ \hline
		& --- & --- & k=1 & k=2 & k=3 & k=4 \\ \hline
		\textbf{C1.1} 			& 41.66  & 49.60 &  81.20 & 81.60 & 82.00 & \textbf{83.80}  \\ \hline
		\textbf{C1.2} 			& 37.33  & 52.20 &  89.81 & 91.10 & 92.80 & \textbf{93.22}  \\ \hline
		\textbf{C2.1} 			&  44.10 & 58.22 &  77.00 & 78.20 & 78.60 & \textbf{79.60}  \\ \hline
		\textbf{C2.2} 		    &  40.66 & 58.40 &  89.60 & 90.20 & 91.60 &  \textbf{92.10} \\ \hline
		%		\bottomrule
	\end{tabular}
	\label{tab:ComparisonPAMandFOCUS}
	\vspace{-.1cm}
\end{table}

Table~\ref{tab:ComparisonPAMandFOCUS} reports the success rate  for PAM and \FC, considering different configurations and values of k representing the number of neighbor apps. The cut-off value $N$ was set to 30, attempting to investigate the systems' performance for a long list of recommendations. The table shows an evident outcome: \FC always achieves a much better success rate than that of PAM and \UM by all the configurations. For instance, with \textbf{C1.2}, \FC gets 89.81\%, 91.10\%, 92.80\%, and 93.22\% as success rate by $k$=1, $k$=2, $k$=3, and $k$=4, respectively, while PAM and \UM get 52.20\% and 37.33\%, respectively. With \textbf{C2.2}, \FC gets a maximum success rate of 92.10\%, which is superior than 58.40\% and 40.66\% obtained by PAM and \UM, respectively. We further confirm the claim by Fowkes and Sutton~\cite{Fowkes:2016:PPA:2950290.2950319}, \ie PAM outperforms \UM also in our setting. Concerning the execution time,
for the given dataset, both \FC and \UM provide a recommendation in less than 0.01 seconds. Specifically, such a time is of 3.8$\times$10$^{-4}$ for \UM (the fastest), 8$\times$10$^{-3}$ seconds for \FC, and 1.6 seconds for PAM.

The performance gain obtained by \FC is understandable in the light of 
the following arguments. \UM works on the basis of clustering techniques and it 
is dependent on the similarity among groups of APIs. In other words, \UM 
computes similarity at the sequence level, \ie invocations that are usually 
found together. PAM is a complex system, which consists of six building blocks, 
\ie probabilistic model, inference, learning, inferring new patterns, candidate 
generation, and mining interesting patterns. The system uses a probability 
distribution to define a distribution over all possible API patterns present in 
client code, based on a set of API patterns. It also employs a generative model 
to infer the most probable patterns from ARFF files. Finally, the system 
generates candidate patterns by relying on the highest support first rule, \ie 
searching for the best candidate earlier. Due to these technical details, both 
\UM and PAM can recommend APIs that commonly appear in different code snippets. 
In contrast, \FC is able to consider similarity both at the project level and 
the declaration level. Therefore, given an active project, \FC mines API calls 
from the most similar declarations in the most similar projects. As a result, 
this allows \FC to outperform both \UM and PAM in finding invocations that fit 
well to a given context.

%\revised{The performance gain obtained by \FC is understandable in the light of the following arguments. PAM employs a generative model that infers the most probable API patterns from ARFF files. The algorithm aims to generate candidate patterns by relying on the highest support first rule, \ie searching for the best candidate earlier. Though this technique works properly with the dataset in the original work~\cite{Fowkes:2016:PPA:2950290.2950319}, its performance decreases on the dataset with big ARFF files in our work. 
%%It depends on the dimension of the considered ARFF files which are bigger than the original ones. %Furthermore, we tune some PAM parameters in order to get the recommendations in a suitable time \ie we reduced the maximum number of iterations performed for each file from 1000 to 10. 
%%In particular, the performance of PAM decreases when large ARFF files are fed input. %This means that there is an obvious threat to validity.
%\UM was evaluated by using a limited dataset, \ie it offers a small number of API instances that are used to train the tool. Furthermore, \UM can identify a few API clusters given the above-mentioned dataset. In contrast, \FC is capable of overcoming such limitations by relying on a content-based collaborative filtering technique that is not greatly affected by data dimension.}

It is worth noting that \FC gets a considerably high performance, given that the dataset is fairly small. The maximum success rate obtained by \textbf{C1.2} and \textbf{C2.2} is  93.22\% and 92.10\%, respectively. Compared to our previous work~\cite{Nguyen:2019:FRS:3339505.3339636}, where a set of 200 \GH projects was considered to compare \FC with PAM, we see that \FC substantially improves its recommendations when more data is incorporated into the training. A feature of the considered datasets which may affect the results obtained by \FC is the level of dependencies in Android apps compared to that of the \GH projects. In particular, by counting the number of unique APIs in each app/project for both the Android dataset and the GitHub dataset we see that the former contains more APIs compared to the latter. Many apps have more than 400 unique APIs, meanwhile, most of the GitHub projects have less than 200 unique APIs. This is further supported by previous work~\cite{viennot_measurement_2014,ruiz_understanding_2012}, which gives evidence that Android projects make heavy use of third-party libraries as well as native libraries.
	
%As stated in previous works, Android projects make heavy use of third-party libraries as well as native libraries \cite{viennot_measurement_2014,ruiz_understanding_2012}. Similarly, software reuse has been investigated in the \GH platform \cite{mockus_large-scale_2007,zhang_detecting_2017}. Although they don't directly assess the degree of dependency among repositories, such studies witness that the reuse of external software artifacts is widely spread even in the \GH community. Thus, \FC performances are not substantially affected by the different dataset \ie \GH, Android.
%One possible feature of the considered datasets which could affect FOCUS results is the level of dependency among the Android apps with respect to the GitHub projects. 
%

%could be different among the 

%Following the experiments on \code{MV$_L$} and \code{MV$_S$} from {\bf RQ$_1$}, we believe that this attributes to the limited background data available for the evaluation, since we only consider $200$ projects.

\vspace{.2cm}
\begin{shadedbox}
	\small{\textbf{Answer to RQ$_1$.} While UP-Miner is the fastest tool in terms of recommendation time, FOCUS is the most effective one as it substantially outperforms both UP-Miner and PAM with respect to the prediction performance, while keeping the recommendation time below 0.01 seconds. %the baselines %\MAX{how much?} %with respect to 
    Moreover, \FC mines better on Android apps compared to \GH projects.} % and timing efficiency
\end{shadedbox}

%In summary, by considering both \textbf{RQ$_2$} and \textbf{RQ$_3$}, we come to the conclusion that 
% and e, accuracy and execution time
%different real-world settings 
%First, we consider the success rates as shown in Table~\ref{tab:SuccessRate}.
%. This aims at investigating the results provided by \FC with respect to different amounts of input data, corresponding to

\vspace{-.2cm}

\begin{table*}[t!]
	\begin{minipage}{.5\linewidth} 
		\footnotesize
		\scriptsize
		\centering
		\vspace{-.2cm}
		\caption{Success rate (\%) for $k=\{2,3,4,6,10\}$ and $N=\{1,5,10,15,20\}$.}
		\vspace{-.3cm}
%		\label{tab:first_table}
		\label{tab:SuccessRate}
		\medskip
			\begin{tabular}{|l||c|c|c|c|c||c|c|c|c|c|} \hline
			& \multicolumn{5}{c||}{\textbf{C1.1}} & \multicolumn{5}{c|}{\textbf{C1.2}}  \\ \cline{2-11}%\hline
%			& \multicolumn{5}{c||}{k} & \multicolumn{5}{c||}{k}  \\ \cline{2-11}
			\textbf{N} & \textbf{k=2} & \textbf{k=3} & \textbf{k=4} & \textbf{k=6} & \textbf{k=10} & \textbf{k=2} & \textbf{k=3} & \textbf{k=4} & \textbf{k=6} & \textbf{k=10}  \\ \hline
			\rowcolor{verylightgray}
			\textbf{1}    &  67.46  &  69.10  &  71.34  &  74.00 & 75.76 & 85.69 & 87.53 & 88.19 & 89.38   &  89.96 \\ \hline
			\textbf{5}    &  76.84  &  78.42  &  80.38  &  82.53 & 84.80 & 91.11 & 92.42 & 92.84 & 93.88   &  94.53 \\ \hline 
			\rowcolor{verylightgray}
			\textbf{10}   &  80.46  &  82.30  &  83.42  &  85.38 & 87.84 & 92.80 & 93.92 & 94.50 & 94.92   &  96.03 \\ \hline
			\textbf{15}   &  81.76  &  84.05  &  84.84  &  87.15 & 89.15 & 93.69 & 94.61 & 95.23 & 95.61   &  96.50 \\ \hline 
			\rowcolor{verylightgray}
			\textbf{20}   &  82.80  &  84.88  &  86.23  &  87.88 & 90.11  & 94.07 & 94.96 & 95.61 & 96.11  &  96.92 \\ \hline
			%		\midrule
			& \multicolumn{5}{c||}{\textbf{C2.1}} & \multicolumn{5}{c|}{\textbf{C2.2}} \\ \hline
			%		& \multicolumn{5}{c||}{k} & \multicolumn{5}{c||}{k}  \\ \cline{2-11}
			\rowcolor{verylightgray}
			\textbf{1}    &  66.30  &  68.11  &  70.50  &  72.65 & 75.42 & 82.84 & 85.50 & 86.92 & 88.15   &  88.96 \\ \hline
			\textbf{5}    &  77.03  &  78.15  &  77.80  &  79.57 & 82.03 & 90.07 & 91.00 & 91.84 & 92.11   &  93.42 \\ \hline 
			\rowcolor{verylightgray}
			\textbf{10}   &  79.46  &  80.57  &  80.26  &  81.96 & 84.57 & 91.65 & 92.50 & 93.19 & 94.00   &  95.11 \\ \hline
			\textbf{15}   &  80.76  &  82.07  &  81.57  &  83.76 & 86.23 & 92.11 & 93.30 & 93.80 & 94.73   &  96.00 \\ \hline
			\rowcolor{verylightgray}
			\textbf{20}   &  81.73  &  84.00  &  84.92  &  87.34 & 89.07 & 92.65 & 93.88 & 94.26 & 94.92   &  96.15 \\ \hline
		\end{tabular}
	\end{minipage}\hfill
	\begin{minipage}{.5\linewidth}
		\scriptsize
		\vspace{-.2cm}
		\caption{Performance gain (\%) among the configurations.}
		\vspace{-.4cm}
		\centering % `center' is defined as an environment
		\label{tab:PerformanceGain}
		\medskip
		\begin{flushright}
			\begin{tabular}{|>{\bfseries} c | *{6}{R}|}		 \hline
				\multicolumn{6}{|c|}{\textbf{Gain of C1.2 w.r.t. C1.1}} \\ \hline
%				\multicolumn{6}{|c|}{\texttt{k}} \\ \hline
				N   & k=2 & k=3 & k=4 & k=6 & k=10  \\ \hline
				1 & 27.02 & 26.67 & 23.62 & 20.78   &  18.74 \\ \hline
				5 & 18.57 & 17.85 & 15.50 & 13.75   &  11.47 \\ \hline 
				10 & 15.34 & 14.12 & 13.28 & 11.30   &  9.32 \\ \hline
				15 & 14.59 & 12.56 & 12.25 & 9.71   &  8.24 \\ \hline 
				20 & 13.61 & 11.88 & 10.88 & 9.37   &  7.56 \\ \hline
				\multicolumn{6}{|c|}{\textbf{Gain of C2.2 w.r.t. C2.1}} \\ \hline
				1 & 25.14 & 25.53 & 23.29 & 21.34   &  17.95 \\ \hline
				5 & 16.93 & 16.44 & 18.05 & 15.76   &  13.89 \\ \hline 
				10 & 15.34 & 14.81 & 16.11 & 14.69   & 12.46 \\ \hline
				15 & 14.05 & 13.68 & 14.99 & 13.10   & 11.33 \\ \hline 
				20 & 13.36 & 11.76 & 11.00 & 8.68   &  7.95 \\ \hline
			\end{tabular}		
		\end{flushright}
	\end{minipage}
	\vspace{-.2cm}	
\end{table*}

\subsection{\rqsecond} \label{sec:RQ2}

In this research question, we are interested in understanding the 
\emph{completeness} and \emph{accuracy} of \FC's recommendations at different 
project's development stages. For the former, we analyze the corresponding 
success rate and performance gain, while for the latter, we take into 
consideration the obtained \emph{precision} and \emph{recall} values. 
Furthermore, we investigate the system's ability to recommend APIs in \emph{the 
long tail}.

% To assess their accuracy

%Table \ref{tab:SuccessRate} reports the success rate obtained . 
%, also with respect to varying the number of neighborhood projects and the cut-off value, \ie $k$ and $N$
% of \textbf{C1.1} and \textbf{C1.2} are $24.59\%$ and $30.65\%$, respectively

\smallskip
\noindent
$\rhd$~\textbf{Success rate.} Table~\ref{tab:SuccessRate} compares the success rates obtained by the considered experimental settings. For the smallest cut-off value N, \ie~N=1, %corresponding to the case that the developer expects a very brief list of items, 
\FC is still able to provide matches. For instance, with \textbf{C1.1} when k=2, the system gets 67.46\% as success rate, and this score increases linearly along N: \FC gets a success rate of 76.84\% and 82.80\% when N=5 and N=20, respectively. By Configuration \textbf{C1.2},  compared to \textbf{C1.1}, we see a sharp increase in performance by all the cut-off values. Take as an example, with k=2, we get 91.11\% as success rate for N=5, and the score goes up to 94.07\% when N=20. This demonstrates that \FC is capable of providing good match even when the developer wants to see a fairly short ranked list. Similarly, by \textbf{C2.1} and \textbf{C2.2}, \FC enhances its success rate alongside k and N. %the two values of 

% $N=20$, a success rate of

%the score improves considerably when we incorporate more neighbor apps to compute recommendations, \ie k. Take as an example, success rate increases from 69.10\% by k=3 to 75.76\% by k=10.

Next, we investigate the effect of changing the number of neighbor apps used in computing recommendations, \ie k, on the final outcome by comparing the results columnwise. It is evident that when incorporating more neighbors for computing recommendations, \FC yields a better success rate. For instance, with \textbf{C1.1}, considering the success rate obtained by k=2 and k=3, we see that there is always a gain in performance: for N=5, \FC obtains 76.84\% and 78.42\%, respectively. This score improves substantially when we use more neighbor projects to compute recommendations. Take as an example, \FC has a success rate of 71.34\% when k=4 and 75.76\% when k=10. In summary, \FC is more accurate if additional apps are considered for computing the missing ratings in the 3D matrix.

%\revised{Next, the effect when the cut-off value $N$ is increased, the corresponding success rates improve linearly. For example, when $N=20$, \FC obtains $40.98\%$ success rate for \textbf{C1.1} and $47.70\%$ for \textbf{C1.2}.}

\smallskip
\noindent
$\rhd$~\textbf{Performance gain.} Referring to Table~\ref{tab:Configurations}, we see that \textbf{C1.1} and \textbf{C1.2} as well as \textbf{C2.1} and \textbf{C2.2} are pairwise comparable. For instance, %to compute recommendations, 
both \textbf{C1.1} and \textbf{C1.2} share the same amount of method declarations ($\delta$), they only differ in the number of invocations used in the testing declaration ($\pi$). Thus, to investigate the effect of changing $\pi$ on the recommendations, we consider each pair of related configurations. The results in Table~\ref{tab:SuccessRate} indicate a sharp rise in performance when the configurations change from \textbf{C1.1} to \textbf{C1.2}. Take as an example, when k=2 and N=1, that means we consider only the first item in the ranked list, \FC obtains 85.69\% as success rate which is much better than 67.46\%, the score yielded by \textbf{C1.1}. When k=10 and N=20, the maximum success rate for \textbf{C2.1} and \textbf{C2.2} is 90.11\% and 96.92\%, respectively. This suggests that incorporating more invocations, \eg four instead of one invocation, helps \FC significantly enhance its overall performance. In practice, this means that given a declaration, the system is able to provide more accurate recommendations proportionally to the project's maturity. % The table demonstrates that by incorporating more invocations, \FC obtains a much better. And this score improves alongside k.    In practice, this means that the accuracy of recommendations improves with the maturity of the project. By comparing the results obtained for \textbf{C1.1} and \textbf{C1.2}, we see that when more invocations are incorporated into the query, \FC provides more precise recommendations.  when the developer adds some more invocations

Given the results in Table~\ref{tab:SuccessRate}, we analyze the performance gain in percentage (\%) and report them it in Table~\ref{tab:PerformanceGain}. 
The green color and various levels of density are employed to represent the corresponding magnitude. From the table, it is evident that the color gradually fades when we move from left to right, top to bottom, implying that the enhancement goes down linearly when we increase k and N. For example, the correlation between \textbf{C1.1} and \textbf{C1.2} is as follows: for N=1 the gain is 27.02\% with k=2, and it decreases to 26.67\% and 23.62\% with k=3 and k=4, respectively; when k=10, the gain boils down to 18.74\%. The same trend can be seen for other values of k and N. Likewise, the improvement obtained by \textbf{C2.2} in comparison to \textbf{C2.1} shares a similar pattern: it is big with low k and N, and small with higher k and N. For instance, it reaches 25.14\% for N=1 and k=2 and shrinks to 7.95\% for N=20 and k=10. Overall, this essentially means that while we get performance gain by incorporating more neighbors, at a certain point, such the gain becomes saturated and there will be no further improvement. % Demonstrated by the color density

%to compare the corresponding success rate 
%, \textbf{C1.1} and \textbf{C1.2} or \textbf{C2.1} and \textbf{C2.2}, by each value of k, we

\begin{figure*}[h!]
	\vspace{-.2cm}
	\centering    
	%	\color{blue}
	\begin{tabular}{c c}		
		\vspace{-.1cm}
		\subfigure[\textbf{C1.1}]{\label{fig:PR_D2600_C11}\includegraphics[width=85mm]{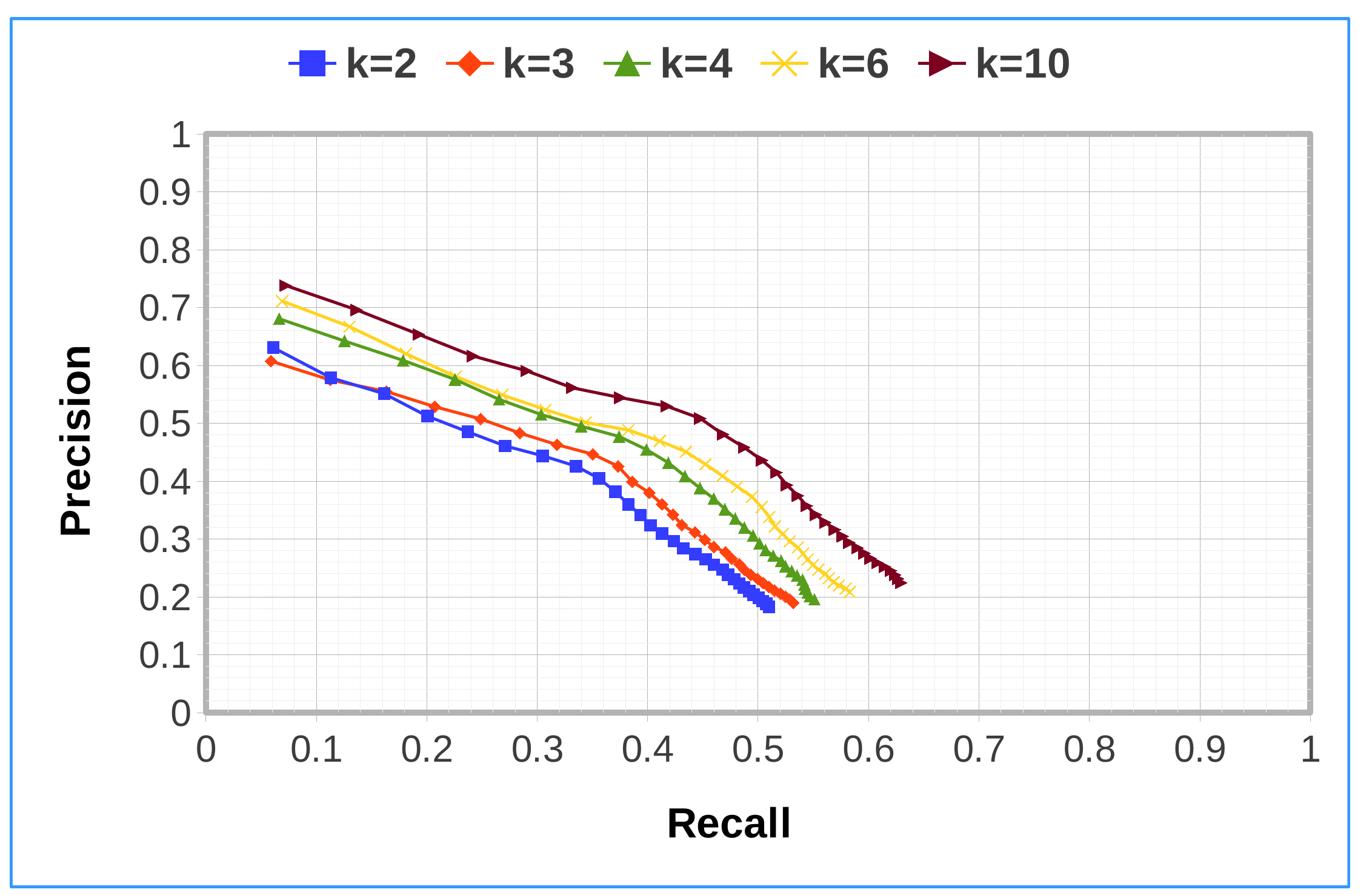}}  &
		\subfigure[\textbf{C1.2}]{\label{fig:PR_D2600_C12}\includegraphics[width=85mm]{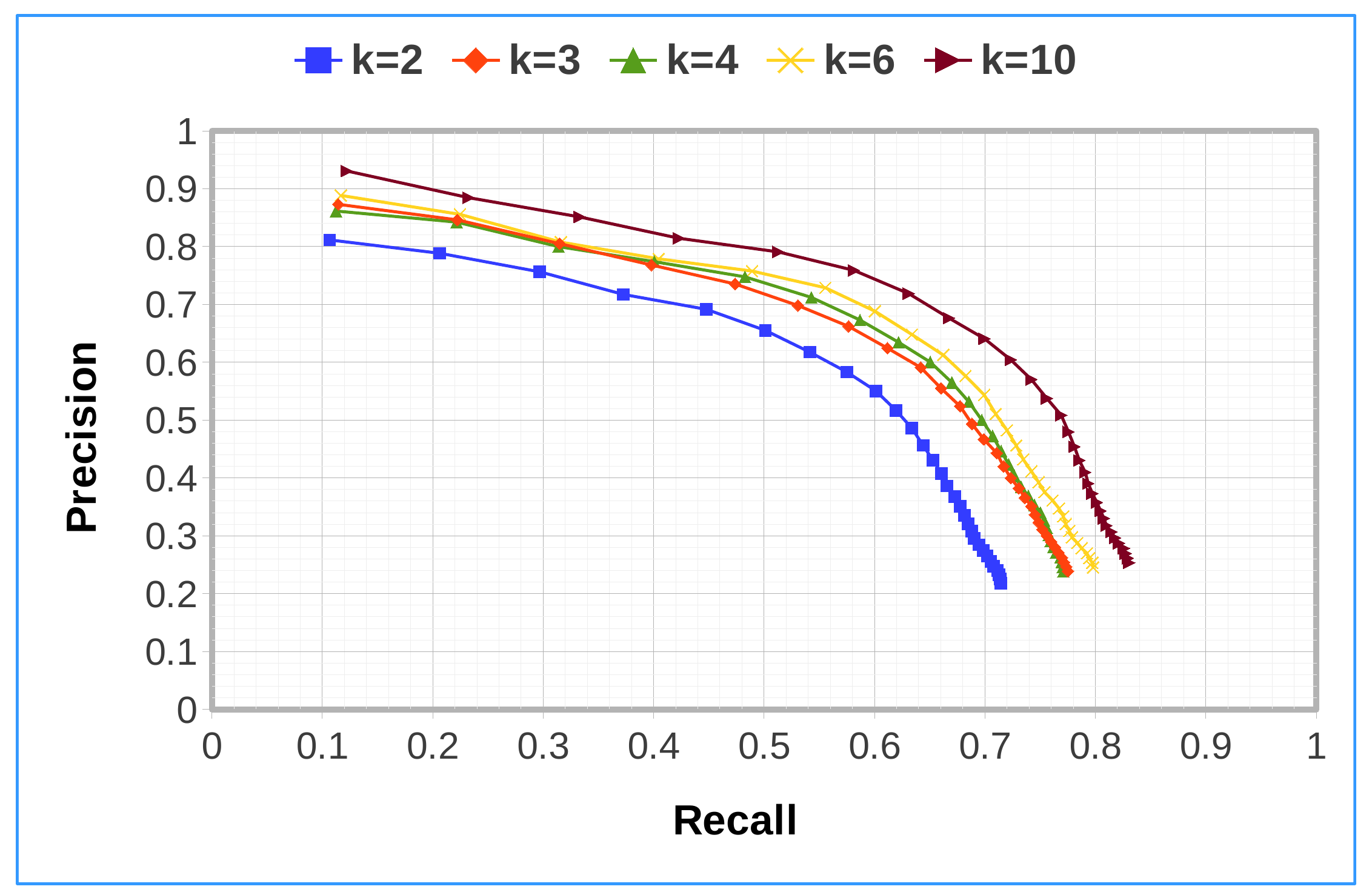}} \\
		\subfigure[\textbf{C2.1}]{\label{fig:PR_D2600_C21}\includegraphics[width=85mm]{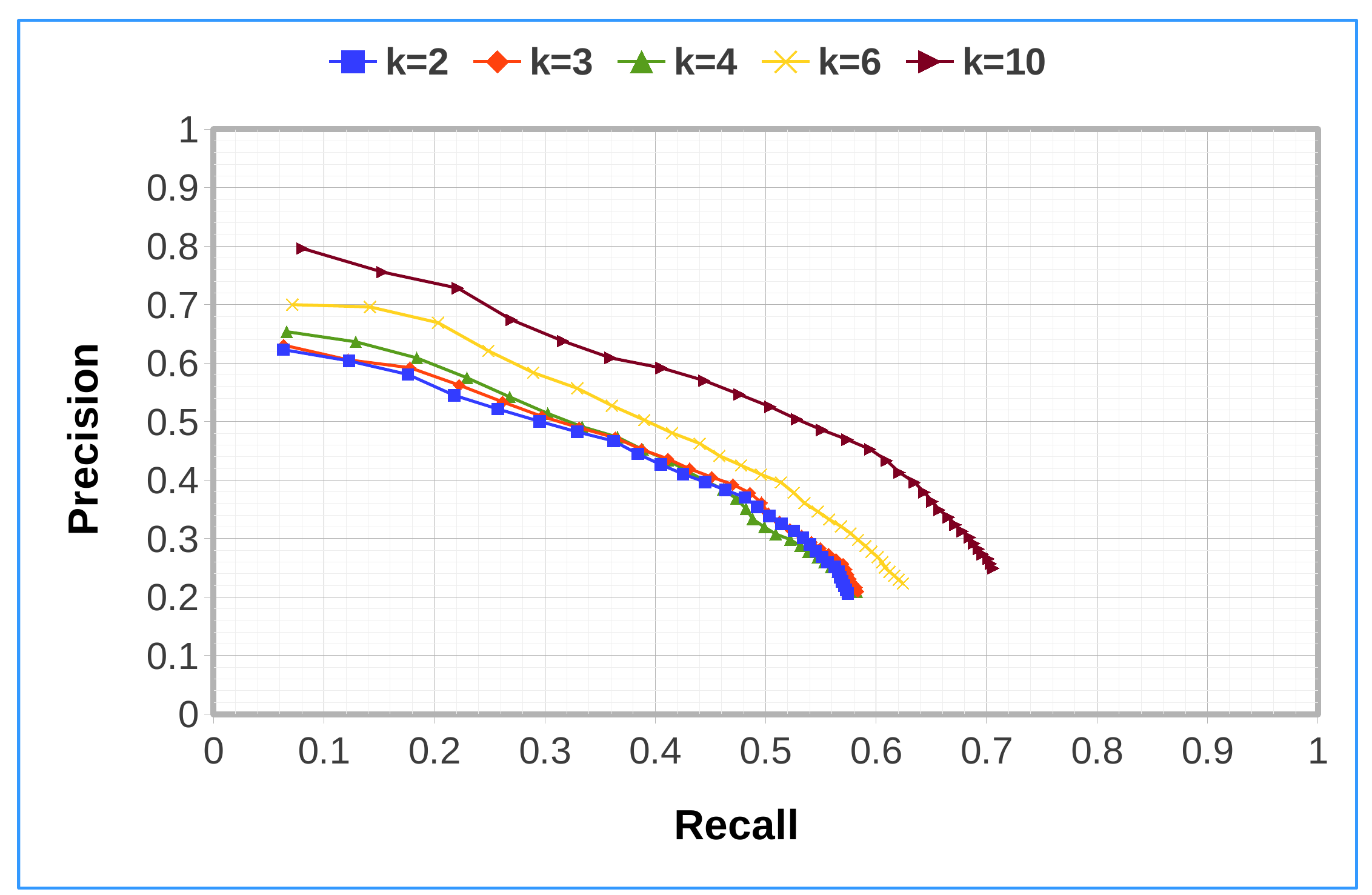}}  &
		\subfigure[\textbf{C2.2}]{\label{fig:PR_D2600_C22}\includegraphics[width=85mm]{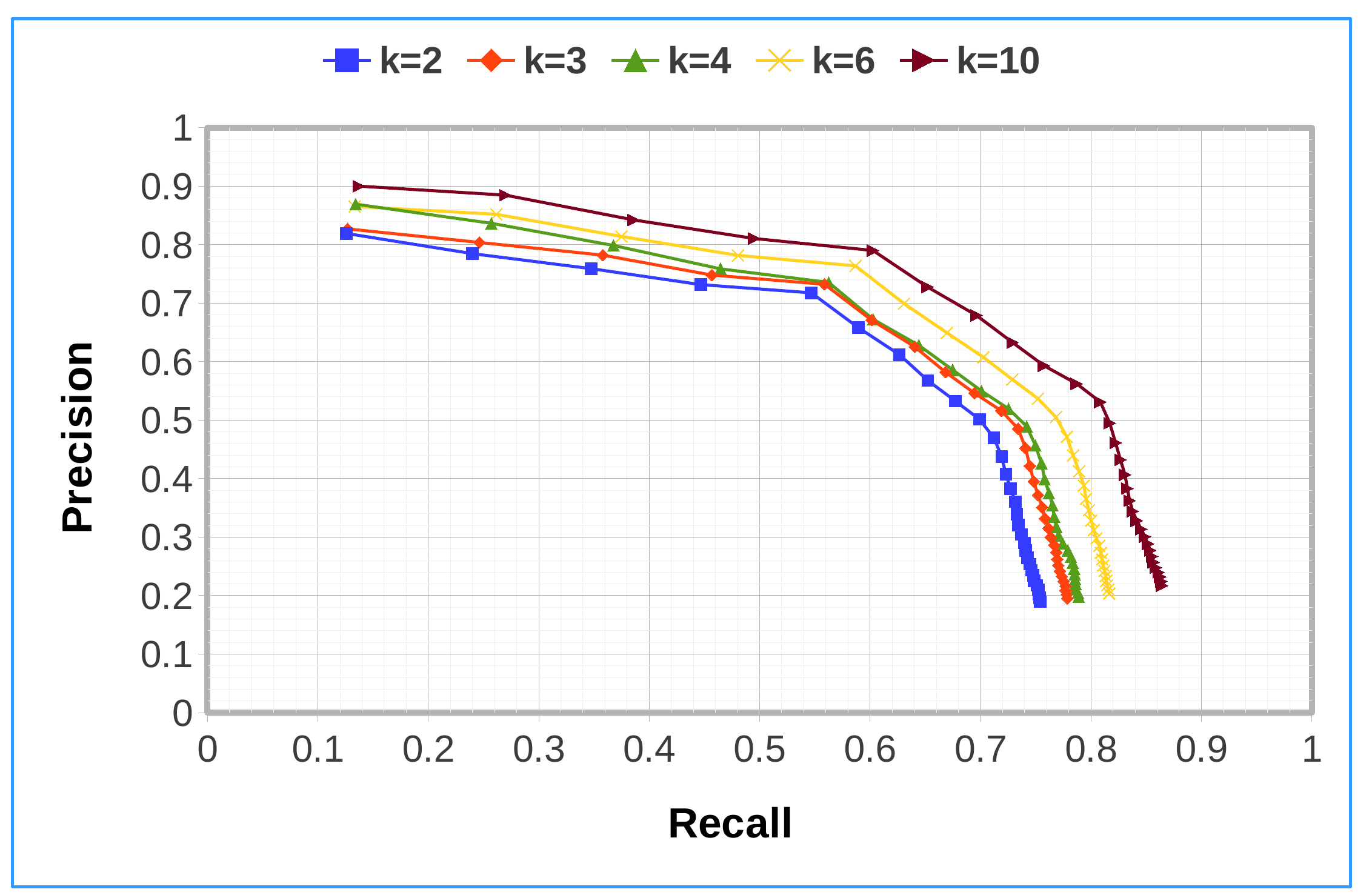}} 
	\end{tabular}
	\vspace{-.1cm}
	\caption{Precision and recall curves obtained for the configurations.}
%	\vspace{.4cm}
	%	\color{black}
\end{figure*}

 \begin{figure*}[t!]
	\centering
	\vspace{-.2cm}
	\includegraphics[width=0.98\linewidth]{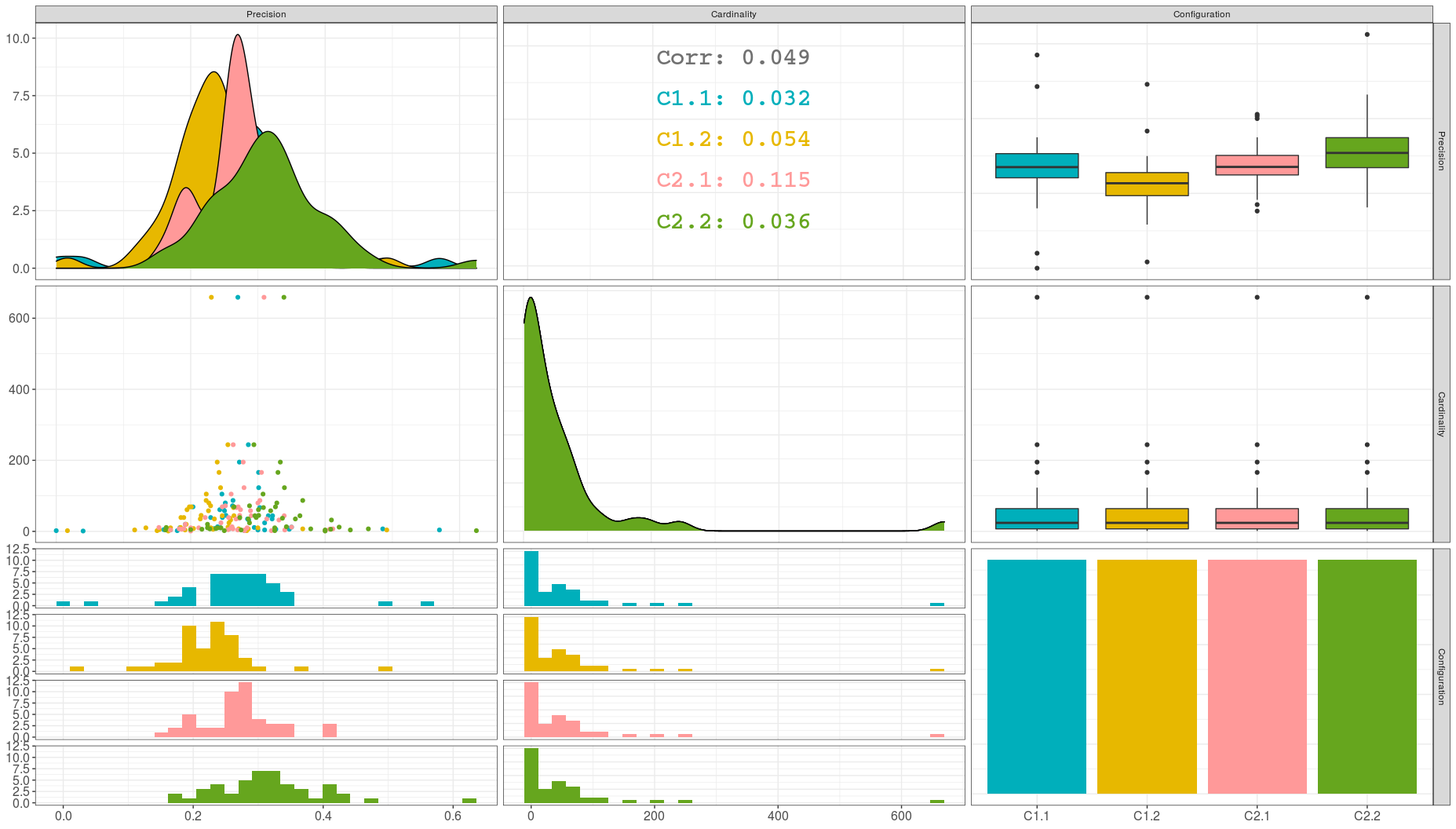}
%	\vspace{-.2cm}
	\caption{Bivariate analysis of Precision and Cardinality.} 
	\label{fig:Bivariate}
%	\vspace{-.2cm}
\end{figure*}

%\begin{figure*}[h!]
%	\vspace{-.2cm}
%	\centering    
%	\begin{tabular}{c c}		
%		\subfigure[\textbf{C1.2}]{\label{fig:PAM_and_FOCUS_C12}\includegraphics[width=80mm]{figs/PAM_and_FOCUS_C12.png}}  &
%		\subfigure[\textbf{C2.2}]{\label{fig:PAM_and_FOCUS_C22}\includegraphics[width=80mm]{figs/PAM_and_FOCUS_C22.png}} 
%	\end{tabular}
%	\vspace{-.2cm}
%	\caption{\revised{Comparison between PAM and FOCUS.}}
%	\vspace{-.2cm}
%\end{figure*}

%precision, recall scores 
%In a diagram, 

\smallskip
\noindent
$\rhd$~\textbf{Accuracy.} We report the accuracy achieved by all configurations using the precision recall curves (PRCs) depicted in Fig.~\ref{fig:PR_D2600_C11}, Fig.~\ref{fig:PR_D2600_C12}, Fig.~\ref{fig:PR_D2600_C21}, and Fig.~\ref{fig:PR_D2600_C22}. The cut-off value N has been varied from 1 to 30, aiming to study \FC's performance further down in the ranked list. %of recommended items. 
%MAX{do the axis start at zero?}
First, we examine the effect of changing k on the precision recall curves. In fact, a system gets a good performance if its precision and recall are high at the same time, and this corresponds to a PRC close to the upper right corner of the diagram. %a PRC stays close to the upper right corner corresponds to high precision and recall, indicating a better accuracy. 
From the figures, it is clear that incorporating more neighbor apps in computing recommendations results in a better accuracy by all configurations. For instance, with \textbf{C1.1}, we see a performance gain when increasing the number of neighbor apps: the best precision and recall are 0.75 and 0.63, respectively, and they are obtained when k=10; while by other value of k, \ie $k=\{2,3,4,6\}$, the system gets a lower precision and recall. Similarly by other configurations, k=10 is also the number of neighbor apps used for computing ratings that contributes to the best accuracy: with \textbf{C1.2}, \FC achieves 0.92 as precision and 0.84 as recall. With \textbf{C2.1} and \textbf{C2.2}, the gain in performance when using 10 apps for computing recommendations becomes more evident, in comparison to other values of k, \ie $k=\{2,3,4,6\}$. This is consistent with the outcomes we got by the success rate scores presented in Table~\ref{tab:SuccessRate}: the system achieves a better performance if it incorporates more similar apps for computing recommendations. 

%Fig.~\ref{fig:PR_D2600_C11}, Fig.~\ref{fig:PR_D2600_C12} depict the precision and recall curves (PRCs) for the above mentioned configurations by varying $N$ from $1$ to $30$. In particular, Fig.~\ref{fig:PrecisionRecall_D11} represents the accuracy when almost the first half of the declarations ($\delta=\Delta/2-1$) together with one (\textbf{C1.1}) and four invocations (\textbf{C1.2}) from the testing declaration $d_a$ are used as query. As a PRC close to the upper right corner indicates a better accuracy, %\cite{DiNoia:2012:LOD:2362499.2362501}, 

In conclusion, we see that the performance of \FC using \textbf{C1.2} is superior to that when using \textbf{C1.1}. Similarly, compared to \textbf{C2.1}, the accuracy obtained by \FC using \textbf{C2.2} improves substantially, \ie by equipping the query with more invocations. These facts further confirm that \FC is able to recommend more relevant invocations when the developer intensifies the declaration by adding more code. As can be seen in Eq.~\ref{eqn:combinedRating}, when more invocations are available, the similarity among declarations can be better determined, resulting in a gain in performance.

\smallskip
\noindent
$\rhd$~\textbf{The long tail.} We counted the APIs that are 
recommended more often by \FC. By carefully checking the top 20 recommended 
items, we realized that most of them reside in the long tail. For example, the 
\emph{java/lang/StringBuilder/toString()} API has been provided 190 times by 
\FC, being the top most recommended item. However, this invocation is only 
ranked 646 in the list of all the APIs in the dataset. Altogether, this is to 
show that while recommending very popular APIs may make sense, \FC goes far 
beyond that by recommending also items in the long tail. This is achieved since 
\FC mines API from highly similar projects, given an active project.
	%\footnote{Due to space limit, a full explanation is provided online: 
	%\url{https://mdegroup.github.io/FOCUS-Appendix/}}}

\vspace{.2cm} 
\begin{shadedbox}
	\small{\textbf{Answer to RQ$_2$.} \FC provides more accurate predictions when more similar projects are used for recommendation. It is capable of suggesting APIs in the long tail. %Given an active declaration, 
		The system improves its accuracy while the developer keeps coding.} 
\end{shadedbox}

\vspace{-.4cm}
\subsection{\rqthird}

Table~\ref{tab:Correlation} depicts the Spearman coefficients for all 
configurations, with respect to different numbers of %cut-off values 
N. The Kendall coefficients ($\tau$) are comparable to the Spearman ones ($\rho$), so 
we omitted them from the table, for the sake of clarity.
%\MAX{maybe we should compute Spearman's correlation (and try Kendall as well) because the relationship does not seem to be linear, and the number of categories might be fairly limited}
% compute similarity, and then we compare

%\begin{table}[h!]
%	%	\footnotesize
%	\vspace{-.3cm}
%	\caption{\revised{Correlation ($\rho$) between cardinality and precision, $N=\{5,10,15,20\}$.}} %Comparison between
%	\centering
%	%	\begin{tabular}{|>{\bfseries} c | *{6}{R}|}		 \hline
%	\vspace{-.2cm}
%	\begin{tabular}{|l|c|c|c|c|}\hline
%		\rowcolor{verylightgray}
%		%		&  \multicolumn{4}{c|}{\FC} \\ \hline
%		N & \textbf{C1.1} & \textbf{C1.2} & \textbf{C2.1} & \textbf{C2.2} \\ \hline
%		%		1     &   &  &  &   \\ \hline
%		5     & 0.067  & 0.145  & 0.080  & 0.040  \\ \hline
%		10    & 0.056  & 0.110 & 0.112 & 0.102 \\ \hline
%		15    & 0.047  & 0.082 & 0.108  & 0.062 \\ \hline
%		20    & 0.043  & 0.068 & 0.119 & 0.036 \\ \hline
%		25    & 0.031  & 0.057 & 0.115 & 0.018  \\ \hline
%		%		\bottomrule
%	\end{tabular}
%	\label{tab:Correlation}
%	\vspace{-.2cm}
%\end{table}

By examining the results in Table~\ref{tab:Correlation} we see that, despite some fluctuations, mainly with \textbf{C2.1} and \textbf{C2.2}, $\rho$ is considerably small, \ie the maximum value is  $\rho$=0.160 for \textbf{C1.2} and N=25. More importantly, most of the scores are close to 0, indicating an extremely low (\eg by $N=\{5,20,25\}$ with \textbf{C1.1}) or almost no correlation (\eg by $N=\{15\}$ with \textbf{C1.1} and \textbf{C2.2}).

\begin{table}[h!]
	\footnotesize
	%	\scriptsize
	\vspace{-.4cm}
	%	\small
	\caption{Correlation ($\rho$) between cardinality and precision, 
	$N=\{5,10,15,20,25\}$.} %Comparison between
	\centering
	%	\begin{tabular}{|>{\bfseries} c | *{6}{R}|}		 \hline
	\vspace{-.2cm}
	\begin{tabular}{|l|c|c|c|c|}\hline
		\rowcolor{verylightgray}
		%		&  \multicolumn{4}{c|}{\FC} \\ \hline
		N &  \textbf{C1.1} & \textbf{C1.2} & \textbf{C2.1} & \textbf{C2.2} \\ 
		\hline
		%		      & $\rho$  & $\rho$  & $\tau$ & $\rho$ & $\tau$ & $\rho$ 
		%&  $\tau$  \\ \hline
		%		1     &   &  &  &   \\ \hline
		5     & 0.059    & 0.150  & 0.068  & -0.157 \\ \hline
		10    & 0.117   & 0.132   & 0.086  & -0.013 \\ \hline
		15    & 0.001   & 0.115   & 0.080  & 0.005  \\ \hline
		20    & 0.046   & 0.150   & 0.156  & 0.054  \\ \hline
%		25    & 0.063   & 0.160   & 0.100  & 0.034  \\ \hline
		%		\bottomrule
	\end{tabular}
	\label{tab:Correlation}
	\vspace{-.1cm}
\end{table}

%To further study the considered relationships in this experiment, 
% the cardinality of a category and the corresponding precision.

As an example, Fig.~\ref{fig:Bivariate} depicts precision and cardinality as well as their correlation for N=25. The variables are shown both on the x-axis and y-axis, however at different parts of the axes. %When moving from the left to the right, top to bottom, we.
This allows us to comprehensively represent the relationship between the two variables for all the four configurations. In particular, on the top-left corner, there is the histogram of precision with respect to cardinality, while the other bar charts at the bottom show the histogram for each of them individually. The middle frame in the top row specifies the correlation coefficients between precision and cardinality for all the configurations. Results show that there is a very weak correlation between the two variables. For instance, the coefficient is 0.032 for \textbf{C1.1}, or 0.036 for \textbf{C2.2}. As a whole, this unfortunately contradicts our initial conjecture: apps belonging to major categories do not get a better recommendation, although they have in principle, more background data. This means that \emph{searching for recommendations just by looking at apps of the same domain(s) does not guarantee that we will gain benefit.} We attempt to ascertain the possible causes in the following.

% \begin{figure*}[t!]
%	\centering
%	\vspace{-.2cm}
%	\includegraphics[width=0.90\linewidth]{figs/BivariateAnalysis4.png}
%	\vspace{-.2cm}
%	\caption{\revised{Bivariate analysis of Precision and Cardinality.}} 
%	\label{fig:Bivariate}
%	\vspace{-.2cm}
%\end{figure*}

According to a previous work~\cite{McMillan:2012:DSS:2337223.2337267}, 
if projects share API calls implementing the same requirements, then the 
projects are considered to be more similar than those that do not have similar 
API usage. We computed similarity among apps using the \emph{Similarity Calculator} component presented in Section~\ref{sec:SimilarityCalculator}.	
%\MAX{two things here (i) cite work by McMillan, (ii) what does it mean similarity=1? I'm just afraid the reader may believe the apps are basically the same in terms of API. In other words, any idea why they are so similar?}	
Such a similarity is measured based on the constituent API function calls of an 
app (cf. Fig.~\ref{fig:Similarity} and Eq.~\ref{eqn:VsmSim}). By carefully 
examining the final results, we realized that generally, similar apps do not 
originate from the same domain. To be concrete, considering a ranked list with 
five items for all 2,600 apps, \ie N=5, the percentage of items that have 
similar apps coming from 1, 2, 3, 4, and 5 categories is 1.14\%, 6.6\%, 
21.42\%, 41.8\%, and 29.0\%, respectively.
%\PN{I am going to add some concrete numbers.}%For example, %we consider the \textbf{vclick.client} app
%Table~\ref{tab:SimilarApps} reports four examples, \ie \textbf{vclick.client}, \textbf{machinekit.appdiscover}, \textbf{compphys.atomify}, and \textbf{blogspot.tasteroids}, together with top five ranked similar items for each. %For every similar app, we attach also the similarity score and the corresponding category. Interestingly, t
%The table shows that given an app, the most similar apps come from different domains. %, but relevant ones. 
For instance, \textbf{machinekit.appdiscover}\footnote{\url{https://bit.ly/3pGKKIL}} belongs to \emph{Libraries \& Demo}, however its highly similar apps are from \emph{Education}, \emph{Books \& Reference}, \emph{Health \& Fitness}, and \emph{Tools}. Since \FC relies on the similarity function (cf. Section~\ref{sec:SimilarityCalculator}), it may retrieve invocations from projects in completely different domains to generate recommendations. This explains why projects of a category with a low number of items still get a good accuracy, resulting in a weak correlation between the cardinality of a category and accuracy. In a nutshell, \emph{there exists no correlation because even apps belonging to different categories still contain similar API usage.} %This is the reason why a similar app of \textbf{compphys.atomify}\footnote{\url{https://play.google.com/store/apps/details?id=com.compphys.atomify\&hl=en}} which belongs to the category \emph{Education}, is \textbf{tarlic.vision}\footnote{\url{https://play.google.com/store/apps/details?id=com.tarlic.vision\&hl=en}} which comes from \emph{Health \& Fitness}, and the two topics seem to have nothing in common.

Though the experiment suggests that we cannot save time by looking into some certain categories, on the bright side, it reveals an interesting feature of \FC: the tool is able to discover API calls from a wide range of apps, regardless of their origins.

\begin{shadedbox}
	\small{\textbf{Answer to RQ$_3$.} There is no direct correlation between the cardinality of a category and prediction accuracy. Moreover, \FC is capable of mining API calls from apps belonging to various application domains.} 
\end{shadedbox}

%, aiming to narrow down the search scope
%once the developer specifies a domain for her app, we can search 

\vspace{-.2cm}
\subsection{\rqfourth}\label{sec:RQ4}

\begin{figure}[h!]
	\centering
	\vspace{-.2cm}
	\includegraphics[width=0.80\linewidth]{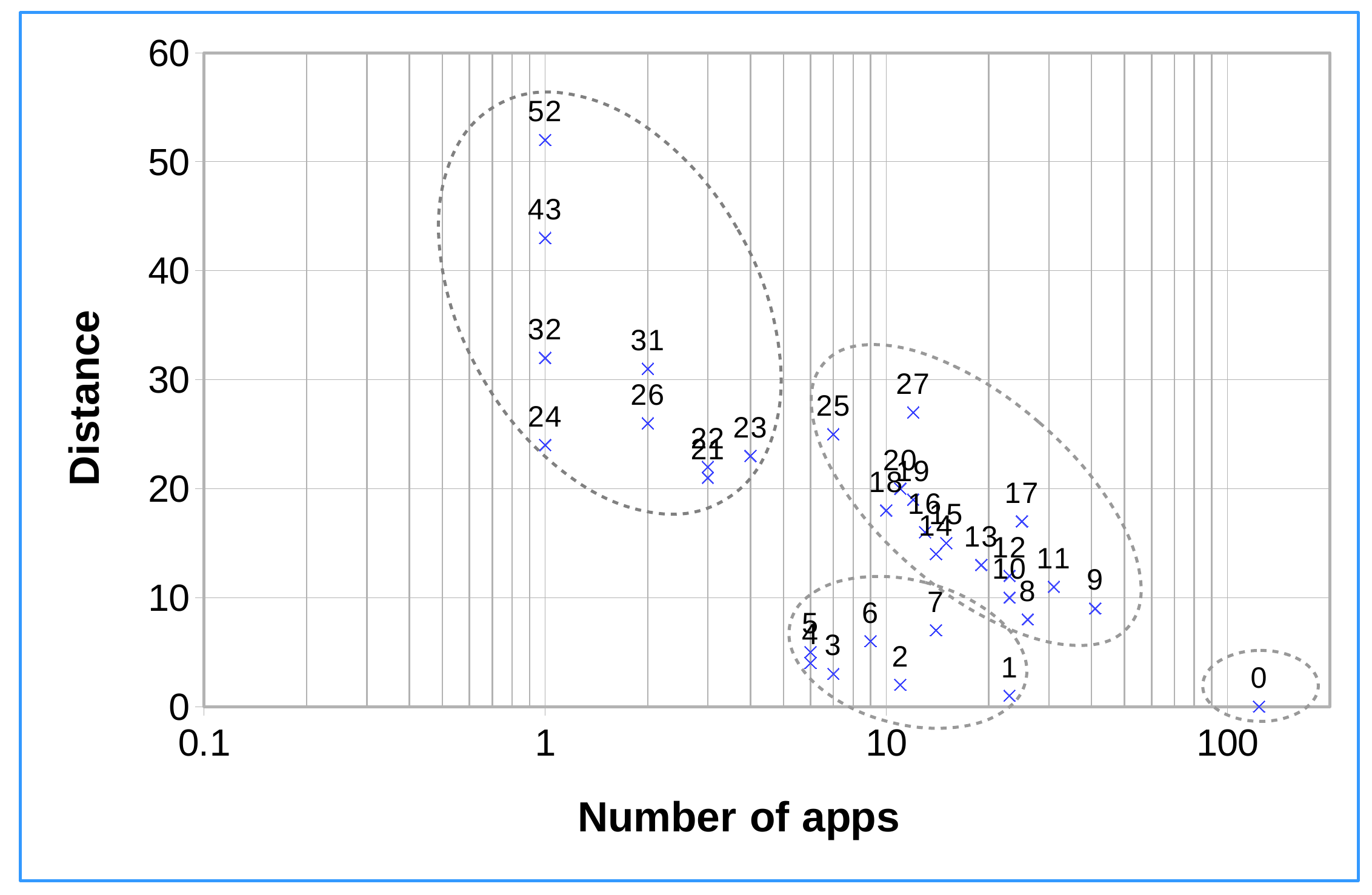}
	\vspace{-.2cm}
	\caption{Levenshtein distance for the set of 500 apps.}  %\PN{The %figure needs to be improved.}
	\label{fig:Levenshtein}
	 	\vspace{-.2cm}
\end{figure}

\smallskip
\noindent
As shown in Section~\ref{sec:CodeRecommendation}, by using the incomplete code in Fig.~\ref{fig:originalCode} together with other testing declarations as 
query to feed \FC, %(cf. Figure~\ref{fig:CodeRemoval}), 
we obtained a relevant snippet depicted in Listing~\ref{lst:RecommendedSnippet}, and this is just one of many good matches we got. To provide a concrete analysis, Fig.~\ref{fig:Levenshtein} depicts the distribution of the 500 apps dataset with respect to the number of projects (x-Axis) and  Levenshtein distance between the testing declaration and the corresponding project (y-Axis). 

To facilitate a better view, we mark the apps as four separate clusters. %The 
%y-Axis represents the distance, and the x-Axis corresponds to the number of 
%projects. %thei results we got by computing the Levenshtein distance for the 
%%set of 500 projects. 
Almost a quarter of the projects or 24\% corresponding to 120 projects get zero as the final result, \ie the distance between the recommended snippet and the original one is zero. %\MAX{missing verb? zero distance to what? to the original?}. 
This means that for each of these projects, the recommended declaration perfectly matches the original one. %In other words, \FC suggests a perfect match for the corresponding testing declarations. 
By the remaining ones, 23 projects among them accounting for 4.6\%, have a distance of one, which also indicates a high level of code similarity. Almost a half of the dataset, \ie 233 apps corresponding to 46.60\%, have a distance being larger than nine. %from the figure, while a small fraction have

%For each category, we computed the precision for all of its constituent apps following Eq.~\ref{eqn:Precision}, and the precision of a category was averaged out over the apps.
%\smallskip
%\noindent

\begin{figure*}[t!]
	\vspace{-.2cm}
	\centering    
	\begin{tabular}{c c}
		%		\vspace{.3cm}
		\subfigure[Q1: Does \FC retrieve code snippets relevant to the context?]{\label{fig:Q1}\includegraphics[width=80mm]{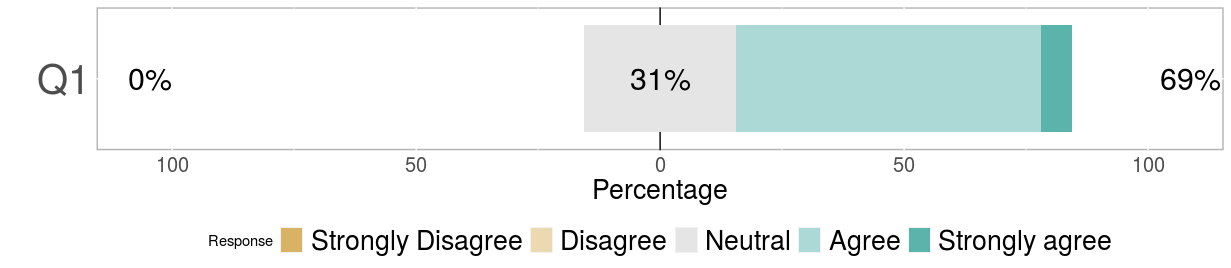}} &		
		\subfigure[Q2: Do the recommended code snippets help you complete the lab assignments?]{\label{fig:Q2}\includegraphics[width=80mm]{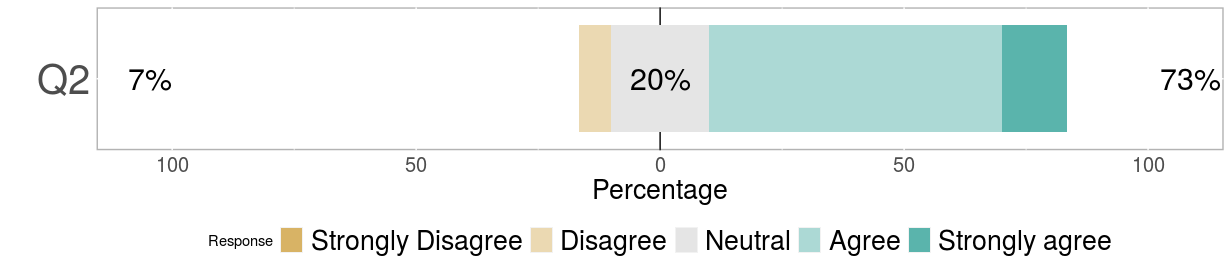}} \\
		\subfigure[Q3: Does \FC retrieve invocations relevant to the context?]{\label{fig:Q3}\includegraphics[width=80mm]{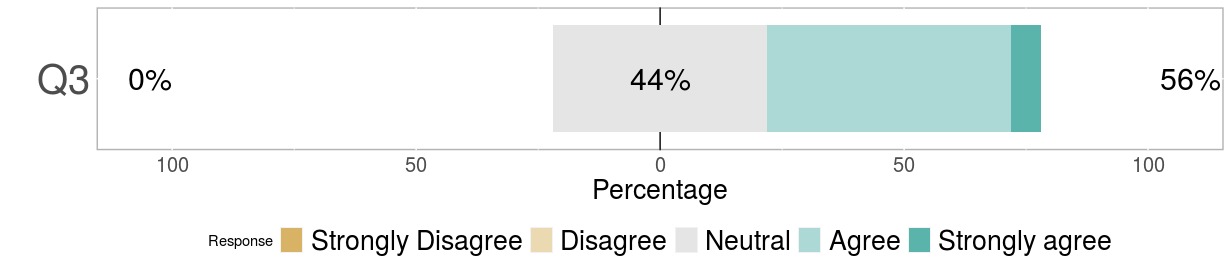}} &
		\subfigure[Q4: Do the recommended invocations help you complete the lab assignments?]{\label{fig:Q4}\includegraphics[width=80mm]{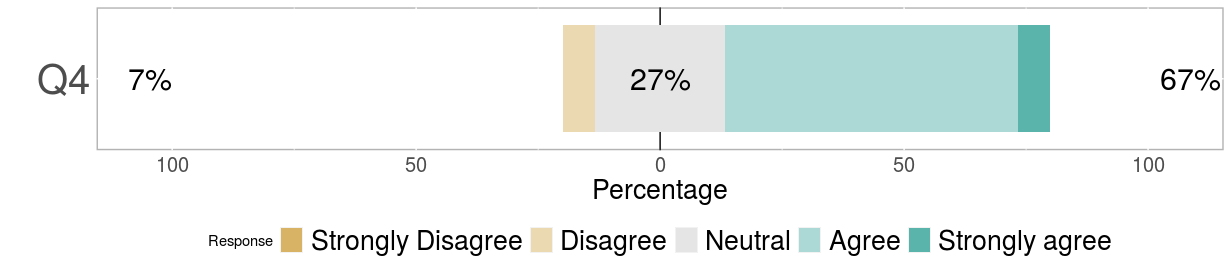}} 
	\end{tabular}
	\caption{Results for evaluating the usefulness of the recommendations.}
	\vspace{-.4cm}
\end{figure*}

Figure~\ref{fig:Levenshtein} shows that, while \FC gains a good recommendation performance %for is able to recommend a good match 
for a considerably large number of apps, it \emph{fails} to retrieve matches for some others, \ie the corresponding Levenshtein distance is large, meaning that the recommended snippets are not relevant to the ground-truth ones. For instance, one project has a distance of 52, or another has a distance of 43. We attempt to find out the rationale behind this outcome. Our main intuition is as follows, by those projects with a large Levenshtein distance, there is a lack of relevant training data. In other words, if there are not enough similar projects, \FC cannot discover API invocations which eventually fit to the active declaration.

To validate the hypothesis, the following test was conducted: we computed the precision scores for all projects, and compared them with the Levenshtein distances using the Spearman's rank correlation coefficient (Similar to RQ$_3$). The resulting score is $\rho$=-0.514, with p-value < 2.2e-16. This can be interpreted as follows: the obtained precision is disproportionate to the Levenshtein distance, or put another way, the higher the precision we get, the shorter the distance, and vice versa. The finding consolidates our assumption: if \FC achieves a high precision, it will be able to recommend more relevant code snippets. Furthermore, as we already proved in RQ$_3$, \FC gets a higher precision if we use more similar apps for computing recommendations. Altogether, we conclude that our proposed approach is able to return relevant code snippets if it is fed with more training data. From the set of apps with a Levenshtein distance of 0, we enumerated the APIs and sorted them in descending order to see which invocations have been recommended most. We got a similar outcome of RQ$_2$ in Section~\ref{sec:RQ2}: FOCUS recommends several APIs which appear late in the ranked list of the most popular invocations.

% favors more projects
% is fed with decent data
%it is able to provide more related invocations, 

% In other words, using code clone detection, we aim at evaluating if the recommended code is similar to the original one.
% Finally, the resulting code snippet is suggested to the developer.
%, the recommended. Rascal also allows us to compute M$^3$ model from both source code folders and binaries, \eg JAR files independently. Thus we implemented a dedicated function %(\ie ~\code{getCodeFromM3}) 
%that
% happen in the same order
% the Levenshtein distance. The Levenshtein distance between two strings s$_1$ and s$_2$ corresponds to the number of
%Similarly, the ground-truth declaration is also encoded. 
%\revised{}

%For this question, we recommend also real code snippet.

\vspace{.3cm}
\begin{shadedbox}
	\small{\textbf{Answer to RQ$_4$.} \FC can provide relevant source code snippets to a testing declaration, as long as we feed it with a rich training dataset, \ie
there are more projects similar to the one being considered.} %enough %\MAX{can we be more precise?} training data.}
\end{shadedbox}

\subsection{\rqfifth}

%\MAX{This paragraph should be moved later, maybe even in section 5, at the beginning of RQ5 results}

As explained in Section \ref{sec:Evaluation}, \numParticipants participants took part in the user study. Among them, 30\% and 50\% of them have three years and more than four years of programming experience, respectively. Most of them use a code search engine in a daily basis. Moreover, 80\% of the participants agree that the tasks are clear and easy.

First, we analyzed whether the use of \FC could help participants produce more correct code. The median number of passed tests was 2 out of 3 both with and without \FC. We compared (pairwise, by participant) the percentage of passed test cases using a Wilcoxon signed-rank test. The test did not indicate a statistically significant difference (\emph{p}-value=0.88), \ie the correctness of the produced implementations did not change with and without \FC. Also the Cliff's $d$ effect size is negligible (\emph{d}=0.01).

We therefore looked at the perceived usefulness of the 
recommendations, in terms of API invocations and code snippets. Results of the 
questionnaire related to the task assignments in Table~\ref{tab:evaluator_task_assignement} are shown in Fig.~\ref{fig:Q1}, 
Fig.~\ref{fig:Q2}, Fig.~\ref{fig:Q3}, and Fig.~\ref{fig:Q4}.

Concerning the first question ``\emph{Q1: Does \FC retrieve code snippets relevant to the context?}'' following Fig.~\ref{fig:Q1}, 69\% of the participants \emph{agree} and \emph{strongly agree} with the fact that the snippets are relevant, while the remaining 31\% of them have no concrete judgment on the results, \ie neutral. This means that most of the developers find that the recommended code snippets fit their programming tasks.
	
By the second question: ``\emph{Q2: Do the recommended code snippets help you complete the lab assignments?}'' as shown in Fig.~\ref{fig:Q2}, most of the participants find that the snippets recommended by \FC are helpful to solve the tasks. In particular, 73\% of them agree and strongly agree with the question. 

With the third question: ``\emph{Q3: Does \FC retrieve invocations relevant to the context?}'' we are interested in understanding whether \FC can fetch invocations related to the given context. The results in Fig.~\ref{fig:Q3} suggest that more than a half of the evaluators, \ie 56\% think that the provided APIs are relevant, while 44\% of them have no concrete judgment.

Finally, the results in Fig.~\ref{fig:Q4}, corresponding to the last question: ``\emph{Q4: Do the recommended invocations help you complete the lab assignments?}'' show that 7\% of the participants disagree that the APIs are useful, while 27\% of them feel neutral about the results. Still, most of them, \ie~67\%, appreciate the recommended APIs, which are helpful to solve their tasks.

% that the invocations are useful
% In particular, 69\% of them agree and strongly agree with the question.
%recommended by \FC are helpful to solve the tasks. In particular, 69\% of them agree and strongly agree with the question.
% that the retrieved code snippets are relevant to the contexts
%is positively received by the developers

Altogether, the Likert scores indicate that \FC provides decent recommendations: both the suggested APIs and code snippets are meaningful to the given contexts.

\vspace{.3cm}
\begin{shadedbox}
	\small{\textbf{Answer to RQ$_5$.} 
			The majority of the study participants positively perceived the context-specific relevance and the usefulness of the recommendations (APIs and code snippets) provided by \FC.} %which are useful to complete the assigned tasks
\end{shadedbox}

	\section{Discussions}
	\label{sec:Discussions}
	%Discussions go here ...

%\MAX{should we merge lessons and conclusions? Depending how substantial will be the lessons}

%\PN{I am still working on this section.}

In this section, we discuss the experience gained from the experiments in Section~\ref{sec:Lessons}. The threats that might hamper the validity of our findings are discussed in Section~\ref{sec:Threats}.

\vspace{-.4cm}
\subsection{Lessons learned} \label{sec:Lessons} %\revised{Lessons learned by conducting the experiments.}

%\revised{The comparison of \FC with \UM and PAM in RQ$_1$ shows that while \UM is efficient, PAM takes considerable more time to produce the final recommendations.} %This indicates the suitability of \FC for being applied in an usage scenario like the one outlined in Section~\ref{sec:MotivatingExample}. 
With \FC, \emph{by choosing a suitable number of similar projects used for computing recommendations, we get a satisfactory prediction accuracy, while still maintaining a reasonable computational complexity.} We suppose that for a training set with big projects, \ie those that have a large number of declarations and APIs, resulting in a very large 3D tensor, it is necessary to apply matrix factorization techniques. This allows us to represent the tensor in a lower dimensional latent space, so as to efficiently handle high dimension data as well as to increase the prediction performance.

The recommended snippet in Listing~\ref{lst:RecommendedSnippet} is of high quality as it matches well with the developer's context. Nevertheless, there is no guarantee that such a good quality will be held for all the possible cases, as this depends a lot on the training data. Thus, we plan to implement a module to ask developers to rate/provide feedback on the given recommendations. By doing this, we would be able to collect information about the quality of a recommended snippet which can then be used to reinforce the learning of \FC.

% which are also big in size
%In this way, we are able to gain a trade-off between accuracy and efficiency. In the experiments, we used a small number of neighbor projects to compute recommendations. More importantly, since we employed a sparse matrix to store the matrix, we substantially optimize the storage and thus the calculation. Overall, this is to say by the current implementation, FOCUS is able to efficiently compute the recommendations, thereby maintaining a trade-off between computational complexity and effectiveness.
% yielded by \FC

%we should take into account the computational complexity, 
%While \FC gets a better performance with more 

% in a development environment
%\revised{As we found out from RQ$_3$, .}
%The results indicate that the ability to compute similarity is important. 
%need to select a suitable number, in order to
%Moreover, this also means that \FC can improve its performance if there is more data for training. In practice, this means we should incorporate as much as data as possible. 
% Discuss the cases when \FC succeeds and fails: when there is a set of highly similar projects, and when there are no similar projects.
%The results obtained from RQ$_3$ reveal that 
%Second, as long as there are similar projects, .
%RQ$_4$ shows that, \FC will fail if there is not enough (good) training data. 

Through the results obtained for RQ$_2$, we confirm the importance of computing similarity among OSS projects~\cite{NGUYEN2019110460,McMillan:2012:DSS:2337223.2337267}. \FC improves its prediction performance substantially, given that we incorporate more similar projects to compute the missing ratings. This implies that, given a project for which we would like to obtain recommendations, if we cannot find any similar projects, then it is not possible to recommend relevant API calls. This, along with the results obtained for RQ$_4$, let us conclude that \emph{\FC relies on the availability of enough training data from similar projects, in order to provide relevant API calls as well as code snippets.}  The results of RQ$_2$ also imply that it would become more difficult for \FC to recommend code samples when the context is not fully available, or worse, is missing. Under the circumstances, we can apply code synthesis approaches~\cite{8952168} to generate code for a location where there is no concrete context. Moreover, we believe that it is important to improve the way \FC computes the similarity among projects and declarations, for instance by optimizing the global mapping using the Hungarian algorithm~\cite{10.1145/3180155.3182550}. We consider all these issues as future improvements for \FC.

% Although the generated code often contains errors, it can be used as a seed to search its clones that were manually written by programmers 
%Zhou et al. [2] present an interesting way to attack this problem. 
%\MAX{is ``background'' the appropriate term? Shouldn't be ``enough training data from similar projects''? or something similar}

%In practice, we expect to have more similar apps within the same category. However, our experiments show that given an app, we can find very similar apps in other categories. 
%\MAX{any reference?}

The RQ$_3$ results reveal two findings about \FC. First, \FC can provide good recommendations, regardless of the categories used for training. Second, given a project in a category, we can find very similar projects in different categories. The outcome provides valuable insight into the meaning of \emph{category} extracted from Google Play: The categories specified by developers provide a rough abstract description of an app, rather than an informative summary of what the app does. \emph{This essentially means that these categories do not have much to do with similarity in API usages.} Recently, attempts have been made to automatically assign a category to projects/apps~\cite{6181075,CAPILUPPI2020106279}. Among others, supervised learning techniques perform computation by exploiting labeled data, \eg the apps and their corresponding categories specified by developers. However, we suppose that someone may fail if they try to classify apps according to API calls together with the categories specified in Google Play. %In other words, \emph{it is infeasible to train a Machine Learning classifier that groups apps with similar API usage into different Google Play categories.} 
Thus, currently \FC makes use of similarity techniques working at the API level without considering any induced categorization. 

Nevertheless, for a huge amount of training data, we anticipate that the categorization of apps may help increase the efficiency as follows. We can perform preprocessing steps to group similar apps in terms of API usages into the same cluster by means of unsupervised clustering techniques~\cite{10.1007/978-3-319-60438-1_47}. Given the presence of such clusters, every time there is an active app, \FC looks only in the cluster(s) that contain(s) similar apps by computing similarity with some of the most representative apps in the clusters. In this way, we can considerably narrow down the search scope for the active app, thereby speeding up the search.

%\MAX{how 
%is this related to what \FC does?}

%In our empirical study, we focused on Java and Kotlin projects, %since currently Rascal (on which \FC relies) supports these languages. %Nevertheless, \FC can be adapted to work on projects developed with other languages, as long as Rascal could support them, or there is a different code analyzer employed to parse source code.
In our empirical study, we focused on Java and Kotlin projects. However, FOCUS can be used to recommend APIs and source code for projects written in other languages, such as PHP, and C/C++, since Rascal also supports them~\cite{BASTEN20157}. Moreover, there are various reverse engineering tools that can extract declarations and invocations from source code. For example, the Eclipse JDT parser\footnote{\url{https://www.vogella.com/tutorials/EclipseJDT/article.html}} has been widely used to parse Java source code in the related studies~\cite{Wang2013Mining,Zhong2009MAPO}. Similarly, \emph{dotPeek}\footnote{\url{https://www.jetbrains.com/decompiler/}} uses a combination of debug information and web services to reconstruct C\# source code. For Android programming, Kerberoid~\cite{10.1145/3319535.3363255} has been proposed to decompile apps source code. Altogether, this implies that our tool becomes independent from Rascal as it can work with data parsed by different tools.

%\MAX{should we also say ``or a different code analyzer is employed''}

\FC is a data-driven approach, being dependent on the availability of decent training data. We believe that a lack of proper training data will lead to a reduction of the accuracy. In other words, the system may fail, especially if the testing project is at an early stage, \ie there are few declarations and APIs, or when there are not enough training projects. One of the most practical countermeasures is to try to collect as many projects as we can to create a rich knowledge base for our tool. In this way, we increase the chance to find highly similar projects, given a project being developed.

%To be more concrete, building a  classifier based on, since two apps may have very similar API usage, however they belong to different categories. 
%We came across an interesting result: the categories do not have much to do with similarity. 

\vspace{-.2cm}
\subsection{Threats to validity} \label{sec:Threats}

\smallskip
\noindent
$\rhd$~\textbf{Internal validity.} Threats to {\em internal validity} are related to confounding factors, internal to our study, that could have influenced the results. One probable threat can be seen through the results obtained for the smaller dataset which consists of 500 apps in RQ$_1$. %\MAX{maybe we should distinguish where this happens (RQ1 and RQ3) and where not}  %\code{SH$_L$} and \code{SH$_S$}. 
As we already mentioned, this dataset was used %exhibit lower precision/recall with respect to \code{MV$_L$} and \code{MV$_S$} due to the limited size of the training sets. However, these datasets were needed 
to compare \FC with \UM and PAM due to the limited scalability of PAM. Moreover, we also 
deliberately made use of such a small dataset in RQ$_4$ to study the extent to 
which \FC is able to recommend relevant source code.
%\footnote{\revised{More examples are available at the following link: 
%\url{https://mdegroup.github.io/FOCUS-Appendix/}.}} 
The intuition is 
that if it performs well on a limited amount of training data, it will be also 
effective on a larger one. 

%This also concerns any confounding factor that could influence the use. 
% In particular, the results labeled by a participant were then double-checked by another one. In case there is a disagreement between any two evaluators, one more researcher was asked to investigate the pair again to reach a final decision;

%such datasets, and in particular \code{SH$_S$}, were created to allow us to compare FOCUS with PAM, otherwise less feasible on larger datasets.
\smallskip
\noindent
In the comparison between \UM, PAM, and \FC we used the implementation of PAM which was published online by its authors. Since the original implementation of \UM is no longer available, we made use of the source code re-implemented by the authors of PAM. To mitigate the threats that may affect internal validity, we also evaluated the systems using exactly the same dataset and evaluation metrics. Furthermore, we ran several trials and counter-checks to validate the evaluation outcomes.

\smallskip
\noindent
Concerning the user study, we (\emph{i}) limited the extent to which 
results depend on personal skills by involving \numParticipants Master's 
students having similar development background and experience; and (\emph{ii}) 
did not  disclose the goals of the experiment to avoid hypothesis guessing. 
Another threat is that, to simplify the setting, developers used \FC 
recommendations as HTML pages instead of having them in the 
IDE.\footnote{See for instance 
\url{https://bit.ly/3d3i2hY}} However, 
we evaluated the perceived usefulness of the recommendations, not of the tool. 
 % 
%to avoid any possible bias %completely automating the evaluation without any 
%%manual interventions

%To avoid introducing a bias 
%Indeed, the implemented tools could be defective. To contrast and mitigate this threat, we strictly followed the descriptions in the original papers to re-implement the tools.

\smallskip
\noindent
$\rhd$~\textbf{External validity.} The main threat to {\em external validity} is that our proposed approach is currently limited to Java and Kotlin programs.
As stated in \Cref{sec:ProposedApproach}, however, \FC makes few assumptions on the underlying language and only requires information about method declarations and invocations to build the 3D matrix. This information could be extracted from programs written in any object-oriented programming language, and we wish to generalize \FC to other languages in the future. Also, in the future  \FC may benefit from an in-field evaluation in an industrial setting.

\smallskip
\noindent
$\rhd$~\textbf{Construct validity.} The main threat to {\em construct validity} 
concerns the simulated setting used to evaluate the approaches, as opposed to 
performing a user study. We mitigated this threat by introducing four 
configurations that simulate different stages of the development process. In a 
real development setting, however, the order in which one writes statements 
might not fully reflect our simulation. Also, in a realistic usage setting, 
there may be cases in which an API recommender turns out to be  more useful 
(\eg when recommending API usages for which the developer has limited 
skills/knowledge), and cases (obvious code completion, or recommending usage 
scenarios for commonly-used APIs), where it is less useful. Such a 
threat has been mitigated with the user study in which participants evaluated the recommendations provided by \FC.

%This makes a further evaluation involving developers highly desirable.

In the user study, we evaluated the perceived usefulness of the recommendations, which may or may not correspond to the actual usefulness. The outcome of the performed task did not show any significant difference in terms of correctness of the produced artifacts. This could possibly depend on the study setting (\ie offline) in which participants had the time to properly implement the task and, possibly, search for plausible solutions and/or API documentation.

%	\section{Threats to Validity}
%	\label{sec:Threats}
%	\input{src/Threats}

	\section{Related Work}
	\label{sec:RelatedWork}
	%\vspace{.1cm}
%\noindent
%$\blacksquare$~\textbf{\revised{Recommender systems in software engineering}} %\subsection{\revised{Recommender systems in software engineering}}
The adoption of recommender systems in software engineering aims at supporting developers in navigating large information spaces and getting instant recommendations, which can give some guidance to undertake the particular development task at hand \cite{Schafer:2007:CFR:1768197.1768208}. In this section, we present an overview of some representative recommender systems by focusing on those specifically conceived to support software development activities.

% summarize related work about API usage recommendation and relate our contributions to the literature.

%\MAX{I'm not sure this is needed, also because the related work in this area is huge and we cannot afford to cite everything. If anything we can put it at the beginning}

%\subsection{API Usage Pattern Recommendation}
\vspace{.1cm}
\noindent
$\blacksquare$~\textbf{API usage pattern recommendation.} Acharya~\etal~\cite{Acharya2007Mining} present a framework to extract API patterns as partial orders from client code. To this aim, control-flow-sensitive static API traces are extracted from source code and sequential patterns are computed.  
Our approach is able to recommend both a list of API calls and related source code.
%Our approach proposes a representation for API patterns, suggestions regarding API usage are still missing.

Zhong \etal implemented MAPO, a tool to retrieve API usage patterns from client projects~\cite{Zhong2009MAPO}. MAPO extracts API usages from source files, and groups API methods into clusters. Afterwards, it mines API usage patterns from the clusters, ranks them according to their similarity with the current development context, and eventually suggests code snippets to developers. Similarly, UP-Miner~\cite{Wang2013Mining} extracts API usage patterns by relying on \emph{SeqSim}, a clustering strategy that reduces patterns redundancy as well as improves coverage. %\UM employs the BIDE algorithm~\cite{Wang2004BIDE} to mine API call sequences. 
While these approaches are based on clustering techniques, and they consider all client projects in the mining regardless of their similarity with the current project, \FC narrows down the search scope by looking into similar projects. In this work, we see that \FC clearly outperforms \UM.

%\emph{Fowkes} \etal introduce 
%has been proposed as an approach 
PAM (Probabilistic API Miner) mines API usage patterns based on a parameter-free probabilistic algorithm~\cite{Fowkes:2016:PPA:2950290.2950319}. The tool 
uses the structural Expectation-Maximization (EM) algorithm to infer the most probable API patterns from client code. %, which are then ranked according to their probability.
PAM outperforms both MAPO and UP-Miner (lower redundancy and higher precision). Through a comparison of \FC with PAM, we see that our approach obtains a better performance with respect to various metrics. %in \Cref{sec:Evaluation}.

%\emph{Niu} \etal~\cite{Niu2017API} rely on the concept of object usage (method invocations on a given API class) to extract API usage patterns using API class or method names as queries. The proposed approach outperforms UP-Miner and Codota,\footnote{\url{https://www.codota.com/}} a commercial recommendation engine with respect to coverage, performance, and ranking relevance. Meanwhile, \FC relies on a context-aware CF technique---which favors recommendations from similar projects and uses the whole development context to query API method calls.

NCBUP-miner (Non Client-based Usage Patterns)~\cite{Saied2015Could} is a technique that identifies unordered API usage patterns from the API source code, based on both structural (methods that modify the same object) and semantic (methods that have the same vocabulary) relations.
The same authors also propose MLUP~\cite{Saied2015Mining}, which is based on vector representation and clustering, but in this case client code is also considered.

XSnippet~\cite{10.1145/1167515.1167508} suggests relevant code snippets starting from the developer's context. The system invokes different queries that consider both the parents of the class and the lexical visible type. Then, queries computed in such a way are passed to a module that mines relevant paths by relying on a graph-based structure. The ranking module eventually ranks the obtained snippets by employing six different heuristics.
% XSnippet's performances are evaluated by conducting a user study as well as a direct comparison with Prospector, another existing code assistant system. The results show that XSnippet success in recommending code snippets in terms of coverage and pertinence with respect to the developer's context.}

DeepAPI~\cite{Gu2016DeepAPI} is a deep-learning method used to generate API usage sequences given a query in natural language.
The learning problem is encoded as a machine translation problem, where queries are considered the source language and API sequences the target language.
Only commented methods are considered during the search.
The same authors~\cite{Gu2018DeepCode} present CODEnn (COde-Description Embedding Neural Network), where, instead of API sequences, code snippets are retrieved to the developer based on semantic aspects such as API sequences, comments, method names, and tokens.

With respect to the aforementioned approaches, FOCUS uses CF techniques to recommend API calls and usage patterns from a set of similar projects.
In the end, not only relevant API invocations are recommended, but also code snippets are returned to the developer as usage examples.

%\vspace{.1cm}
%\noindent
%$\blacksquare$~\textbf{Third-party library recommendation.} In a recent work~\cite{NGUYEN2019110460}, we presented CrossRec, a recommender system to assist open source software developers in selecting suitable third-party libraries. The system exploits a collaborative filtering technique to recommend libraries to developers by relying on the set of dependencies, which have been included in the project being developed. CrossRec makes use of a 2D ratings matrix to perform recommendation, while \FC employs a 3D context-aware ratings matrix. We suppose that more dimensions can be added to the rating matrix, in order to incorporate more input features. 

%. is a recommender system used to suggest API usage

%\subsection{API-Related Code Search Approaches}
\vspace{.1cm}
\noindent
$\blacksquare$~\textbf{API-related code search engines.} Strathcona~\cite{Holmes2005Using} is an Eclipse plug-in that extracts the structural context of code and uses it as a query to request API usages from a remote repository. The system performs the match by employing six heuristics associated to class inheritance, method calls, and field types. In a similar fashion, the technique proposed by \emph{Buse and Weimer}~\cite{Buse2012Synthesizing} synthesizes API examples for a given data type. An algorithm based on data-flow analysis, k-Medoids clustering and pattern abstraction is designed. Its outcome is a set of syntactically correct and well-typed code snippets where example length, exception handling, variables initialization and naming, abstract uses are considered.

MUSE (Method USage Examples) is an approach proposed by \emph{Moreno} \etal~\cite{Moreno:2015:IUT:2818754.2818860} to recommending code examples a given API method. MUSE extracts API usages from client code, simplifies code examples with static slicing, and detects clones to group similar snippets. It also ranks examples according to certain properties, \ie reusability,  understandability, and popularity). % and documents them

SWIM (Synthesizing What I Mean)~\cite{Raghothaman2016SWIM} seeks API structured call sequences (control and data-flows are considered), and then synthesizes API-related code snippets according to a query in natural language.
The underlying learning model is also built with the EM algorithm. 
Similarly, Raychev \etal~\cite{Raychev2014Code} propose a code completion approach based on natural language processing, which receives as input a partial program and outputs a set of API call sequences filling the gaps of the input. 
Both invocations and invocation arguments are synthesized considering multiple types of an API.

\emph{Thummalapenta and Xie} propose SpotWeb \cite{Thum2008a}, an approach that provides starting points (hotspots) for understanding a framework, and highlights where examples finding could be more challenging (coldspots). 
%McMillan \etal \cite{McMil2011a} propose Portfolio, a tool that finds relevant functions implementing high-level requirements. 
Other tools exploit StackOverflow discussions to suggest %context-specific 
code snippets and documentation~\cite{Cord2012a,Ponz2014b,PonzanelliSBMOP17,Rabman:2014, Rigb2013a,Taku2011a,TreudeR16}.

%that might be helpful to solve a particular development task.  Furthermore, recommender systems have been conceived to provide developers with suitable recommendations, 
	
	\vspace{-.2cm}
	\section{Conclusions}% and Future Work
	\label{sec:Conclusions}
	
We presented \FC, a recommender system to provide developers with suitable API function calls and code snippets while they are programming. 
%a context-aware collaborative-filtering system to assist developers in selecting suitable API function calls and usage patterns. 
A thorough evaluation has been conducted \emph{(i)} on an Android dataset to study the 
approach's performance, and \emph{(ii)} in a user study with \numParticipants participants to assess the perceived usefulness of \FC recommendations.

We succeeded in integrating \FC into the Eclipse IDE, 
and  we made available online the developed tool together with the parsed 
metadata~\cite{focus-zenodo}. This aims at providing the research community at 
large with a sound replication package, which then allows one to seamlessly 
reproduce the experiments presented in our paper. % The ultimate aim is to 
%contribute to the advancement of the research topic.

Future research in this area includes \emph{(i)} replicating the 
empirical evaluation on further projects, by also, possibly, supporting further 
programming languages, and \emph{(ii)} %evaluating the usefulness of \FC by conducting a user study. 
updating the code base of the Eclipse Scava project\footnote{\url{https://www.eclipse.org/scava/}} (which embraces 
all the development outcomes produced in the context of the EU CROSSMINER 
project) with the FOCUS tool as presented in this paper. 

\ifCLASSOPTIONcompsoc
  \section*{Acknowledgments}
\else
  \section*{Acknowledgment}
\fi

%The research described has been carried out as part of the CROSSMINER Project, which has received funding from the European Union's Horizon 2020 Research and Innovation Programme under grant agreement No. 732223. 

The research described has been carried out as part of the CROSSMINER Project, which has received funding from the European Union's Horizon 2020 Research and Innovation Programme under grant agreement No. 732223. We thank Gian Luca Scoccia for providing us with the tool to extract APK files from Apkpure. We also thank the students who kindly participated in the user study, despite difficulties caused by the unprecedented pandemic. Finally, we are grateful to the anonymous reviewers for their valuable comments and suggestions that helped us improve the paper.


\ifCLASSOPTIONcaptionsoff
  \newpage
\fi



%\bibliographystyle{IEEEtran}

%\begin{thebibliography}{1}
%\end{thebibliography}

\bibliographystyle{IEEEtranS}
\bibliography{main}

% Generated by IEEEtranS.bst, version: 1.14 (2015/08/26)
\begin{thebibliography}{10}
\providecommand{\url}[1]{#1}
\csname url@samestyle\endcsname
\providecommand{\newblock}{\relax}
\providecommand{\bibinfo}[2]{#2}
\providecommand{\BIBentrySTDinterwordspacing}{\spaceskip=0pt\relax}
\providecommand{\BIBentryALTinterwordstretchfactor}{4}
\providecommand{\BIBentryALTinterwordspacing}{\spaceskip=\fontdimen2\font plus
\BIBentryALTinterwordstretchfactor\fontdimen3\font minus
  \fontdimen4\font\relax}
\providecommand{\BIBforeignlanguage}[2]{{%
\expandafter\ifx\csname l@#1\endcsname\relax
\typeout{** WARNING: IEEEtranS.bst: No hyphenation pattern has been}%
\typeout{** loaded for the language `#1'. Using the pattern for}%
\typeout{** the default language instead.}%
\else
\language=\csname l@#1\endcsname
\fi
#2}}
\providecommand{\BIBdecl}{\relax}
\BIBdecl

\bibitem{dex2jar}
\BIBentryALTinterwordspacing
``\BIBforeignlanguage{en-US}{dex2jar},'' library Catalog: tools.kali.org.
  [Online]. Available: \url{https://tools.kali.org/reverse-engineering/dex2jar}
\BIBentrySTDinterwordspacing

\bibitem{rodin_aarssen_2017_891122}
\BIBentryALTinterwordspacing
R.~Aarssen, ``cwi-swat/clair: v0.1.0,'' Sep. 2017. [Online]. Available:
  \url{https://doi.org/10.5281/zenodo.891122}
\BIBentrySTDinterwordspacing

\bibitem{Acharya2007Mining}
M.~Acharya, T.~Xie, J.~Pei, and J.~Xu, ``{Mining API Patterns As Partial Orders
  from Source Code: From Usage Scenarios to Specifications},'' in \emph{6th
  Joint Meeting of the European Software Engineering Conference and the ACM
  SIGSOFT Symposium on The Foundations of Software Engineering}.\hskip 1em plus
  0.5em minus 0.4em\relax New York: ACM, 2007, pp. 25--34.

\bibitem{Basten2015M3}
B.~Basten, M.~Hills, P.~Klint, D.~Landman, A.~Shahi, M.~J. Steindorfer, and
  J.~J. Vinju, ``{M3: A General Model for Code Analytics in Rascal},'' in
  \emph{1st International Workshop on Software Analytics}.\hskip 1em plus 0.5em
  minus 0.4em\relax Piscataway: IEEE, 2015, pp. 25--28.

\bibitem{BASTEN20157}
B.~Basten, J.~[van~den Bos], M.~Hills, P.~Klint, A.~Lankamp, B.~Lisser,
  A.~[van~der Ploeg], T.~[van~der Storm], and J.~Vinju, ``Modular language
  implementation in rascal – experience report,'' \emph{Science of Computer
  Programming}, vol. 114, pp. 7 -- 19, 2015, lDTA (Language Descriptions,
  Tools, and Applications) Tool Challenge.

\bibitem{Buse2012Synthesizing}
R.~P.~L. Buse and W.~Weimer, ``{Synthesizing API Usage Examples},'' in
  \emph{34th International Conference on Software Engineering}.\hskip 1em plus
  0.5em minus 0.4em\relax Piscataway: IEEE, 2012, pp. 782--792.

\bibitem{CAPILUPPI2020106279}
A.~Capiluppi, D.~{Di Ruscio}, J.~{Di Rocco}, P.~T. Nguyen, and N.~Ajienka,
  ``Detecting java software similarities by using different clustering
  techniques,'' \emph{Information and Software Technology}, vol. 122, p.
  106279, 2020.

\bibitem{Chen:2005:CCF:2154509.2154540}
A.~Chen, ``{Context-Aware Collaborative Filtering System: Predicting the User's
  Preference in the Ubiquitous Computing Environment},'' in \emph{First
  International Conference on Location- and Context-Awareness}.\hskip 1em plus
  0.5em minus 0.4em\relax Berlin, Heidelberg: Springer, 2005, pp. 244--253.

\bibitem{Cord2012a}
J.~Cordeiro, B.~Antunes, and P.~Gomes, ``{Context-Based Recommendation to
  Support Problem Solving in Software Development},'' in \emph{Third
  International Workshop on Recommendation Systems for Software
  Engineering}.\hskip 1em plus 0.5em minus 0.4em\relax Piscataway: IEEE, 2012,
  pp. 85--89.

\bibitem{10.1145/3293882.3330571}
M.~Fazzini, Q.~Xin, and A.~Orso, ``Automated api-usage update for android
  apps,'' in \emph{Proceedings of the 28th ISSTA}, ser. ISSTA 2019.\hskip 1em
  plus 0.5em minus 0.4em\relax New York, NY, USA: Association for Computing
  Machinery, 2019, p. 204–215.

\bibitem{Fowkes:2016:PPA:2950290.2950319}
J.~Fowkes and C.~Sutton, ``{Parameter-free Probabilistic API Mining Across
  GitHub},'' in \emph{24th ACM SIGSOFT International Symposium on Foundations
  of Software Engineering}.\hskip 1em plus 0.5em minus 0.4em\relax New York:
  ACM, 2016, pp. 254--265.

\bibitem{8595172}
F.~{Geiger}, I.~{Malavolta}, L.~{Pascarella}, F.~{Palomba}, D.~{Di Nucci}, and
  A.~{Bacchelli}, ``A graph-based dataset of commit history of real-world
  android apps,'' in \emph{2018 IEEE/ACM 15th International Conference on
  Mining Software Repositories (MSR)}, 2018, pp. 30--33.

\bibitem{Cliff:2005}
R.~J. Grissom and J.~J. Kim, \emph{Effect sizes for research: A broad practical
  approach}, 2nd~ed.\hskip 1em plus 0.5em minus 0.4em\relax Lawrence Earlbaum
  Associates, 2005.

\bibitem{Gu2018DeepCode}
X.~Gu, H.~Zhang, and S.~Kim, ``{Deep Code Search},'' in \emph{40th
  International Conference on Software Engineering}.\hskip 1em plus 0.5em minus
  0.4em\relax New York: ACM, 2018, pp. 933--944.

\bibitem{Gu2016DeepAPI}
X.~Gu, H.~Zhang, D.~Zhang, and S.~Kim, ``{Deep API Learning},'' in \emph{24th
  ACM SIGSOFT International Symposium on Foundations of Software
  Engineering}.\hskip 1em plus 0.5em minus 0.4em\relax New York: ACM, 2016, pp.
  631--642.

\bibitem{hills2014php}
M.~Hills and P.~Klint, ``Php air: Analyzing php systems with rascal,'' in
  \emph{2014 Software Evolution Week-IEEE Conference on Software Maintenance,
  Reengineering, and Reverse Engineering (CSMR-WCRE)}.\hskip 1em plus 0.5em
  minus 0.4em\relax IEEE, 2014, pp. 454--457.

\bibitem{Holmes2005Using}
R.~Holmes and G.~C. Murphy, ``{Using Structural Context to Recommend Source
  Code Examples},'' in \emph{27th International Conference on Software
  Engineering}.\hskip 1em plus 0.5em minus 0.4em\relax New York: ACM, 2005, pp.
  117--125.

\bibitem{10.1145/3319535.3363255}
H.~Jang, B.~Jin, S.~Hyun, and H.~Kim, ``Kerberoid: A practical android app
  decompilation system with multiple decompilers,'' in \emph{Proceedings of the
  2019 ACM SIGSAC Conference on Computer and Communications Security}, ser. CCS
  '19.\hskip 1em plus 0.5em minus 0.4em\relax New York, NY, USA: Association
  for Computing Machinery, 2019, p. 2557–2559.

\bibitem{koch199980}
R.~Koch, \emph{The 80/20 Principle: The Secret of Achieving More with Less},
  ser. A Currency book.\hskip 1em plus 0.5em minus 0.4em\relax Doubleday, 1999.

\bibitem{Kohavi:1995:SCB:1643031.1643047}
R.~Kohavi, ``{A Study of Cross-validation and Bootstrap for Accuracy Estimation
  and Model Selection},'' in \emph{14th International Joint Conference on
  Artificial Intelligence}.\hskip 1em plus 0.5em minus 0.4em\relax San
  Francisco: Morgan Kaufmann Publishers Inc., 1995, pp. 1137--1143.

\bibitem{levenshtein1966bcc}
V.~Levenshtein, ``{Binary Codes Capable of Correcting Deletions, Insertions and
  Reversals},'' \emph{Soviet Physics Doklady}, vol.~10, p. 707, 1966.

\bibitem{McMillan:2012:DSS:2337223.2337267}
C.~McMillan, M.~Grechanik, and D.~Poshyvanyk, ``Detecting similar software
  applications,'' in \emph{Proceedings of the 34th International Conference on
  Software Engineering}, ser. ICSE '12.\hskip 1em plus 0.5em minus 0.4em\relax
  Piscataway, NJ, USA: IEEE Press, 2012, pp. 364--374.

\bibitem{Moreno:2015:IUT:2818754.2818860}
L.~Moreno, G.~Bavota, M.~Di~Penta, R.~Oliveto, and A.~Marcus, ``{How Can I Use
  This Method?}'' in \emph{37th International Conference on Software
  Engineering}.\hskip 1em plus 0.5em minus 0.4em\relax Piscataway: IEEE, 2015,
  pp. 880--890.

\bibitem{nasehi2012makes}
S.~M. Nasehi, J.~Sillito, F.~Maurer, and C.~Burns, ``{What Makes a Good Code
  Example?: A Study of Programming Q\&A in StackOverflow},'' in \emph{28th IEEE
  International Conference on Software Maintenance}.\hskip 1em plus 0.5em minus
  0.4em\relax Piscataway: IEEE, 2012, pp. 25--34.

\bibitem{NGUYEN2019110460}
P.~T. Nguyen, J.~{Di Rocco}, D.~{Di Ruscio}, and M.~{Di Penta}, ``{CrossRec:
  Supporting Software Developers by Recommending Third-party Libraries},''
  \emph{Journal of Systems and Software}, p. 110460, 2019.

\bibitem{Nguyen:2019:FRS:3339505.3339636}
P.~T. Nguyen, J.~Di~Rocco, D.~Di~Ruscio, L.~Ochoa, T.~Degueule, and
  M.~Di~Penta, ``{FOCUS: A Recommender System for Mining API Function Calls and
  Usage Patterns},'' in \emph{Proceedings of the 41st International Conference
  on Software Engineering}, ser. ICSE '19.\hskip 1em plus 0.5em minus
  0.4em\relax Piscataway, NJ, USA: IEEE Press, 2019, pp. 1050--1060.

\bibitem{focus-zenodo}
\BIBentryALTinterwordspacing
P.~T. Nguyen, J.~{Di Rocco}, C.~{Di Sipio}, D.~{Di Ruscio}, and M.~{Di Penta},
  ``{TSE FOCUS replication package},'' Jan. 2021. [Online]. Available:
  \url{https://doi.org/10.5281/zenodo.4415618}
\BIBentrySTDinterwordspacing

\bibitem{10.1007/978-3-319-60438-1_47}
P.~T. Nguyen, K.~Eckert, A.~Ragone, and T.~Di~Noia, ``{Modification to
  K-Medoids and CLARA for Effective Document Clustering},'' in
  \emph{Foundations of Intelligent Systems}.\hskip 1em plus 0.5em minus
  0.4em\relax Cham: Springer International Publishing, 2017, pp. 481--491.

\bibitem{Oppenheim:1992}
A.~N. Oppenheim, \emph{Questionnaire Design, Interviewing and Attitude
  Measurement}.\hskip 1em plus 0.5em minus 0.4em\relax Pinter Publishers, 1992.

\bibitem{Parnas1971Information}
D.~L. Parnas, ``{Information Distribution Aspects of Design Methodology},''
  Departement of Computer Science, Carnegie Mellon University, Pittsburgh,
  Tech. Rep., 1971.

\bibitem{Ponz2014b}
L.~Ponzanelli, G.~Bavota, M.~Di~Penta, R.~Oliveto, and M.~Lanza, ``{Mining
  StackOverflow to Turn the IDE into a Self-confident Programming Prompter},''
  in \emph{11th Working Conference on Mining Software Repositories}.\hskip 1em
  plus 0.5em minus 0.4em\relax New York: ACM, 2014, pp. 102--111.

\bibitem{PonzanelliSBMOP17}
L.~Ponzanelli, S.~Scalabrino, G.~Bavota, A.~Mocci, R.~Oliveto, M.~Di~Penta, and
  M.~Lanza, ``{Supporting Software Developers with a Holistic Recommender
  System},'' in \emph{39th International Conference on Software
  Engineering}.\hskip 1em plus 0.5em minus 0.4em\relax Piscataway: IEEE, 2017,
  pp. 94--105.

\bibitem{Raghothaman2016SWIM}
M.~Raghothaman, Y.~Wei, and Y.~Hamadi, ``{SWIM: Synthesizing What I Mean: Code
  Search and Idiomatic Snippet Synthesis},'' in \emph{38th International
  Conference on Software Engineering}.\hskip 1em plus 0.5em minus 0.4em\relax
  New York: ACM, 2016, pp. 357--367.

\bibitem{Rabman:2014}
M.~Rahman, S.~Yeasmin, and C.~Roy, ``{Towards a Context-Aware IDE-Based Meta
  Search Engine for Recommendation about Programming Errors and Exceptions},''
  in \emph{Conference on Software Maintenance, Reengineering, and Reverse
  Engineering}.\hskip 1em plus 0.5em minus 0.4em\relax Piscataway: IEEE, 2014,
  pp. 194--203.

\bibitem{Raychev2014Code}
V.~Raychev, M.~Vechev, and E.~Yahav, ``{Code Completion with Statistical
  Language Models},'' in \emph{35th ACM SIGPLAN Conference on Programming
  Language Design and Implementation}.\hskip 1em plus 0.5em minus 0.4em\relax
  New York: ACM, 2014, pp. 419--428.

\bibitem{Rigb2013a}
P.~C. Rigby and M.~P. Robillard, ``{Discovering Essential Code Elements in
  Informal Documentation},'' in \emph{35th International Conference on Software
  Engineering}.\hskip 1em plus 0.5em minus 0.4em\relax Piscataway: IEEE, 2013,
  pp. 832--841.

\bibitem{robillard2009makes}
M.~P. Robillard, ``{What Makes APIs Hard to Learn? Answers from Developers},''
  \emph{IEEE software}, vol.~26, no.~6, pp. 27--34, 2009.

\bibitem{Robillard:2013:AAP:2498733.2498776}
M.~P. Robillard, E.~Bodden, D.~Kawrykow, M.~Mezini, and T.~Ratchford,
  ``{Automated API Property Inference Techniques},'' \emph{IEEE Transactions on
  Software Engineering}, vol.~39, no.~5, pp. 613--637, 2013.

\bibitem{ruiz_understanding_2012}
I.~J.~M. Ruiz, M.~Nagappan, B.~Adams, and A.~E. Hassan,
  ``\BIBforeignlanguage{en}{Understanding reuse in the {Android} {Market}},''
  in \emph{\BIBforeignlanguage{en}{2012 20th {IEEE} {International}
  {Conference} on {Program} {Comprehension} ({ICPC})}}.\hskip 1em plus 0.5em
  minus 0.4em\relax Passau, Germany: IEEE, Jun. 2012, pp. 113--122.

\bibitem{10.1145/1167515.1167508}
N.~Sahavechaphan and K.~Claypool, ``Xsnippet: Mining for sample code,''
  \emph{SIGPLAN Not.}, vol.~41, no.~10, p. 413–430, Oct. 2006.

\bibitem{Saied2015Mining}
M.~A. Saied, O.~Benomar, H.~Abdeen, and H.~Sahraoui, ``{Mining Multi-level API
  Usage Patterns},'' in \emph{22nd International Conference on Software
  Analysis, Evolution, and Reengineering}.\hskip 1em plus 0.5em minus
  0.4em\relax Piscataway: IEEE, 2015, pp. 23--32.

\bibitem{Saied2015Could}
M.~A. Saied, H.~Abdeen, O.~Benomar, and H.~Sahraoui, ``{Could We Infer
  Unordered API Usage Patterns Only Using the Library Source Code?}'' in
  \emph{23rd International Conference on Program Comprehension}.\hskip 1em plus
  0.5em minus 0.4em\relax Piscataway: IEEE, 2015, pp. 71--81.

\bibitem{6181075}
B.~{Sanz}, I.~{Santos}, C.~{Laorden}, X.~{Ugarte-Pedrero}, and P.~G. {Bringas},
  ``On the automatic categorisation of android applications,'' in \emph{2012
  IEEE Consumer Communications and Networking Conference (CCNC)}, 2012, pp.
  149--153.

\bibitem{Schafer:2007:CFR:1768197.1768208}
J.~B. Schafer, D.~Frankowski, J.~Herlocker, and S.~Sen, \emph{Collaborative
  Filtering Recommender Systems}.\hskip 1em plus 0.5em minus 0.4em\relax
  Berlin, Heidelberg: Springer Berlin Heidelberg, 2007, pp. 291--324.

\bibitem{Schafer2007Collaborative}
------, ``{The Adaptive Web: Methods and Strategies of Web Personalization},''
  P.~Brusilovsky, A.~Kobsa, and W.~Nejdl, Eds.\hskip 1em plus 0.5em minus
  0.4em\relax Berlin, Heidelberg: Springer, 2007, ch. {Collaborative Filtering
  Recommender Systems}, pp. 291--324.

\bibitem{8543433}
G.~L. {Scoccia}, S.~{Ruberto}, I.~{Malavolta}, M.~{Autili}, and P.~{Inverardi},
  ``An investigation into android run-time permissions from the end users'
  perspective,'' in \emph{2018 IEEE/ACM 5th International Conference on Mobile
  Software Engineering and Systems (MOBILESoft)}, 2018, pp. 45--55.

\bibitem{Taku2011a}
W.~Takuya and H.~Masuhara, ``{A Spontaneous Code Recommendation Tool Based on
  Associative Search},'' in \emph{3rd International Workshop on Search-Driven
  Development: Users, Infrastructure, Tools, and Evaluation}.\hskip 1em plus
  0.5em minus 0.4em\relax New York: ACM, 2011, pp. 17--20.

\bibitem{Thum2008a}
S.~Thummalapenta and T.~Xie, ``{SpotWeb: Detecting Framework Hotspots and
  Coldspots via Mining Open Source Code on the Web},'' in \emph{23rd IEEE/ACM
  International Conference on Automated Software Engineering}.\hskip 1em plus
  0.5em minus 0.4em\relax Washington: IEEE, 2008, pp. 327--336.

\bibitem{10.1007/s10664-009-9108-x}
S.~Thummalapenta, L.~Cerulo, L.~Aversano, and M.~Di~Penta, ``An empirical study
  on the maintenance of source code clones,'' \emph{Empirical Softw. Engg.},
  vol.~15, no.~1, p. 1–34, Feb. 2010.

\bibitem{TreudeR16}
C.~Treude and M.~P. Robillard, ``{Augmenting API Documentation with Insights
  from Stack Overflow},'' in \emph{38th International Conference on Software
  Engineering}.\hskip 1em plus 0.5em minus 0.4em\relax New York: ACM, 2016, pp.
  392--403.

\bibitem{uddin2015api}
G.~Uddin and M.~P. Robillard, ``{How API Documentation Fails},'' \emph{IEEE
  Software}, vol.~32, no.~4, pp. 68--75, 2015.

\bibitem{10.1145/2645710.2645744}
S.~Vargas and P.~Castells, ``Improving sales diversity by recommending users to
  items,'' in \emph{Proceedings of the 8th ACM Conference on Recommender
  Systems}, ser. RecSys '14.\hskip 1em plus 0.5em minus 0.4em\relax New York,
  NY, USA: Association for Computing Machinery, 2014, p. 145–152.

\bibitem{viennot_measurement_2014}
N.~Viennot, E.~Garcia, and J.~Nieh, ``\BIBforeignlanguage{en}{A measurement
  study of google play},'' in \emph{\BIBforeignlanguage{en}{The 2014 {ACM}
  international conference on {Measurement} and modeling of computer systems -
  {SIGMETRICS} '14}}.\hskip 1em plus 0.5em minus 0.4em\relax Austin, Texas,
  USA: ACM Press, 2014, pp. 221--233.

\bibitem{Wang2013Mining}
J.~Wang, Y.~Dang, H.~Zhang, K.~Chen, T.~Xie, and D.~Zhang, ``{Mining Succinct
  and High-coverage API Usage Patterns from Source Code},'' in \emph{10th
  Working Conference on Mining Software Repositories}.\hskip 1em plus 0.5em
  minus 0.4em\relax Piscataway: IEEE, 2013, pp. 319--328.

\bibitem{wilcoxon}
F.~Wilcoxon, ``Individual comparisons by ranking methods,'' \emph{Biometrics
  Bulletin}, vol.~1, no.~6, pp. 80--83, 1945.

\bibitem{WONG20152839}
T.-T. Wong, ``{Performance Evaluation of Classification Algorithms by K-fold
  and Leave-one-out Cross Validation},'' \emph{Pattern Recognition}, vol.~48,
  no.~9, pp. 2839--2846, 2015.

\bibitem{10.1145/3180155.3182550}
\BIBentryALTinterwordspacing
H.~Zhong and N.~Meng, ``Towards reusing hints from past fixes: An exploratory
  study on thousands of real samples,'' in \emph{Proceedings of the 40th
  International Conference on Software Engineering}, ser. ICSE '18.\hskip 1em
  plus 0.5em minus 0.4em\relax New York, NY, USA: Association for Computing
  Machinery, 2018, p. 885. [Online]. Available:
  \url{https://doi.org/10.1145/3180155.3182550}
\BIBentrySTDinterwordspacing

\bibitem{Zhong2009MAPO}
H.~Zhong, T.~Xie, L.~Zhang, J.~Pei, and H.~Mei, ``{MAPO: Mining and
  Recommending API Usage Patterns},'' in \emph{23rd European Conference on
  Object-Oriented Programming}.\hskip 1em plus 0.5em minus 0.4em\relax Berlin,
  Heidelberg: Springer, 2009, pp. 318--343.

\bibitem{8952168}
S.~{Zhou}, B.~{Shen}, and H.~{Zhong}, ``Lancer: Your code tell me what you
  need,'' in \emph{2019 34th IEEE/ACM International Conference on Automated
  Software Engineering (ASE)}, 2019, pp. 1202--1205.

\end{thebibliography}

\balance
%\bibliography{TSE-MetamodelClassification}


%\end{IEEEbiography}


%\newpage



%\vfill


\end{document}